\documentclass[a4paper,11pt]{article}
\pdfoutput=1
\usepackage{jheppub}
\usepackage[T1]{fontenc}
\usepackage{verbatim}   
\usepackage{color}      
\usepackage{subfigure}  
\usepackage{multirow}
\usepackage{dsfont}
\usepackage{comment}
\usepackage{pdflscape}
\usepackage{rotating}
\usepackage{feynmp} 
\usepackage{float}	

\begin{document}

\title{Higgs production in association with a top-antitop pair in the Standard Model Effective Field Theory at NLO in QCD}

\author[a]{Fabio Maltoni,}
\author[a]{Eleni Vryonidou}
\author[b]{and Cen Zhang}
\affiliation[a]{Centre for Cosmology, Particle Physics and Phenomenology (CP3),\\
	Universit\'e catholique de Louvain, B-1348 Louvain-la-Neuve, Belgium} 
\affiliation[b]{Department of Physics, Brookhaven National Laboratory, Upton, NY 11973, USA}

\emailAdd{fabio.maltoni@uclouvain.be}
\emailAdd{eleni.vryonidou@uclouvain.be}
\emailAdd{cenzhang@bnl.gov}
\preprint{CP3-16-39, MCnet-16-29}
\abstract{
We present the results of the computation of the next-to-leading order QCD corrections to the  production  cross section of a Higgs
boson in association with a top-antitop pair at the LHC, including the three relevant dimension-six operators ($O_{t \varphi }, O_{\varphi G}, O_{tG}$) of 
the standard model effective field theory.  These operators also contribute to the production of Higgs bosons in loop-induced processes at the LHC, 
such as inclusive Higgs, $Hj$ and $HH$ production, and modify the Higgs decay branching ratios for which we also provide predictions. We perform a detailed study of the cross sections and their uncertainties at the total as well as differential level and of the structure of the effective field theory at NLO including renormalisation group effects. Finally, we show how the combination of information coming from measurements of these production processes will allow to constrain the three operators at the current and future LHC runs. Our results lead to a significant improvement of the accuracy and precision of the  deviations expected from higher-dimensional operators in the SM in both the top-quark and the Higgs-boson sectors and provide a necessary ingredient for performing a global EFT fit to the LHC data at NLO accuracy. }
\maketitle

\section{Introduction}

The top quark and the Higgs boson are the two heaviest elementary particles
known to date.  Their interaction is expected to reveal evidence
of beyond the standard model (SM) physics and possibly determine the fate of
the universe \cite{Buttazzo:2013uya}.
The LHC provides us the first chance to measure the interactions of these
two particles through the associated production of a Higgs with a single top quark \cite{Maltoni:2001hu,Farina:2012xp,Demartin:2015uha} or a top quark pair \cite{Ellis:2013yxa,Khatibi:2014bsa,Demartin:2014fia,Kolodziej:2015qsa,Buckley:2015vsa,
Plehn:2015cta,Moretti:2015vaa,Cao:2016wib,Cirigliano:2016nyn,Gritsan:2016hjl}.  At the LHC, precise measurements require accurate input from the theory side.  Predictions for the $t\bar tH$ process 
in the SM are known at next-to-leading order (NLO) in QCD
\cite{Beenakker:2001rj,Beenakker:2002nc,Reina:2001sf,Reina:2001bc,Dawson:2002tg,Dawson:2003zu,Frederix:2011zi,Garzelli:2011vp},
with off-shell effects \cite{Denner:2015yca}, and at NLO in electroweak
\cite{Yu:2014cka,Frixione:2014qaa,Frixione:2015zaa,Hartanto:2015uka}.
Next-to-leading logarithmic matched to NLO \cite{Kulesza:2015vda} as well as
approximate next-to-next-to-leading order QCD predictions
\cite{Broggio:2015lya} have also become available recently.

In addition to the SM prediction, we expect that a precise understanding of the patterns of deviations from the SM will become equally
important at the LHC Run II, given the complicated nature of the corresponding
measurements
\cite{Khachatryan:2014qaa,Aad:2014lma,Aad:2015iha,Aad:2015gra,CMS-PAS-HIG-15-008,CMS-PAS-HIG-16-004},
and the attainable precision on the top-quark couplings expected for Run II
measurements \cite{atlasprojection}.  A powerful and predictive framework to analyse 
possible deviations is provided by the SM effective field theory (SMEFT)
\cite{Weinberg:1978kz,Buchmuller:1985jz,Leung:1984ni}, i.e.~the SM augmented by
higher-dimensional operators. In this framework, radiative corrections to the
SM deviations can be consistently included in a model-independent way, thus
allowing for systematically improving the predictions.
In fact, NLO corrections for a set of top-quark processes have recently started to become
available in such a framework, including top-quark decay processes,
flavor-changing neutral production, top-pair production, single-top production,
and $t\bar t$ associated production with a $Z$-boson and with a photon
\cite{Zhang:2013xya,Zhang:2014rja,Degrande:2014tta,Franzosi:2015osa,Zhang:2016omx,Bylund:2016phk,Rontsch:2014cca,Rontsch:2015una}.
Several Higgs decay results have also become available recently
\cite{Hartmann:2015oia,Ghezzi:2015vva,Hartmann:2015aia,Gauld:2015lmb}.

\begin{figure}[t!b]
	\begin{center}
		\includegraphics[width=.41\linewidth]{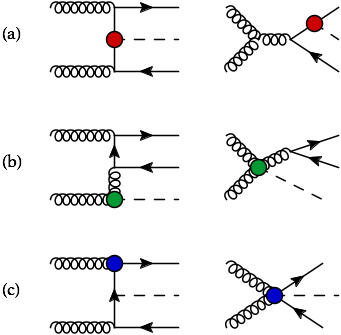}
	\end{center}
	\caption{Example diagrams for $t\bar tH$ production. The inserted operators are:
		(a) $O_{t\varphi}$  (b) $O_{\varphi G}$  (c) $O_{tG}$.}
	\label{fig:tthdiagram}
\end{figure}

\begin{figure}[t!b]
	\begin{center}
		\includegraphics[width=.7\linewidth]{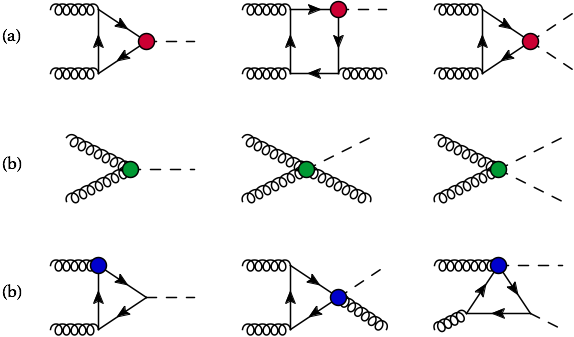}
	\end{center}
	\caption{Example diagrams for $H$, $Hj$, and $HH$ production. The inserted operators are:
		(a) $O_{t\varphi}$ (b) $O_{\varphi G}$ (c) $O_{tG}$. LO contributions from
		$O_{\varphi G}$ are at the tree level, while those from the
	other two operators are induced by a top-quark loop.}
	\label{fig:loopdiagram}
\end{figure}

The goal of this work is to improve the predictions of such deviations 
in $t\bar tH$ production in SMEFT by computing
the NLO QCD corrections.  Besides, we will also present SMEFT results for
processes that are top-loop induced in the SM, such as $p p \to H$, $p p \to
Hj$ and Higgs pair production $p p \to HH$. Selected Feynman diagrams at the
leading order (LO) are shown in Figure~\ref{fig:tthdiagram} and
\ref{fig:loopdiagram} for the $t\bar tH$ and loop-induced processes,
respectively.  The relevant effective operators in these processes, i.e.~those
modifying $ttH$, $ttg$, and $ggH$ vertices, are both physically interesting and
practically important, because they connect the top-quark sector with the
Higgs-boson sector in the SMEFT at dimension-six.  Studying these processes and
interactions will allow us to investigate how much we can learn about the
top quark from Higgs measurements, and vice versa.  In particular, the
chromo-magnetic dipole operator $O_{tG}$, which gives rise to a dipole
interaction in the $gtt$ vertex and introduces $ggtt$, $gttH$, and $ggttH$
vertices, is often left out in Higgs operator analyses (see, for example,
\cite{Corbett:2012dm,Corbett:2012ja,Masso:2012eq,Dumont:2013wma,Banerjee:2013apa,
Ellis:2014jta,Edezhath:2015lga,Corbett:2015ksa,Butter:2016cvz}), because it is often 
considered as part of top-quark measurements.  Here we will show that the
current $t\bar tH$ and $ p p \to H$ measurements already provide useful
information about the chromo-dipole moment, comparable to what we can learn
from top-pair production, and that future measurements will improve the limits.  
This observation implies that Higgs measurements are becoming
sensitive to this interaction and therefore it should not be neglected.
Furthermore,  the extraction of the Higgs self-coupling from $pp\to HH$
measurements relies on a precise knowledge of the top-Higgs interactions.  Here
we compute for the first time the contribution from the chromo-dipole moment
$O_{tG}$ to Higgs pair production. As it will be shown in the following,  this
operator gives a large contribution to this process, even taking into account
the current constraints  from $t\bar t$ production on the size of its
coefficient.

Let us briefly discuss the motivations for having NLO SMEFT predictions. 

\begin{itemize}
	\item First, the impact of QCD corrections on the central values, which can be
	 conventionally estimated by a $K$ factor (the ratio of NLO central prediction to
		LO), is often large at the LHC, and for an inclusive measurement this will
		improve the exclusion limit on the effective operators.  In
		addition, NLO corrections improve not only the accuracy of the
		predictions by modifying the central value, but also the
		precision by reducing the theoretical uncertainties due to
		missing higher-order corrections, which leads to a further
		improvement on the limits.  For example, the current limit on the
		chromo-magnetic dipole operator from $t\bar t$ is improved
		by a factor of 1.5 by including QCD corrections
		\cite{Franzosi:2015osa}, and the effects are even larger in the
		flavor-changing neutral sector
		\cite{Degrande:2014tta,Liu:2005dp,Gao:2009rf,Zhang:2011gh,Li:2011ek,Wang:2012gp}. 

	\item Second, QCD corrections often change the distributions due to
		effective operators in a nontrivial way, not captured by the LO
		scale uncertainty.  As the distribution measurements start to
		play an important role in the EFT global analyses both in the
		top-quark and in the Higgs sector
		\cite{Buckley:2015nca,Buckley:2015lku,
		Ellis:2014jta,Edezhath:2015lga,Corbett:2015ksa,Butter:2016cvz},
		reliable predictions for the distributions are needed as theory
		inputs.  In fact, Ref.~\cite{Zhang:2016omx} has shown that in
		an operator fit, missing QCD corrections to the shapes could
		lead to a biased interpretation in terms of new physics models.
		For this reason our final goal is to use the NLO predictions in a
		global EFT fit, including differential measurements, to extract
		maximal information on the operator coefficients.  

	\item Third, unlike in the SM where all the gauge couplings are known,
		the SMEFT has many operator coefficients, and several of them
		remain to be constrained. Higher order effects are important in
		that respect as they can be enhanced by the ratio of two
		operator coefficients, $C_2/C_1$, if operator $O_1$ contributes
		at the tree level, while $O_2$ at the loop level, and if $C_2$
		is loosely constrained.  An example is $gg\to H$, where at the
		tree level only the operator $O_{\phi G}$ (corresponding to a
		$ggH$ vertex) enters, but at the loop level additional
		operators, for example the $O_{t\phi}$ which shifts the SM top
		Yukawa coupling, will come in through a top loop.  The process
		can be then used to determine the top Yukawa through a top-loop
		effect.  Similar loop-induced scenarios have been used to
		obtain information on many other operators (see, for example,
		\cite{Degrande:2012gr,Greiner:2011tt,Zhang:2012cd,
		Mebane:2013cra,Mebane:2013zga,deBlas:2015aea,Elias-Miro:2013eta}).
		Even though this is not particularly relevant in the $t\bar tH$
		process, the same tree/loop degeneracy in the $ggH$ and the
		$ttH$ vertices will occur, and having NLO corrections allow us
		to control these effects. 

	\item Fourth, NLO is important to understand the structure of the
		effective theory, mainly because of the renormalisation group
		(RG) and operator mixing effects
		\cite{Jenkins:2013zja,Jenkins:2013wua,Alonso:2013hga}, and the
		corresponding theoretical uncertainties due to missing
		higher-order terms, which we will discuss in this work.  At NLO
		we start to have control on these effects. Furthermore, it is
		often the case that new operators enter when certain processes
		are considered at NLO. 

	\item Finally, in the future, the sensitivity of measurements to the
		effective deviations may be improved, by making use of the
		accurate EFT predictions and designing optimised experimental
		strategies in a top-down way.  However, given that any  
		useful predictions at the LHC must be at least at NLO, possibly with parton shower
		(PS) simulation, this improvement will be difficult without a
		consistent EFT at NLO prediction. 
\end{itemize}

As discussed also in our previous work, an important feature of our computation is that NLO
predictions matched to the parton shower can be obtained in an automatic way.
The results we will provide are important not only because predictions are
improved in accuracy and in precision, but also because NLO+PS event generation
can be directly used in an experimental simulation, allowing for a more
dedicated investigation of potential deviations, with possibly optimised
selections and improved sensitivities to  EFT signals.  Our approach is based
on the {\sc MadGraph5\_aMC@NLO} ({\sc MG5\_aMC}) framework
\cite{Alwall:2014hca}, and is part of the ongoing efforts of automating NLO EFT
simulations for colliders~\cite{Zhang:2016snc}.    

This paper is organised as follows. In Section \ref{sec:operators}, we discuss
the operators used in this work, while in Section \ref{sec:setup} we present
our calculation setup. Section \ref{sec:unc} is dedicated to a discussion of
the relevant theoretical uncertainties. Section \ref{sec:results} presents our
results for the total cross sections and differential distributions, and in
Section \ref{sec:rg} we discuss RG effects relevant for our study. Finally, in
Section \ref{sec:fit} we obtain constraints on the dimension-six operators using
LHC results before we conclude in Section \ref{sec:conc}.

\section{Effective operators}
\label{sec:operators}
In an EFT approach, potential deviations from the SM
can be consistently and systematically described by adding
higher-dimensional operators of the SM fields.  By employing global
analyses~\cite{Durieux:2014xla,Buckley:2015nca,Buckley:2015lku}, experimental
results can be used to determine the size of the deviations due to each
effective operator.  The established deviations can then be consistently
evolved up to high scales, and matched to possible new physics scenarios.  In
the absence of convincing evidence for new resonance states\footnote{A generic EFT 
might be useful also in the presence of new resonances, see for example
\cite{Gripaios:2016mmi}.}, the SMEFT provides the most model-independent
approach to a global interpretation of measurements.

The Lagrangian of SMEFT can be written as
\begin{equation} 
	\mathcal{L}_\mathrm{EFT}=\mathcal{L}_\mathrm{SM}+
	\sum_i\frac{C_{i}}{\Lambda^2}O_{i}+\mathcal{O}(\Lambda^{-4}) +h.c. ,
\end{equation} 
where $\Lambda$ is the scale of new physics.
In the $t\bar tH$ production there are two kinds of relevant operators: those with
four fermion fields and those with two or no fermion fields.  The main focus of this work is on the
second kind, including the following three operators:
\begin{flalign}
	&O_{t\phi} = y_t^3 \left( \phi^\dagger\phi \right)\left( \bar Qt \right)
	\tilde\phi \,,
	\\
	&O_{\phi G} = y_t^2 \left( \phi^\dagger\phi \right) G_{\mu\nu}^A G^{A\mu\nu}\,,
        \\
	&O_{tG} = y_t g_s (\bar Q\sigma^{\mu\nu}T^A t)\tilde\phi G_{\mu\nu}^A\,.
\end{flalign}
Our convention is such that for each operator we will add its Hermitian conjugate, 
even if it is already Hermitian.  We assume $C_{t\phi}$ and $C_{tG}$ are real,
so that the Lagrangian respects the CP symmetry. $C_{\phi G}$ is always real
because the operator is Hermitian. We do not include the CP-odd $O_{\phi \tilde{G}}$ operator. 
In this work we focus on these three operators\footnote{
	The operator $O_{\varphi\Box}=\left( \varphi^\dagger\varphi \right)\Box
	\left( \varphi^\dagger\varphi \right)$ also contributes by universally 
	rescaling the Higgs couplings.  We omit it here because the NLO
	corrections for this operator would be the same as in the SM.}.  The first one rescales the
top Yukawa coupling in the SM, and also gives rise to a new $ttHH$ coupling
which contributes to Higgs pair production.  The third one represents the
chromo-dipole moment of the top quark.  It modifies the $gtt$ vertex in the
SM and produces new four-point vertices, $ggtt$ and $gttH$, as well as a five-point
$ggttH$ vertex.  The second one is a
loop-induced interaction between the gluon and Higgs fields.  Even though it
does not involve a top-quark field explicitly, it needs to be included for
consistency, in particular at NLO, because the $O_{tG}$ mixes into this
operator, and this operator in addition mixes into $O_{t\phi}$, through RG
running.  All three operators contribute to the $t\bar tH$ process at the tree
level.  In addition, they can also be probed by $H$, $Hj$, and $HH$ production
processes, where $O_{\phi G}$ contributes at the tree level while the other two
contribute at the loop level.  In particular, the degeneracy in $gg\to H$
between the tree/loop-level contributions from $O_{\phi G}$ and $O_{t\phi}$ has
been discussed in the context of $Hj$
\cite{Grojean:2013nya,Buschmann:2014sia,Banfi:2013yoa} and $HH$ production
\cite{Cao:2015oaa}.  We note here that the Higgs pair production involves
additional operators that modify the Higgs self coupling. We do not include
those here as the QCD part factorises and their effect has been extensively
discussed in the literature
\cite{Contino:2012xk,Goertz:2014qta,Azatov:2015oxa,Grober:2015cwa}.  Rather, we
focus on the top-quark operators, in particular $O_{tG}$, which has not been
considered in previous SMEFT analyses.

We normalise these operators so that their tree level contributions to the
$pp\to t\bar tH$ cross section are of the same order [$O(\alpha_s^2 \alpha)
$]\footnote{This is not the case for single Higgs production for which $O_{\phi
G}$ enters at $O( \alpha) $, while $O_{t\phi}$ and $O_{tG}$ enter at
$O(\alpha_s^2 \alpha) $.}.  Then we can define the ``NLO QCD'' corrections,
i.e.~higher-order contributions at $O(\alpha_s^3 \alpha) $.  In addition,
with this parametrisation, the relevant mixing terms are always
$\mathcal{O}(\alpha_s)$.  Had we chosen a different normalisation of operators,
this would not be true.

We briefly discuss the mixing structure of these operators.  The complete RG
structure has been given in
\cite{Jenkins:2013zja,Jenkins:2013wua,Alonso:2013hga}.  In this work we will
consider the QCD induced mixing, which is relevant for our calculation,
i.e.~$\mathcal{O}(\alpha_s)$ terms with our normalisation.  The mixing matrix
for $(O_{t\phi},O_{\phi G},O_{tG})$ has a triangle form:
\begin{equation}
	\frac{dC_i(\mu)}{d\log\mu}=\frac{\alpha_s}{\pi}\gamma_{ij}C_j(\mu),\quad
	\gamma= \left(
	\begin{array}{ccc}
		-2 & 16 & 8 \\
		0  & -7/2 & 1/2 \\
		0 & 0 & 1/3
	\end{array}
\right)\,.
\label{eq:rg}
\end{equation}
These operators involve three, two and one Higgs fields, respectively.  In
$t\bar tH$ production with only QCD corrections, only one of the Higgs fields
can be dynamic, so in this sense their ``dimensions'' are 4, 5, 6 respectively.
The triangle form of the matrix implies that only a ``higher-dimensional'' operator
can mix into a ``lower-dimensional one'', i.e. $O_{tG}$ mixes into
$O_{\phi G}$, and both of them mix into $O_{t \phi}$, but not the other way
around.

Four-fermion operators that contribute to top-pair production
would also play a role in this process.  However the $t\bar t$ cross section
measurement is sufficient to constrain this effect.  
There are four linear combinations of operators that enter, defined
as $C_{u,d}^{1,2}$ in Ref.~\cite{Zhang:2010dr}.
Using the notations of Ref.~\cite{Grzadkowski:2010es},
they can be written as
\begin{flalign}
	&C_u^1=C_{qq}^{(1)1331}+C_{uu}^{1331}+C_{qq}^{(3)1331},
	\\
	&C_u^2=C_{qu}^{(8)1133}+C_{qu}^{(8)3311},
	\\
	&C_d^1=C_{qq}^{(3)1331}+\frac{1}{4}C_{ud}^{(8)3311},
	\\
	&C_d^2=C_{qu}^{(8)1133}+C_{qd}^{(8)3311}\,.
\end{flalign}
Consider $\sigma_{t\bar t}$ and $\sigma_{t\bar tH}$ at 8 TeV (at LO):
\begin{flalign}
	&\sigma_{t\bar t}^{8TeV}[pb]=158\left[1+0.0101\left( C_u^1+C_u^2+0.64C_d^1+0.64C_d^2\right)+0.65 C_{tG} \right]
	\\
	&\sigma_{t\bar tH}^{8TeV}[pb]=0.110\left[1+0.055\left(
	C_u^1+C_u^2+0.61C_d^1+0.61C_d^2\right)+2.02 C_{tG} \right]\,,
	\label{eq:sens}
\end{flalign}
 assuming
$\Lambda=1$ TeV.  We can see that approximately only one linear combination of
the four-fermion operator coefficients, i.e.~$C_4 \simeq
C_u^1+C_u^2+0.6C_d^1+0.6C_d^2$ enters both cross
sections, a pattern we do not expect to change at NLO. This behaviour
 is expected because the Higgs only couples to the top quark
with the same coupling regardless of which operator triggers the process,
so adding one Higgs does not resolve the degeneracy between four-fermion operators.
It does, however, increase the relative sensitivity to these operators,
because having a Higgs boson in the final state largely increases the center of
mass energy of the process, which in turn increases the relative contribution
from the four-fermion operators.  On the other hand, the $C_{tG}$ contribution
is independent of $C_4$, and this is because the Higgs particle
can be emitted not only from the top quark but also from
the operator $O_{tG}$ with a four-point contact $gttH$ or a five-point $ggttH$ vertex,
leading to a different topology than the four-fermion operator cases.  Since we are
going to focus on the $t\bar tH$ process, as a first approximation we assume
that the $t\bar t$ measurement is already constraining $C_4$. A complete 
treatment at NLO including four-fermion operators requires a dedicated study which we
defer to future work.
  Alternatively,
one could make use of the ratio of the two cross sections:
\begin{flalign}
	\frac{\sigma_{tth}^{8TeV}}{\sigma_{tt}^{8TeV}(\sqrt{\hat{s}}>1000\mathrm{GeV})}
	=0.047 [1+1.43 C_{tG}+0.0016(C_u^1+C_u^2)+0.0028(C_d^1+C_d^2)],
\end{flalign}
which will decrease the dependence on the four-fermion operators by one order of
magnitude. (A similar ratio has been proposed for $ttZ$ \cite{Schulze:2016qas}.)
In any case, the fact that four-fermion operators enter as one linear combination in $t\bar t$
 and $t\bar t H$ will allow us to focus on $O_{tG}$, $O_{t\phi}$ and $O_{\phi G}$ in the $t\bar tH$
process.  We should also mention that, by making use of Tevatron data and
asymmetry measurements, all the relevant four-fermion operators can in principle
be constrained from $t\bar t$ measurements
\cite{Degrande:2012gr,Buckley:2015nca,Buckley:2015lku,Rosello:2015sck}.

The final goal of top-quark measurements is to construct a global fit, where
by combining $t\bar t$ and $t\bar tH$ processes we will be able to set bounds
on $C_4$ and the other coefficients separately.  Intuitively, however, it is 
useful to have an approximate understanding of which operators are constrained
by which process.  $t\bar t$ has often been thought of as the key process to
bound $O_{tG}$, but what we are proposing here is to use $t\bar t$ to constrain
four-fermion operators, while $O_{tG}$ will be better constrained by $t\bar tH$
together with the other two operators, $O_{t\phi}$ and $O_{\phi G}$.  As we
will see in Section \ref{sec:fit}, even though this is not exactly the case with the
current measurements, it is likely to happen in the future for high luminosity 
LHC (HL-LHC), because the $t\bar tH$ process is more sensitive to $O_{tG}$ than the
$t\bar t$ process, as can be seen from Eq.~(\ref{eq:sens}), and its measurement
has more room to improve.

Finally, another operator that contributes to the processes is 
$O_G=g_sf^{ABC}G_\mu^{A\nu}G_\nu^{B\rho}G_\rho^{C\mu}$,
which would enter by modifying the gluon self-interactions. As this is not a
top-quark operator, we will not consider it further here, assuming also that
its contribution is sufficiently suppressed due to constraints from the
accurately measured $t\bar t$ and dijet cross sections.
		
\section{Calculation setup}
\label{sec:setup}
Our computation is performed within the {\sc MG5\_aMC} framework
\cite{Alwall:2014hca}, as described in
\cite{Franzosi:2015osa,Zhang:2016omx,Bylund:2016phk}.  All the elements
entering the NLO computations are available automatically starting from the
SMEFT Lagrangian \cite{Alloul:2013bka, Degrande:2011ua,Degrande:2014vpa,
Hirschi:2011pa,Frederix:2009yq,Hirschi:2015iia}.  NLO results can be matched to
parton shower programs, such as  {\sc PYTHIA8} \cite{Sjostrand:2014zea} and
{\sc HERWIG++} \cite{Bahr:2008pv},  through the {\sc MC@NLO}
\cite{Frixione:2002ik} formalism. 

Special care needs to be taken for the UV counterterms, which are
required for the virtual corrections.
With $\overline{MS}$ the coefficient renormalisation is
\begin{flalign}
	C^0_i=Z_{ij}C_j,\qquad
	\delta
	Z_{ij}=\frac{\alpha_s}{2\pi}\Delta(\mu_{EFT})\frac{1}{\epsilon}\gamma_{ij}\,,
	\label{eq:zij}
\end{flalign}
where $\mu_{EFT}$ is the scale at which we define the EFT, and
\begin{equation}
  \Delta(x)\equiv \Gamma(1+\epsilon)\left( \frac{4\pi\mu^2}{x^2}
  \right)^\epsilon\,.
\end{equation}
The $\Delta(\mu_{EFT})$ is inserted to separate the running of operator coefficients
from the running of $\alpha_s$.  In other words, these coefficients are always
defined at scale $\mu_{EFT}$.  They run and mix whenever $\mu_{EFT}$ changes.

The $\delta Z_{ij}$ and fields/coupling renormalisations
determine all the counterterms.
However, in practice, we define the $O_{t\phi}$ and $O_{\phi G}$ operators as
\begin{flalign}
	&\bar O_{t\phi} = y_t^3 \left( \phi^\dagger\phi
	-v^2/2\right)\left( \bar Qt \right) \tilde\phi \\ 
	&\bar O_{\phi G} = y_t^2 \left(
	\phi^\dagger\phi -v^2/2\right) G_{\mu\nu}^A G^{A\mu\nu}\,,
\end{flalign}
to remove the dimension-four terms due to Higgs fields taking vevs,
so that there is no need to redefine the SM fields and masses etc..
In principle this does not change any physics; in practice the
UV counterterms need to be treated slightly differently.  For example,
$O_{tG}$ mixes into $O_{\phi G}$, meaning that 
at one loop $O_{tG}$ needs the following counterterms to cancel the divergence
\begin{flalign}
	\frac{1}{4}\frac{\alpha_s}{\pi}\frac{1}{\epsilon}
	y_t^2\left( \phi^\dagger\phi \right)G_{\mu\nu}^AG^{A\mu\nu}
	+h.c.=\frac{\alpha_s}{4\pi}\frac{1}{\epsilon}
	\left[ \bar O_{\phi G}+m_t^2G_{\mu\nu}^AG^{A\mu\nu}+h.c. \right]\ .
\end{flalign}
With our redefinition, $\delta Z_{\phi G,tG}$ will only produce the first
counterterm.  The second term is of dimension four, corresponding to a
modification to the SM counterterms for the gluon field and $g_s$.
These are, however, only the divergent part.  Finite pieces need to be added, for the
gluon fields to be renormalised on-shell and for $g_s$ to be subtracted at the
zero-momentum transfer:
\begin{flalign}
  &\delta Z_2^{(g)}=\delta Z_{2,SM}^{(g)}-C_{tG}\frac{2\alpha_s m_t^2}{\pi\Lambda^2}
  \Delta(m_t) \frac{1}{\epsilon_{UV}}\ ,
  \label{eq:ct1}
  \\
  &\delta Z_{g_s}=\delta Z_{g_s,SM}+C_{tG}\frac{\alpha_s m_t^2}{\pi\Lambda^2}
  \Delta(m_t) \frac{1}{\epsilon_{UV}}\ .
\end{flalign}
Similarly the top-quark field and top-Yukawa renormalisation are related
to $O_{tG}\to O_{t\phi}$ mixing:
	\begin{flalign}
  &\delta Z_{m_t}=\delta Z_{y_t} = \delta Z_{m_t,SM} - C_{tG}\frac{4\alpha_s
  m_t^2}{\pi\Lambda^2} \Delta(m_t) \left(
  \frac{1}{\epsilon_{UV}}+\frac{1}{3} \right)\ ,
  \\
	  &\delta Z_2^{(t)}=\delta Z_{2,SM}^{(t)}-C_{tG}\frac{2\alpha_s
	  m_t^2}{\pi\Lambda^2} \Delta(m_t) \left(
	  \frac{1}{\epsilon_{UV}}+\frac{1}{3} \right)\ .
  \label{eq:ct2}
	\end{flalign}
In summary, we calculate UV counterterms by using $\delta Z_{ij}$ in
Eq.~(\ref{eq:zij}) for $O_{tG}$, $\bar O_{\phi G}$, and $\bar O_{t\phi}$, while
modifying the SM counterterms as described in Eqs.~(\ref{eq:ct1}-\ref{eq:ct2}).
In the following sections we will simply denote $\bar O$ by $O$.

\section{Theoretical uncertainties}
\label{sec:unc}
One of the motivations to perform a NLO computation is that this provides the
possibility to assess reliably the size of the residual uncertainties
from missing higher orders. Discussions about the impact and the methods to estimate theoretical
uncertainties in the SMEFT can be found in
Refs.~\cite{Berthier:2015oma,Berthier:2015gja,Passarino:2138031,Contino:2016jqw,Contino:2137947}.
Here, before moving on to numerical results, we briefly discuss how
theoretical uncertainties are treated in our calculation.

The SMEFT calculation is an expansion in two classes of parameters, the
perturbative expansion parameters, such as $\alpha_s$ and $\alpha$, and the EFT
expansion parameter $C/\Lambda^2$, and so there are two main types of
theoretical uncertainties, those related to missing higher orders in the gauge
couplings and those from higher-order terms in the $C/\Lambda^2$ expansion. In the
former class, we can list:

\begin{itemize}
	\item Uncertainties due to parton-distribution functions. 
	
	This type of uncertainty is also present in the SM calculations and can
	be treated in the same way, i.e. by following the procedures associated
	with the corresponding PDF sets.

	\item Uncertainties due to missing higher orders in the $\alpha_s$
		expansion as in the SM. 
		
	These are typically estimated by varying the
	renormalisation and factorisation scales, as in SM calculations.  
	All results presented in this work are provided with
	uncertainties that are estimated by varying these two scales
	independently by a factor of two up and down from the central value. 
	 The uncertainties due to higher orders in $\alpha$
	are expected to be subdominant.
	
	\item Uncertainties due to missing higher orders in the $\alpha_s$
		expansion of the EFT operators. 
		
	These are additional uncertainties, related to the scale
	$\mu_{EFT}$, at which
	the operators are defined. This uncertainty characterises the uncancelled logarithmic terms in the renormalisation
	group running and mixing of the operators. They can be obtained as
	follows: consider the cross section obtained at the central scale,
	$\mu_{EFT}=\mu_0$
\begin{flalign}
	\sigma(\mu_0)=\sigma_{SM}+\sum_i\frac{1{\rm TeV}^2}{\Lambda^2}C_i(\mu_0)
	\sigma_i(\mu_0) +\sum_{i,j} \frac{1{\rm TeV}^4}{\Lambda^4}
	C_i(\mu_0)C_j(\mu_0)\sigma_{ij}(\mu_0)\,,
		\label{eq:ij}
\end{flalign}
and at a different scale, $\mu_{EFT}=\mu$,
\begin{flalign}
	\sigma(\mu)=&\sigma_{SM}+\sum_i\frac{1{\rm TeV}^2}{\Lambda^2}C_i(\mu)
	\sigma_i(\mu) +\sum_{i,j} \frac{1{\rm TeV}^4}{\Lambda^4}
	C_i(\mu)C_j(\mu)\sigma_{ij}(\mu)
	\\
	=&\sigma_{SM}+\sum_i\frac{1{\rm TeV}^2}{\Lambda^2}
	C_i(\mu_0)\sigma_i(\mu_0;\mu)
	+\sum_{i,j} \frac{1{\rm TeV}^4}{\Lambda^4}
	C_i(\mu_0)C_j(\mu_0) \sigma_{ij}(\mu_0;\mu)\ ,
	\label{eq:eftunc}
\end{flalign}
where we define
\begin{flalign}
	&\sigma_i(\mu_0;\mu)=\Gamma_{ji}(\mu,\mu_0)
	\sigma_j(\mu)\ ,
	\\
	&\sigma_{ij}(\mu_0;\mu)= \Gamma_{ki}(\mu,\mu_0) \Gamma_{lj}(\mu,\mu_0)
	\sigma_{kl}(\mu)\ .
	\label{eq:sigma2}
\end{flalign}
The $\Gamma_{ij}$ describes the running of operator coefficients:
\begin{flalign}
	C_i(\mu)=\Gamma_{ij}(\mu,\mu_0)C_j(\mu_0)\,,
\end{flalign}
and is given by
\begin{flalign}
	&\Gamma_{ij}(\mu,\mu_0)=\exp\left(
	\frac{-2}{\beta_0}\log\frac{\alpha_s(\mu)}{\alpha_s(\mu_0)}
	\gamma_{ij}\right)\ ,
	\\
	&\beta_0=11-2/3n_f\ ,
\end{flalign}
	where $n_f=5$ is the number of active flavors in the running of $\alpha_s$.

	 According to Eq.~(\ref{eq:eftunc}), we will use
	$\sigma_{i,ij}(\mu_0;\mu)$ with $\mu_0=m_t$ and $\mu$ varying between
	$m_t/2$ and $2m_t$ to assess the EFT scale uncertainty.  The physical
	interpretation is that they are the cross sections from $C(\mu_0)$, but
	evaluated at a different EFT scale $\mu$, and then evolved back to
	scale $\mu_0$, with all the mixing and running effects taken into
	account, to allow for a fair comparison with the cross sections
	evaluated directly at $\mu_0$.  A more detailed discussion of
	uncertainties specifically related to RG mixing effects with concrete
	examples will be presented in Section \ref{sec:rg}.

\end{itemize}
Now consider the uncertainties due to missing higher order contributions in the 
$C/\Lambda^{2}$ expansion.  The cross section (or any other observable)
can be written as:
	\begin{flalign} 
		\sigma=\sigma_{SM}&+\sum_i
		\frac{C_i^\mathrm{dim6}}{(\Lambda/1\mathrm{TeV})^2}\sigma_i^{(\mathrm{dim6})}
		+\sum_{i\leq j}
		\frac{C_i^\mathrm{dim6}C_j^\mathrm{dim6}}{(\Lambda/1\mathrm{TeV})^4}\sigma_{ij}^{(\mathrm{dim6})}
		\nonumber\\&
		+\sum_i
		\frac{C_i^\mathrm{dim8}}{(\Lambda/1\mathrm{TeV})^4}\sigma_i^{(\mathrm{dim8})}
		+\mathcal{O}(\Lambda^{-6})\ .
	\end{flalign}
The second term comes from the squared contribution of dimension-six operators,
while the last term comes from the interference between the SM and dimension-eight
operators.  These two terms are formally $\mathcal{O}(\Lambda^{-4})$
contributions, but they should be considered separately
\cite{Bylund:2016phk,Contino:2016jqw}.
The last term must be less than the leading dimension-six contributions, otherwise
the EFT expansion would break down and in that case one should not use the EFT
approach from the very beginning.  On the other hand, the second term can be 
much larger than the first one without invalidating the expansion, as can be
justified for cases where the expansion in $E^2/\Lambda^2$ is under control but
the squared contribution may still be large, due to less constrained operator
coefficients, i.e.~if $C_i^2
\frac{E^4}{\Lambda^4}>C_i\frac{E^2}{\Lambda^2}>1>\frac{E^2}{\Lambda^2}$ is
satisfied.  An example is given in \cite{Contino:2016jqw}. Another possibility
is that the interference term is suppressed due to symmetries, kinematics,
or even simply accidentally \cite{Bylund:2016phk}, while the squared contribution
is not.  In the following we consider the two terms separately.

\begin{itemize}

	\item Impact of the squared contributions
		$\sigma_{ij}^{(\mathrm{dim6})}$ coming from dimension-six
		operators.
	
	These terms can be explicitly calculated and included in the central
	prediction, and in that case there is no uncertainty related with this
	``expansion'', because there are no more terms beyond
	$(C/\Lambda^2)^2$.  In this work we will give full results for
	$\sigma_{ij}^{(\mathrm{dim6})}$.  Of course, once constraints on
	operator coefficients are derived, one might decide to neglect some of
	these terms for simplicity.  Only in this case should one use, for
	example $\sigma_{ii}^{(\mathrm{dim6})}$, as an uncertainty estimate.
	
	\item Impact of missing higher-dimensional operators.

	The contribution $\sigma_{i}^{(1,\mathrm{dim8})}$ cannot be computed
	without analysing dimension-eight operators.
	A corresponding uncertainty should be taken into account.  For example,
	one can use $s_\mathrm{max}/\Lambda^2$ to estimate the relative size of
	dimension-eight interference to dimension-six interference, where
	$s_\mathrm{max}$ is a cutoff on the centre-of-mass energy of the
	process, as applied in the analysis \cite{Contino:2016jqw}.  The
	results we present in this work are not computed with this cutoff, but
	with our setup this is straightforward.
\end{itemize}

\section{Numerical results}
\label{sec:results}

In this section we give results for total cross sections and distributions.
Results are obtained with MMHT2014 LO/NLO PDFs \cite{Harland-Lang:2014zoa}, for
LO and NLO results respectively; input parameters are
\begin{flalign}
	&m_t=172.5\ \mathrm{GeV}\,, \quad
	m_H=125\ \mathrm{GeV}\,, \quad
	m_Z=91.1876\ \mathrm{GeV}\,, \\
	&\alpha_{EW}^{-1}=127.9\,, \quad
	G_F=1.16637\times10^{-5}\ \mathrm{GeV}^{-2}\,.
	\label{eq:input}
\end{flalign}
Central scales for $\mu_R,\mu_F$ are chosen as $m_t+m_H/2$ for the $t\bar tH$
process, and $m_H$ for the other loop-induced processes.
The central scale for $\mu_{EFT}$ is chosen as $m_t$ for all processes.

\subsection{Total cross sections}

Cross sections from dimension-six operators can be parametrised as 
\begin{flalign}
	\sigma=\sigma_{SM}+\sum_i\frac{1{\rm TeV}^2}{\Lambda^2}C_i\sigma_i
	+\sum_{i\leq j}
	\frac{1{\rm TeV}^4}{\Lambda^4}C_iC_j\sigma_{ij}.
	\label{eq:xsecpara}
\end{flalign}
Note that this parametrisation is slightly different from Eq.~(\ref{eq:ij}) because
we have added a ``$i\leq j$'' in the summation.  In other words, for the cross
terms (i.e.~$\sigma_{ij}, i\neq j$), a factor of two from exchanging $i,j$ will be
included in $\sigma_{ij}$.  We will now present results for $\sigma_{SM}$,
$\sigma_i$, and $\sigma_{ij}$.

We quote numbers with three uncertainties.  The first is the standard
scale uncertainty, obtained by independently setting
$\mu_R$ and $\mu_F$ to $\mu/2$, $\mu$ and $2\mu$, where $\mu$ is the central
scale obtaining nine $(\mu_R,\mu_F)$ combinations. The third uncertainty comes
from the MMHT PDF sets.  The second one is the EFT scale uncertainty,
representing the missing higher-order corrections to the operators, obtained by
using Eq.~(\ref{eq:eftunc}).

\newcommand{\xs}[7]{$#1^{+#2+#6+#4}_{-#3-#7-#5}$}
\begin{table}
\renewcommand{\arraystretch}{1.8}
\tiny
 \makebox[\linewidth]{
	\begin{tabular}{llllll}
		\hline\hline
		8 TeV &$\sigma$ LO&$\sigma/\sigma_{SM}$ LO&
		$\sigma$ NLO&$\sigma/\sigma_{SM}$ NLO& K
		\\\hline
$\sigma_{SM}$&\xs{0.127}{0.049}{0.032}{0.002}{0.001}{0.000}{0.000}&\xs{1.000}{0.000}{0.000}{0.000}{0.000}{0.000}{0.000}&\xs{0.132}{0.005}{0.012}{0.002}{0.002}{0.000}{0.000}&\xs{1.000}{0.000}{0.000}{0.000}{0.000}{0.000}{0.000}&1.03\\
$\sigma_{t\phi}$&\xs{-0.015}{0.004}{0.006}{0.000}{0.000}{0.001}{0.001}&\xs{-0.119}{0.000}{0.000}{0.000}{0.000}{0.005}{0.006}&\xs{-0.016}{0.002}{0.001}{0.000}{0.000}{0.000}{0.000}&\xs{-0.123}{0.001}{0.002}{0.000}{0.000}{0.000}{0.002}&1.07\\
$\sigma_{\phi G}$&\xs{0.161}{0.064}{0.042}{0.003}{0.002}{0.021}{0.017}&\xs{1.264}{0.016}{0.015}{0.005}{0.003}{0.168}{0.137}&\xs{0.211}{0.030}{0.030}{0.004}{0.004}{0.009}{0.007}&\xs{1.599}{0.158}{0.091}{0.006}{0.009}{0.066}{0.052}&1.31\\
$\sigma_{tG}$&\xs{0.123}{0.049}{0.032}{0.002}{0.001}{0.000}{0.000}&\xs{0.963}{0.010}{0.009}{0.003}{0.002}{0.000}{0.002}&\xs{0.127}{0.005}{0.012}{0.002}{0.002}{0.000}{0.001}&\xs{0.963}{0.004}{0.003}{0.002}{0.003}{0.004}{0.007}&1.03\\
$\sigma_{t\phi,t\phi}$&\xs{0.0004}{0.0002}{0.0001}{0.0000}{0.0000}{0.0001}{0.0000}&\xs{0.0035}{0.0000}{0.0000}{0.0000}{0.0000}{0.0004}{0.0002}&\xs{0.0005}{0.0000}{0.0001}{0.0000}{0.0000}{0.0000}{0.0000}&\xs{0.0036}{0.0001}{0.0001}{0.0000}{0.0000}{0.0002}{0.0000}&1.07\\
$\sigma_{\phi G,\phi G}$&\xs{0.120}{0.056}{0.035}{0.005}{0.003}{0.028}{0.021}&\xs{0.942}{0.058}{0.048}{0.025}{0.014}{0.218}{0.161}&\xs{0.187}{0.041}{0.035}{0.005}{0.006}{0.018}{0.015}&\xs{1.418}{0.243}{0.148}{0.022}{0.028}{0.140}{0.115}&1.56\\
$\sigma_{tG,tG}$&\xs{0.118}{0.056}{0.035}{0.005}{0.003}{0.002}{0.002}&\xs{0.929}{0.060}{0.051}{0.026}{0.015}{0.015}{0.018}&\xs{0.126}{0.007}{0.015}{0.004}{0.004}{0.001}{0.002}&\xs{0.959}{0.017}{0.024}{0.015}{0.020}{0.008}{0.012}&1.07\\
$\sigma_{t\phi,\phi G}$&\xs{-0.010}{0.003}{0.004}{0.000}{0.000}{0.001}{0.002}&\xs{-0.075}{0.001}{0.001}{0.000}{0.000}{0.011}{0.014}&\xs{-0.013}{0.002}{0.002}{0.000}{0.000}{0.001}{0.001}&\xs{-0.098}{0.006}{0.011}{0.001}{0.000}{0.004}{0.007}&1.34\\
$\sigma_{t\phi,tG}$&\xs{-0.007}{0.002}{0.003}{0.000}{0.000}{0.000}{0.000}&\xs{-0.058}{0.001}{0.001}{0.000}{0.000}{0.003}{0.002}&\xs{-0.008}{0.001}{0.000}{0.000}{0.000}{0.000}{0.000}&\xs{-0.060}{0.001}{0.001}{0.000}{0.000}{0.001}{0.000}&1.07\\
$\sigma_{\phi G,tG}$&\xs{0.131}{0.058}{0.037}{0.004}{0.002}{0.013}{0.011}&\xs{1.026}{0.047}{0.040}{0.019}{0.011}{0.100}{0.083}&\xs{0.175}{0.027}{0.028}{0.004}{0.005}{0.005}{0.004}&\xs{1.328}{0.148}{0.095}{0.016}{0.021}{0.034}{0.031}&1.34\\
\hline
	\end{tabular}}
	 \makebox[\linewidth]{
	\begin{tabular}{llllll}
		\hline
		13 TeV &$\sigma$ LO&$\sigma/\sigma_{SM}$ LO&
		$\sigma$ NLO&$\sigma/\sigma_{SM}$ NLO& K
		\\\hline
$\sigma_{SM}$&\xs{0.464}{0.161}{0.111}{0.005}{0.004}{0.000}{0.000}&\xs{1.000}{0.000}{0.000}{0.000}{0.000}{0.000}{0.000}&\xs{0.507}{0.030}{0.048}{0.007}{0.008}{0.000}{0.000}&\xs{1.000}{0.000}{0.000}{0.000}{0.000}{0.000}{0.000}&1.09\\
$\sigma_{t\phi}$&\xs{-0.055}{0.013}{0.019}{0.000}{0.001}{0.002}{0.003}&\xs{-0.119}{0.000}{0.000}{0.000}{0.000}{0.005}{0.006}&\xs{-0.062}{0.006}{0.004}{0.001}{0.001}{0.001}{0.001}&\xs{-0.123}{0.001}{0.001}{0.000}{0.000}{0.001}{0.002}&1.13\\
$\sigma_{\phi G}$&\xs{0.627}{0.225}{0.153}{0.007}{0.005}{0.081}{0.067}&\xs{1.351}{0.011}{0.011}{0.002}{0.001}{0.175}{0.145}&\xs{0.872}{0.131}{0.123}{0.013}{0.016}{0.037}{0.035}&\xs{1.722}{0.146}{0.089}{0.004}{0.005}{0.073}{0.068}&1.39\\
$\sigma_{tG}$&\xs{0.470}{0.167}{0.114}{0.005}{0.004}{0.000}{0.002}&\xs{1.014}{0.006}{0.006}{0.001}{0.001}{0.000}{0.004}&\xs{0.503}{0.025}{0.046}{0.007}{0.008}{0.001}{0.003}&\xs{0.991}{0.004}{0.010}{0.000}{0.001}{0.003}{0.006}&1.07\\
$\sigma_{t\phi,t\phi}$&\xs{0.0016}{0.0005}{0.0004}{0.0000}{0.0000}{0.0002}{0.0001}&\xs{0.0035}{0.0000}{0.0000}{0.0000}{0.0000}{0.0004}{0.0003}&\xs{0.0019}{0.0001}{0.0002}{0.0000}{0.0000}{0.0001}{0.0000}&\xs{0.0037}{0.0001}{0.0000}{0.0000}{0.0000}{0.0002}{0.0001}&1.17\\
$\sigma_{\phi G,\phi G}$&\xs{0.646}{0.274}{0.178}{0.018}{0.010}{0.141}{0.107}&\xs{1.392}{0.079}{0.066}{0.025}{0.014}{0.304}{0.231}&\xs{1.021}{0.204}{0.178}{0.024}{0.029}{0.096}{0.085}&\xs{2.016}{0.267}{0.178}{0.021}{0.027}{0.190}{0.167}&1.58\\
$\sigma_{tG,tG}$&\xs{0.645}{0.276}{0.178}{0.020}{0.010}{0.011}{0.015}&\xs{1.390}{0.082}{0.069}{0.028}{0.016}{0.023}{0.031}&\xs{0.674}{0.036}{0.067}{0.016}{0.019}{0.004}{0.007}&\xs{1.328}{0.011}{0.038}{0.014}{0.018}{0.008}{0.014}&1.04\\
$\sigma_{t\phi,\phi G}$&\xs{-0.037}{0.009}{0.013}{0.000}{0.000}{0.006}{0.007}&\xs{-0.081}{0.001}{0.001}{0.000}{0.000}{0.012}{0.015}&\xs{-0.053}{0.008}{0.008}{0.001}{0.001}{0.003}{0.004}&\xs{-0.105}{0.006}{0.009}{0.000}{0.000}{0.006}{0.007}&1.42\\
$\sigma_{t\phi,tG}$&\xs{-0.028}{0.007}{0.010}{0.000}{0.000}{0.001}{0.001}&\xs{-0.060}{0.000}{0.000}{0.000}{0.000}{0.002}{0.003}&\xs{-0.031}{0.003}{0.002}{0.000}{0.000}{0.000}{0.000}&\xs{-0.061}{0.000}{0.000}{0.000}{0.000}{0.000}{0.001}&1.10\\
$\sigma_{\phi G,tG}$&\xs{0.627}{0.252}{0.166}{0.014}{0.008}{0.053}{0.047}&\xs{1.349}{0.054}{0.046}{0.016}{0.009}{0.114}{0.100}&\xs{0.859}{0.127}{0.126}{0.017}{0.022}{0.021}{0.020}&\xs{1.691}{0.137}{0.097}{0.013}{0.017}{0.042}{0.039}&1.37\\
\hline
	\end{tabular}}
	 \makebox[\linewidth]{
	\begin{tabular}{llllll}
		\hline
		14 TeV &$\sigma$ LO&$\sigma/\sigma_{SM}$ LO&
		$\sigma$ NLO&$\sigma/\sigma_{SM}$ NLO& K
		\\\hline
$\sigma_{SM}$&\xs{0.558}{0.191}{0.132}{0.005}{0.004}{0.000}{0.000}&\xs{1.000}{0.000}{0.000}{0.000}{0.000}{0.000}{0.000}&\xs{0.614}{0.039}{0.058}{0.008}{0.009}{0.000}{0.000}&\xs{1.000}{0.000}{0.000}{0.000}{0.000}{0.000}{0.000}&1.10\\
$\sigma_{t\phi}$&\xs{-0.066}{0.016}{0.023}{0.001}{0.001}{0.003}{0.004}&\xs{-0.119}{0.000}{0.000}{0.000}{0.000}{0.005}{0.007}&\xs{-0.075}{0.008}{0.006}{0.001}{0.001}{0.001}{0.001}&\xs{-0.123}{0.001}{0.001}{0.000}{0.000}{0.001}{0.002}&1.14\\
$\sigma_{\phi G}$&\xs{0.758}{0.268}{0.184}{0.008}{0.006}{0.098}{0.081}&\xs{1.359}{0.011}{0.010}{0.002}{0.001}{0.176}{0.144}&\xs{1.064}{0.160}{0.149}{0.015}{0.018}{0.045}{0.040}&\xs{1.731}{0.143}{0.087}{0.003}{0.004}{0.073}{0.066}&1.40\\
$\sigma_{tG}$&\xs{0.567}{0.198}{0.136}{0.006}{0.005}{0.000}{0.001}&\xs{1.017}{0.006}{0.005}{0.001}{0.001}{0.001}{0.001}&\xs{0.609}{0.029}{0.054}{0.008}{0.009}{0.003}{0.002}&\xs{0.992}{0.006}{0.014}{0.000}{0.000}{0.006}{0.003}&1.07\\
$\sigma_{t\phi,t\phi}$&\xs{0.0020}{0.0007}{0.0005}{0.0000}{0.0000}{0.0002}{0.0001}&\xs{0.0036}{0.0000}{0.0000}{0.0000}{0.0000}{0.0003}{0.0002}&\xs{0.0022}{0.0002}{0.0002}{0.0000}{0.0000}{0.0003}{0.0000}&\xs{0.0036}{0.0000}{0.0000}{0.0000}{0.0000}{0.0005}{0.0000}&1.12\\
$\sigma_{\phi G,\phi G}$&\xs{0.817}{0.342}{0.223}{0.022}{0.012}{0.179}{0.134}&\xs{1.465}{0.083}{0.069}{0.025}{0.014}{0.320}{0.240}&\xs{1.293}{0.256}{0.223}{0.029}{0.036}{0.122}{0.105}&\xs{2.105}{0.268}{0.182}{0.021}{0.027}{0.198}{0.170}&1.58\\
$\sigma_{tG,tG}$&\xs{0.819}{0.345}{0.224}{0.024}{0.012}{0.014}{0.017}&\xs{1.468}{0.087}{0.073}{0.028}{0.016}{0.025}{0.030}&\xs{0.852}{0.046}{0.081}{0.019}{0.024}{0.007}{0.005}&\xs{1.388}{0.014}{0.048}{0.014}{0.018}{0.011}{0.008}&1.04\\
$\sigma_{t\phi,\phi G}$&\xs{-0.045}{0.011}{0.016}{0.000}{0.000}{0.006}{0.009}&\xs{-0.081}{0.001}{0.001}{0.000}{0.000}{0.012}{0.016}&\xs{-0.065}{0.009}{0.010}{0.001}{0.001}{0.004}{0.005}&\xs{-0.105}{0.006}{0.009}{0.000}{0.000}{0.006}{0.008}&1.44\\
$\sigma_{t\phi,tG}$&\xs{-0.033}{0.008}{0.012}{0.000}{0.000}{0.001}{0.002}&\xs{-0.060}{0.000}{0.000}{0.000}{0.000}{0.002}{0.003}&\xs{-0.038}{0.004}{0.002}{0.001}{0.000}{0.001}{0.000}&\xs{-0.062}{0.000}{0.001}{0.000}{0.000}{0.001}{0.000}&1.13\\
$\sigma_{\phi G,tG}$&\xs{0.783}{0.310}{0.205}{0.016}{0.009}{0.066}{0.056}&\xs{1.403}{0.056}{0.048}{0.016}{0.009}{0.119}{0.101}&\xs{1.070}{0.154}{0.154}{0.021}{0.026}{0.033}{0.024}&\xs{1.741}{0.132}{0.096}{0.012}{0.016}{0.053}{0.039}&1.37\\
\hline\hline
	\end{tabular}}
	\caption{\label{tab:tth1}
	Total cross section in pb for $pp\to t\bar tH$ at 8, 13, and 14 TeV, as
	parametrised in Eq.~(\ref{eq:xsecpara}).}
\end{table}

\newcommand{\xl}[3]{$#1^{+#2}_{#3}$}
\newcommand{\xls}[5]{$#1^{+#2+#4}_{#3#5}$}
\begin{table}
\center
\renewcommand{\arraystretch}{1.8}
\tiny
\begin{tabular}{lll}
	\hline\hline
		8 TeV &$\sigma$ LO&$\sigma/\sigma_{SM}$ LO 
		\\\hline
$\sigma_{SM}$&\xls { 8.08 }{    2.11 }{   -1.60 }{0.000}{-0.000}&\xls { 1.000}{    0.000}{  -0.000}{0.000}{-0.000}\\
$\sigma_{t\phi}$&\xls {  -0.962 }{      0.190 }{     -0.252  }{0.043}{-0.049}&\xls {   -0.119 }{    0.000035 }{   -0.000039  }{0.0053}{-0.0061}\\
$\sigma_{\phi G}$&\xls { 551.0 }{     71.1 }{    -62.8}{50.8}{-42.6}&\xls {  68.2 }{    7.70 }{   -7.64 }{6.29}{-5.27}\\
$\sigma_{tG}$&\xls {5.47 }{      1.43 }{     -1.08 }{0.657}{-1.88}&\xls {  0.677 }{    0.000029 }{   -0.000059  }{0.081}{-0.23}\\
$\sigma_{t\phi,t\phi}$&\xls {  0.0286 }{      0.0075 }{     -0.0057 }{0.00301}{-0.00250}&\xls { 0.00354 }{    0.000000 }{   -0.000001 }{0.00037}{-0.00031}\\
$\sigma_{\phi G,\phi G}$&\xls {9289}{     24.2 }{   -130}{1792}{-1382} &\xls{1149 }{  263 }{ -236 }{222}{-171}\\
$\sigma_{tG,tG}$&\xls {  0.924 }{      0.2415 }{     -0.1826}{5.44}{-0.0}&\xls{  0.1144 }{    0.000000 }{   -0.000002 }{0.673}{-0.0}\\
$\sigma_{t\phi,\phi G}$&\xls {  -32.57 }{      3.673 }{     -4.146 }{3.86}{-4.83}&\xls { -4.030 }{    0.447 }{   -0.449 }{0.478}{-0.597}\\
$\sigma_{t\phi,tG}$&\xls {  -0.326 }{      0.0643 }{     -0.0851 }{0.125}{-0.0407}&\xls { -0.0403 }{    0.000005 }{   -0.000003   }{0.0154}{-0.0050}\\
$\sigma_{\phi G,tG}$&\xls { 185.08 }{     22.96 }{    -20.38   }{0.0}{-393}&\xls {  22.90 }{    2.495 }{   -2.491  }{0.0}{-48.7}\\\hline
	\end{tabular}

\tiny
\begin{tabular}{lll}
	\hline
		13 TeV &$\sigma$ LO&$\sigma/\sigma_{SM}$ LO 
		\\\hline
$\sigma_{SM}$& \xls { 19.6 }{    5.47}{   -4.17 }{0.000}{-0.000} & \xls { 1.000}{    0.000}{  -0.000}{0.000}{-0.000}\\
$\sigma_{t\phi}$ &\xls { -2.34 }{      0.439 }{     -0.576  }{0.104}{-2.46} &\xls {    -0.119 }{    0.000004 }{   -0.000006}{0.0053}{-0.0061}\\
$\sigma_{\phi G}$&\xls {  1307 }{    183.9 }{   -166.0}{120}{-101}&\xls {   66.7 }{    7.29 }{   -7.24}{6.16}{-5.16}\\
$\sigma_{tG}$&\xls {   13.28 }{      3.71 }{     -2.83}{1.99}{-4.90}&\xls {   0.678 }{    0.000051 }{   -0.000018  }{0.102}{-0.250}\\
$\sigma_{t\phi,t\phi}$&\xls { 0.0695}{      0.0194 }{     -0.0148 }{0.00732}{-0.00607}&\xls { 0.00355 }{    0.0000 }{   -0.0000 }{0.00037}{-0.00031}\\
$\sigma_{\phi G,\phi G}$&\xls {22515 }{    377 }{   -732} {4340}{-3350}&\xls{ 1150 }{  264 }{ -236}{222}{-171}\\
$\sigma_{tG,tG}$&\xls {  2.253 }{      0.631 }{     -0.481}{13.2}{-0.0}&\xls {  0.115 }{    0.000050 }{   -0.000062 }{0.676}{-0.0}\\
$\sigma_{t\phi,\phi G}$&\xls {    -76.8 }{      9.38 }{    -10.3}{9.11}{-11.4}&\xls { -3.923 }{    0.446 }{   -0.453 }{0.47}{-0.58}\\
$\sigma_{t\phi,tG}$&\xls {  -0.799 }{      0.171 }{     -0.224 }{0.332}{-0.134}&\xls {  -0.04078 }{    0.000062 }{   -0.000050  }{0.017}{-0.007}\\
$\sigma_{\phi G,tG}$&\xls {  450 }{     63.3 }{    -57.3 }{0.0}{-954}&\xls {   23.0 }{    2.50 }{   -2.49}{0.0}{-48.7}\\\hline
	\end{tabular}

\begin{tabular}{lll}
	\hline
		14 TeV &$\sigma$ LO&$\sigma/\sigma_{SM}$ LO 
		\\\hline
$\sigma_{SM}$&\xls {   22.4 }{    6.41 }{   -4.87}{0.000}{-0.000}&\xls { 1.000}{    0.000}{  -0.000}{0.000}{-0.000}\\
$\sigma_{t\phi}$&\xls {   -2.66 }{    0.576 }{   -0.757}{0.118}{-0.136}&\xls {   -0.118 }{    0.000065 }{   -0.000080}{0.00529}{-0.00608}\\
$\sigma_{\phi G}$&\xls { 1509 }{  224 }{ -203}{139}{-117}&\xls {   67.3 }{    7.47 }{   -7.50}{6.20}{-5.20}\\
$\sigma_{tG}$&\xls {   15.1 }{    4.296 }{   -3.27}{2.06}{-5.33}&\xls {    0.673 }{    0.000628 }{   -0.000506}{0.092}{-0.238}\\
$\sigma_{t\phi,t\phi}$&\xls {    0.0791 }{    0.0225 }{   -0.0171}{0.00832}{-0.0069}&\xls {    0.00352 }{    0.000003 }{   -0.000002}{0.00037}{-0.00031}\\
$\sigma_{\phi G,\phi G}$&\xls {2564 }{  546 }{ -962}{4947}{-3813} &\xls{ 1143}{  263 }{ -235}{221}{-170}\\
$\sigma_{tG,tG}$&\xls {    2.55 }{    0.727 }{   -0.553}{15.1}{-0.0}&\xls {    0.114 }{    0.000074 }{   -0.000060}{0.673}{-0.0}\\
$\sigma_{t\phi,\phi G}$&\xls {  -89.5}{   12.85 }{  -14.1}{10.6}{-13.2}&\xls {   -3.99 }{    0.479 }{   -0.478}{0.473}{-0.59}\\
$\sigma_{t\phi,tG}$&\xls {   -0.895 }{    0.194 }{   -0.254}{0.340}{-0.204}&\xls {   -0.0399 }{    0.000046 }{   -0.000057}{0.0152}{-0.0091}\\
$\sigma_{\phi G,tG}$&\xls {  515 }{   75.2 }{  -67.8}{0.0}{-1089}&\xls {   22.94 }{    2.506 }{   -2.490}{0.0}{-48.6}\\\hline\hline
	\end{tabular}
		\caption{\label{tab:h1}
	Total cross section in pb for $pp\to H$ at 8, 13, and 14 TeV, as
	parametrised in Eq.~(\ref{eq:xsecpara}). Only the
renormalisation and factorisation and EFT  scale uncertainties are shown. }
\end{table}

\begin{table}
\center
\renewcommand{\arraystretch}{1.8}
\tiny
\begin{tabular}{lll}
	\hline\hline
		8 TeV &$\sigma$ NLO&$\sigma/\sigma_{SM}$ NLO 
		\\\hline
$\sigma_{SM}$&\xl{     13.54 }{    2.812 }{   -2.192} & \xl { 1.000}{    0.000}{  -0.000}\\
$\sigma_{t\phi}$&\xl {   -1.577 }{      0.252 }{     -0.322}& \xl{   -0.117 }{    0.000369 }{   -0.000305}\\
$\sigma_{\phi G}$&\xl{ 1036 }{    140 }{   -113}& \xl{   76.5 }{    4.797 }{   -4.632} \\
$\sigma_{t\phi,t\phi}$&\xl{    0.0459 }{      0.0092 }{     -0.0072}& \xl{    0.00339 }{    0.000018 }{   -0.000022} \\
$\sigma_{\phi G,\phi G}$&\xl {19802}{   1319 }{  -1120}& \xl{ 1463 }{  184 }{ -171}\\\hline
	\end{tabular}

\tiny
\begin{tabular}{lll}\hline
		13 TeV &$\sigma$ NLO&$\sigma/\sigma_{SM}$ NLO 
		\\\hline
$\sigma_{SM}$& \xl {     31.29 }{    7.08 }{   -5.68}& \xl { 1.000}{    0.000}{  -0.000}\\
$\sigma_{t\phi}$ &\xl {   -3.646 }{      0.653 }{     -0.809}& \xl{   -0.117 }{    0.000416 }{   -0.000339}\\
$\sigma_{\phi G}$&\xl { 2395 }{    365 }{   -309}& \xl{   76.5 }{    4.89 }{   -4.67}\\
$\sigma_{t\phi,t\phi}$&\xl {    0.106 }{      0.0231 }{     -0.0187}& \xl{    0.00339 }{    0.000020 }{   -0.000025}\\
$\sigma_{\phi G,\phi G}$&\xl{45785 }{   3817}{  -3493}& \xl{ 1463 }{  188 }{ -173} \\\hline
	\end{tabular}

\begin{tabular}{lll}\hline
		14 TeV &$\sigma$ NLO&$\sigma/\sigma_{SM}$ NLO 
		\\\hline
$\sigma_{SM}$&\xl{     35.33 }{    8.07 }{   -6.50}&  \xl { 1.000}{    0.000}{  -0.000}\\
$\sigma_{t\phi}$&\xl{   -4.117 }{      0.747 }{     -0.922}&\xl{   -0.117 }{    0.000425 }{   -0.000344}\\
$\sigma_{\phi G}$&\xl { 2704 }{    418 }{   -356}&\xl{   76.5 }{    4.91}{   -4.68}\\
$\sigma_{t\phi,t\phi}$&\xl {    0.120 }{      0.0263 }{     -0.0215}&\xl{    0.00339 }{    0.000020 }{   -0.000025}\\
$\sigma_{\phi G,\phi G}$&\xl {51700 }{   4410 }{  -4070}&\xl{ 1463 }{  189 }{ -173}\\\hline\hline
	\end{tabular}
		\caption{\label{tab:h1NLO}
	Total cross section in pb for $pp\to H$ at 8, 13, and 14 TeV at NLO in the infinite top mass limit. Only the renormalisation and factorisation scale uncertainties are shown. }
\end{table}

\begin{table}
\center
\renewcommand{\arraystretch}{1.8}
\tiny
\begin{tabular}{lll}\hline\hline
		8 TeV &$\sigma$ LO&$\sigma/\sigma_{SM}$ LO 
		\\\hline
$\sigma_{SM}$&\xls{ 4.168 }{   2.05 }{   -1.28 }{0.000}{-0.000}&\xls{ 1.000}{    0.000}{  -0.000}{0.000}{-0.000}\\
$\sigma_{t\phi}$&\xls{  -0.495 }{      0.152 }{     -0.244  }{0.022}{-0.025}&\xls{   -0.119 }{    0.000027 }{   -0.000045  }{0.0053}{-0.0061}\\
$\sigma_{\phi G}$&\xls{ 270.6 }{    86.1 }{    -61.0}{25.0}{-20.9}&\xls{  64.9 }{    7.62 }{   -7.55 }{5.99}{-5.02}\\
$\sigma_{tG}$&\xls{2.85}{       1.40 }{     -0.874 }{0.549}{-1.13}&\xls{  0.683 }{    0.000212 }{   -0.000071  }{0.132}{-0.271}\\
$\sigma_{t\phi,t\phi}$&\xls{  0.0147 }{      0.0072 }{     -0.0045 }{0.00154}{-0.00128}&\xls{ 0.00352 }{    0.000001 }{   -0.000001 }{0.00037}{-0.00031}\\
$\sigma_{\phi G,\phi G}$&\xls{4875}{    909 }{   -725}{940} {-724}&\xls{1170 }{  269 }{ -241 }{226}{-173}\\
$\sigma_{tG,tG}$&\xls{  0.496 }{      0.246 }{     -0.153}{0.214}{-0.318}&\xls{   0.119 }{    0.000276 }{   -0.000206 }{0.051}{-0.076}\\
$\sigma_{t\phi,\phi G}$&\xls{  -16.5 }{      3.82 }{     -5.42}{1.96}{-2.45}&\xls{  -3.97 }{    0.43 }{   -0.43}{0.47}{-0.59}\\
$\sigma_{t\phi,tG}$&\xls{   -0.167 }{      0.051 }{     -0.082 }{0.064}{-0.047}&\xls{ -0.0400 }{    0.000031 }{   -0.000045   }{0.0154}{-0.011}\\
$\sigma_{\phi G,tG}$&\xls{  100.8 }{     33.4 }{    -23.4  }{15.0}{-44.5}&\xls{ 24.17 }{    2.63 }{   -2.61  }{3.59}{-10.7}\\\hline
	\end{tabular}

\tiny
\begin{tabular}{lll}\hline
		13 TeV &$\sigma$ LO&$\sigma/\sigma_{SM}$ LO 
		\\\hline
$\sigma_{SM}$&\xls{  11.49 }{    5.091 }{   -3.296 }{0.000}{-0.000}&\xls{ 1.000}{    0.000}{  -0.000}{0.000}{-0.000}\\
$\sigma_{t\phi}$&\xls{  -1.378 }{      0.393 }{     -0.604  }{0.0615}{-0.0707}&\xls{  -0.1200 }{    0.000379 }{   -0.000321}{0.00535}{-0.0061}\\
$\sigma_{\phi G}$&\xls{  792 }{    220.4 }{   -163}{73.0}{-61.2}&\xls{   68.9 }{    7.89 }{   -7.89}{6.36}{-5.33}\\
$\sigma_{tG}$&\xls{   7.91 }{      3.50}{     -2.27}{0.984}{-2.71}&\xls{   0.689 }{    0.000223 }{   -0.000253  }{0.0857}{-0.235}\\
$\sigma_{t\phi,t\phi}$&\xls{ 0.0410 }{      0.0180 }{     -0.0117 }{0.0043}{-0.0036}&\xls{ 0.003574 }{    0.000008 }{   -0.000010 }{0.00038}{-0.00031}\\
$\sigma_{\phi G,\phi G}$&\xls{13663 }{   2021 }{-1697}{2636}{-2032}  &\xls{ 1189 }{  274 }{ -245}{229}{-177}\\
$\sigma_{tG,tG}$&\xls{  1.400 }{      0.625 }{     -0.404}{0.454}{-0.890}&\xls{   0.122 }{    0.000257 }{   -0.000193 }{0.0395}{-0.0774}\\
$\sigma_{t\phi,\phi G}$&\xls{  -46.5 }{      9.62 }{    -13.1}{5.51}{-6.89}&\xls{ -4.04 }{    0.454 }{   -0.453 }{0.48}{-0.60}\\
$\sigma_{t\phi,tG}$&\xls{ -0.470 }{      0.134 }{     -0.207 }{0.180}{-0.084}&\xls{   -0.0409 }{    0.000079 }{   -0.000068  }{0.0157}{-0.0073}\\
$\sigma_{\phi G,tG}$&\xls{282 }{     85.2 }{    -61.3 }{50.4}{-124}&\xls{  24.6 }{    3.02 }{   -2.91}{4.39}{-10.8}\\\hline
	\end{tabular}

\tiny
\begin{tabular}{lll}\hline
		14 TeV &$\sigma$ LO&$\sigma/\sigma_{SM}$ LO 
		\\\hline
$\sigma_{SM}$&\xls{  13.35 }{    5.79 }{   -3.78  }{0.000}{-0.000}&\xls{ 1.000}{    0.000}{  -0.000}{0.000}{-0.000}\\
$\sigma_{t\phi}$&\xls{  -1.589 }{      0.451 }{     -0.693  }{0.0709}{-0.0815}&\xls{ -0.1190 }{    0.000132 }{   -0.000159 }{0.0053}{-0.0061}\\
$\sigma_{\phi G}$&\xls{ 905 }{    254 }{   -186}{83.4}{-70.0}&\xls{ 67.74 }{    7.567 }{   -7.483  }{6.25}{-5.24}\\
$\sigma_{tG}$&\xls{9.223 }{      4.03 }{     -2.62 }{1.38}{-3.33}&\xls{   0.691 }{    0.001289 }{   -0.001025 }{0.103}{-0.25}\\
$\sigma_{t\phi,t\phi}$&\xls{  0.0472 }{      0.0205 }{     -0.0134  }{0.00497}{-0.00412}&\xls{ 0.00354 }{    0.000000 }{   -0.000000   }{0.00037}{-0.00031}\\
$\sigma_{\phi G,\phi G}$&\xls{15820 }{   2254 }{  -1907} {3050}{-2350}&\xls{ 1185 }{  275 }{ -245}{229}{-176}\\
$\sigma_{tG,tG}$&\xls{  1.631 }{      0.7185 }{     -0.466  }{-0.87}{0.49}&\xls{ 0.122}{    0.000567 }{   -0.000437  }{0.037}{-0.065}\\
$\sigma_{t\phi,\phi G}$&\xls{  -54.8 }{     11.3 }{    -15.3 }{6.49}{-8.12}&\xls{ -4.10 }{    0.45 }{   -0.45}{0.486}{-0.608}\\

$\sigma_{t\phi,tG}$&\xls{ -0.554 }{      0.158 }{     -0.244 }{0.208}{-0.076}&\xls{-0.0414 }{    0.000156 }{   -0.000199}{0.0156}{-0.0057}\\
$\sigma_{\phi G,tG}$&\xls{323 }{     91.3 }{    -67.4 }{65.9}{-125}&\xls{  24.19 }{    2.63 }{   -2.65 }{4.93}{-9.38}\\\hline\hline
	\end{tabular}

		\caption{\label{tab:h1j}
	Total cross section in pb for $pp\to Hj$ at 8, 13, and 14 TeV at LO for a $p_T^j>30$ GeV cut. Only the $\mu_{R,F}$ and EFT  scale  uncertainties are shown. }
\end{table}

\begin{table}
\center
\renewcommand{\arraystretch}{1.8}
\tiny
	\begin{tabular}{lll}
		\hline\hline
		8 TeV &$\sigma$ LO&$\sigma/\sigma_{SM}$ LO 
		\\\hline
$\sigma_{SM}$&\xls {0.00755}{0.00313}{-0.00206}{0.000}{-0.000} &\xls{ 1.000}{    0.000}{  -0.000}{0.000}{-0.000}\\
$\sigma_{t\phi}$&\xls{0.00167}{ 0.000704}{-0.000459}{0.000086}{-0.000075}&\xls{ 0.221}{ 0.00111}{-0.000876}{0.0113}{-0.0099}\\
$\sigma_{\phi G}$&\xls{-0.348}{ 0.0676}{-0.0903}{0.0273}{-0.0325}&\xls{-46.0}{ 5.04}{-4.93}{3.61}{-4.31}\\
$\sigma_{tG}$&\xls{-0.0111}{ 0.00290}{-0.00432}{0.00183}{-0.0010}&\xls{-1.46}{ 0.0244}{-0.0203}{0.243}{-0.135}\\
$\sigma_{t\phi,t\phi}$&\xls{0.000198}{ 0.000088}{0.000057}{0.0000208}{-0.0000173}&\xls{0.0262}{ 0.00060}{-0.00048}{0.0028}{-0.0023}\\
$\sigma_{\phi G,\phi G}$&\xls{19.42}{ 2.67}{-2.19}{3.75}{-2.89}&\xls{2571 } { 626}{   -544}{497}{-383}\\
$\sigma_{tG,tG}$&\xls{ 0.0127}{ 0.00559}{-0.00359}{0.00133}{-0.00323}&\xls{1.69}{ 0.0289}{-0.0209}{0.176}{-0.427}\\
$\sigma_{t\phi,\phi G}$&\xls{-0.0853}{ 0.0186}{-0.0257}{0.010}{-0.013}&\xls{-11.29}{ 1.54}{-1.62}{1.35}{-1.69}\\
$\sigma_{t\phi,tG}$&\xls{-0.00255}{ 0.000700}{-0.00107}{0.000546}{-0.000323}&\xls{-0.337}{ 0.00113}{-0.00127}{0.072}{-0.043}\\
$\sigma_{\phi G,tG}$&\xls{ 0.987}{ 0.277}{-0.199}{0.143}{-0.202}&\xls{130.7}{ 16.4}{-15.4}{18.9}{-26.8}\\
\hline
	\end{tabular}
\tiny

\begin{tabular}{lll}
	\hline
		13 TeV &$\sigma$ LO&$\sigma/\sigma_{SM}$ LO 
		\\\hline
$\sigma_{SM}$&\xls{ 0.0256 }{     0.00904 }{    -0.00625}{0.000}{-0.000} &\xls{ 1.000}{    0.000}{  -0.000}{0.000}{-0.000} \\
$\sigma_{t\phi}$&\xls{0.00580}{ 0.00209}{-0.00144}{0.000297}{-0.000259}&\xls{ 0.227}{ 0.00114}{-0.000918 }{0.0116}{-0.0101}\\
$\sigma_{\phi G}$&\xls{-1.208}{ 0.231}{-0.291}{0.0948}{-0.113}&\xls{-47.3}{ 6.18}{-6.14}{3.707}{-4.42}\\
$\sigma_{tG}$&\xls{-0.0347}{ 0.00804}{-0.0113}{0.0041}{-0.0013}&\xls{-1.356}{ 0.0271}{-0.0225}{0.161}{-0.051}\\
$\sigma_{t\phi,t\phi}$&\xls{0.000748}{ 0.000290}{-0.000194}{0.000079}{-0.000065}&\xls{0.0293}{ 0.000727}{-0.000584}{0.0031}{-0.0026}\\
$\sigma_{\phi G,\phi G}$&\xls{73.02}{ 7.54}{-6.48}{14.1}{-10.9}&\xls{2856.2} { 743.3}{ -628.5}{552}{-425}\\
$\sigma_{tG,tG}$&\xls{0.0496}{ 0.0198}{-0.01305 }{0.00505}{-0.0126}&\xls{1.940}{ 0.0650}{-0.0477}{0.198}{-0.493}\\
$\sigma_{t\phi,\phi G}$&\xls{-0.303}{ 0.0506}{-0.0641}{0.0362}{-0.0453}&\xls{-11.83}{ 1.39}{-1.41}{1.42}{-1.77}\\
$\sigma_{t\phi,tG}$&\xls{-0.00870}{ 0.00213}{-0.00309}{0.00163}{-0.00120}&\xls{-0.340}{ 0.000238}{-0.000438}{0.064}{-0.047}\\
$\sigma_{\phi G,tG}$&\xls{3.77}{ 0.914}{-0.681 }{0.554}{-0.802}&\xls{147.5}{ 20.83}{-18.86}{20.7}{-31.4}\\
\hline
	\end{tabular}
	
\tiny
\begin{tabular}{lll}
	\hline
		14 TeV &$\sigma$ LO&$\sigma/\sigma_{SM}$ LO 
		\\\hline
$\sigma_{SM}$&\xls{ 0.0305 }{     0.0105 }{    -0.00734}{0.000}{-0.000}&\xls{ 1.000}{    0.000}{  -0.000}{0.000}{-0.000}\\
$\sigma_{t\phi}$&\xls{0.00694}{ 0.00245}{-0.00169}{0.00031}{-0.00036}&\xls{ 0.227}{ 0.00131}{-0.00106 }{0.0117}{-0.0101}\\
$\sigma_{\phi G}$&\xls{-1.508}{ 0.214}{-0.256}{0.118}{-0.141}&\xls{-49.4}{ 6.45}{-6.39}{3.87}{-4.61}\\
$\sigma_{tG}$&\xls{-0.0408}{ 0.00929}{-0.0130}{0.0037}{-0.0008}&\xls{-1.337}{ 0.0271}{-0.0224}{0.122}{-0.0262}\\
$\sigma_{t\phi,t\phi}$&\xls{0.000904}{ 0.000343}{-0.000232}{0.000095}{-0.000079}&\xls{0.0296}{ 0.00076}{-0.00061}{0.0031}{-0.0026}\\
$\sigma_{\phi G,\phi G}$&\xls{88.35}{ 8.72}{-7.55}{17.1}{-13.2}&\xls{2896} {741}{ -641}{560}{-431}\\
$\sigma_{tG,tG}$&\xls{0.0608}{ 0.0241}{-0.0159 }{0.00605}{-0.0148}&\xls{1.994}{ 0.0753}{-0.0556}{0.198}{-0.484}\\
$\sigma_{t\phi,\phi G}$&\xls{-0.367}{ 0.0520}{-0.0670}{0.0439}{-0.0550}&\xls{-12.0}{ 1.46}{-1.56}{1.44}{-1.80}\\
$\sigma_{t\phi,tG}$&\xls{-0.0104}{ 0.00253}{-0.00368}{0.00192}{-0.00174}&\xls{-0.341}{ 0.0014}{-0.002}{0.063}{-0.057}\\
$\sigma_{\phi G,tG}$&\xls{4.60}{ 1.09}{-0.816}{0.640}{-0.923}&\xls{150.7}{ 21.69}{-19.53}{21.0}{-30.3}\\
\hline\hline
	\end{tabular}
		\caption{\label{tab:hh}
	Total cross section in pb for $pp\to HH$ at 8, 13, and 14 TeV at LO. Only the $\mu_{R,F}$ and EFT  scale uncertainties are shown.}
\end{table}

In Table \ref{tab:tth1} we give the LO/NLO results for $t\bar tH$ total cross
section for the LHC at 8, 13 and 14 TeV.  Both LO and NLO cross sections, as
well as their ratios over the SM cross section, are given. In general
the ratios to the SM contribution increase with energy, as expected in an
EFT, except for the $O_{t\varphi}$ contributions which only rescale the SM
Yukawa coupling.  The quadratic terms and cross terms, i.e.~$\sigma_{ij}$
displayed in the last six rows, are in general not small, and we will see that
given the current bounds on these operators they should not be neglected.  The
$K$ factors range between roughly 1 to 1.6, depending on operators, and can be
very different from the SM.  In particular, contributions related to $O_{\varphi G}$
tend to have large $K$ factors.  Improved precision is clearly reflected
by the reduced uncertainties at NLO. The dominant uncertainties come from
$\mu_{R,F}$ scale variation.  However, these uncertainties, together with the
PDF uncertainties, are reduced once taking ratios with respect to
$\sigma_{SM}$. On the contrary, the EFT scale uncertainty is not affected by
taking ratios, so it becomes relatively more important.  This means that in a
measurement where ratios are used, for example as proposed by
\cite{Plehn:2015cta}, or in a global fit where correlations are correctly taken
into account, the EFT uncertainties may be the dominant ones at dimension-six.
We also note here that at the NLO accuracy the effect of the $O_{t\varphi}$
operator is not a simple rescaling of the SM predicition, because the $y_t$ in
the SM is defined with on-shell top mass, while the $C_{t\varphi}$ is defined
with the $\overline{MS}$ scheme. As a result the corresponding $K$ factors are
different from the SM ones.

The corresponding LO results for single $H$ production are shown in Table
\ref{tab:h1}, where we have kept the exact top mass dependence. The PDF
uncertainties are not shown as these are found to be at the percent level and
significantly smaller than the $\mu_{R,F}$  scale uncertainties. We note that
these results suffer from large $\mu_{R,F}$  scale uncertainties as they are at
LO. As in the case of  $t\bar tH$ we find that the $\mu_{R,F}$  scale
uncertainties get significantly reduced by taking the ratio over the SM. This
reduction of the uncertainties in the ratios is not dramatic for the $O_{\phi
G}$ contributions. This is related to the choice of operator normalisation,
which leads to $O_{\phi G}$ entering at $O( \alpha) $ while the SM contribution
is at $O( \alpha_s^2 \alpha) $.  We also find that the ratios over the SM are
not sensitive to the collider energy as the partonic centre-of-mass energy is
fixed at the Higgs mass. Again we find that $\sigma_{ij}$ terms are important.
For this process, we find that the contributions involving the $O_{\phi G}$
operator are large as it enters at tree-level while the other two operators
enter through top-quark loops.

While the computation of the NLO corrections for the $O_{t\phi}$ and $O_{\phi
G}$ operators is possible both in the infinite top mass limit and with the
exact top-mass dependence, this is not the case for the $O_{tG}$ operator for
which a dedicated computation, beyond the scope of this work, is required. As a
first step towards NLO results, we show in Table \ref{tab:h1NLO} results for a
subset of the contributions computed in the infinite top mass limit. The
results  show the well-known large $K$ factors ($\sim 2$) for gluon fusion and
demonstrate the reduction of the $\mu_{R,F}$ uncertainties. 

The results for the $pp\to Hj$ process are shown in Table \ref{tab:h1j}. For
these results a 30 GeV cut has been imposed on the jet transverse momentum. The
relative contributions of the operators at the total cross-section level remain
similar to those for single $H$ as the cut on the jet is rather soft. While
$O_{t\phi}$ just gives a rescaling compared to the SM, the $O_{ \phi G}$ and
$O_{tG}$ operators have a different energy dependence causing also small
differences in the ratios between the three collider energies, as a range of
partonic energies is probed.

Finally we show the results for Higgs pair production in Table \ref{tab:hh}.
This process behaves in a different way than the single Higgs production
process.  The contribution of $O_{t\phi}$ operator is no longer a simple
universal rescaling of the SM, as it not only affects the triangle, box
diagrams and their interference in a different way but also introduces the
$ttHH$ interaction.  Thus the ratio over the SM behaves differently from single
Higgs production and also acquires an energy dependence.  The contributions
involving $O_{\phi G}$ and $O_{tG}$ relative to the SM increase with the
collider energy. The EFT scale uncertainties in the ratios over the SM are
similar in size to the $\mu_{R,F}$ ones for $O_{\phi G}$, while they dominate
for $O_{tG}$ and $O_{t \phi}$. We find that the contribution of the
chromo-dipole is large, and even with the current constraints it could have a
large impact on $HH$ production.  This observation implies that an EFT analysis
of di-Higgs production should consistently take into consideration such an
impact, as it will affect the limits set on other dimension-six operators
entering the process, in particular those modifying the Higgs self-coupling. We
also emphasize that this contribution involves a UV pole that is canceled by
the mixing from $O_{tG}$ to $O_{\varphi G}$, therefore its cross section has a
dependence on $\mu_{EFT}$, and should always be used with care.  The EFT
uncertainty presented in Table \ref{tab:hh} correctly takes into account the
contribution from $O_{\varphi G}$ through mixing effects while changing
$\mu_{EFT}$. A more detailed discussion on this issue will be presented in
Section \ref{sec:rg}.

\subsection{Distributions}

Differential distributions are obtained at LO and NLO for the $pp\to t\bar{t}H$
process.  This can be done also with matching to PS simulation, and with top
quarks decayed while keeping spin correlations \cite{Artoisenet:2012st},
all implemented in the
{\sc MG5\_aMC} framework. Hence our approach can be used directly in a
realistic experimental simulation, with NLO+PS event generation, which
allows for more detailed studies of possible EFT signals.
In this work, for illustration purpose, we only present fixed order NLO
distributions.  

\begin{figure}[b!]
 \begin{minipage}[t]{0.5\linewidth}
\centering
\includegraphics[width=.99\linewidth]{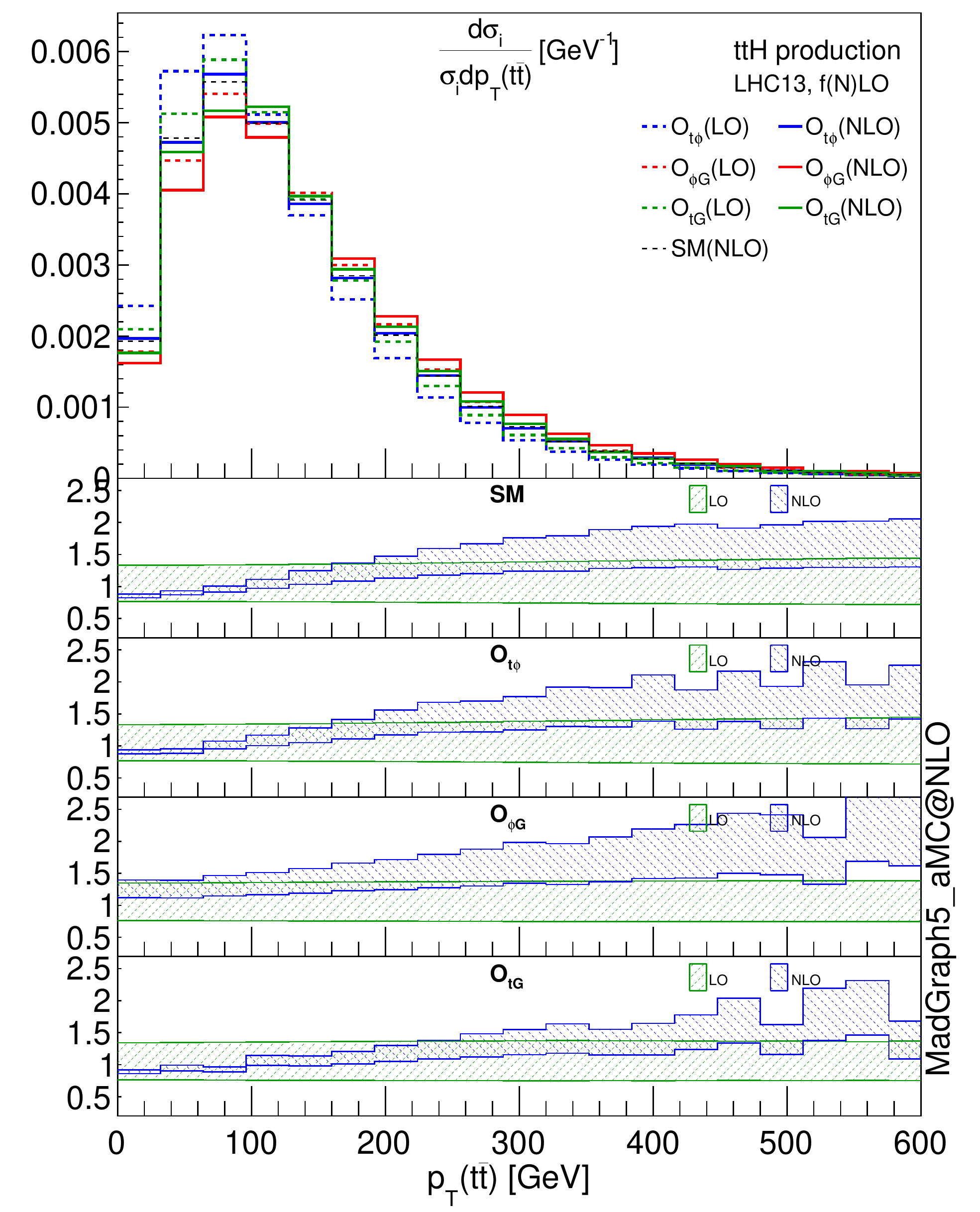}
\end{minipage}
\hspace{0.5cm}
 \begin{minipage}[t]{0.5\linewidth}
 \centering
 \includegraphics[width=.99\linewidth]{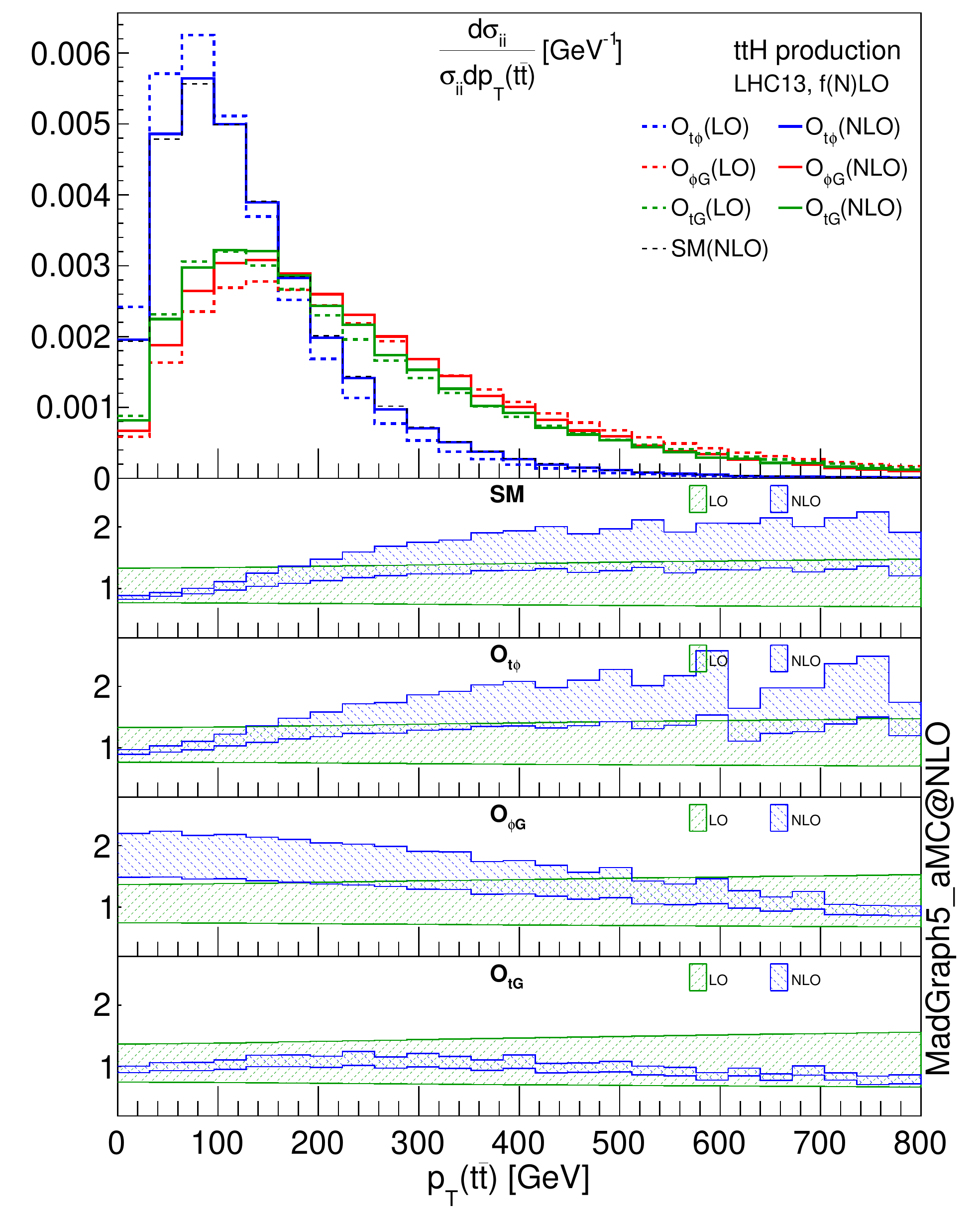}
 \end{minipage}
\caption{\label{fig:pt2} 
Transverse momentum distributions of the $t\bar t$ system, normalised.  Left:
interference contributions from $\sigma_i$. Right: squared contributions
$\sigma_{ii}$.  SM contributions and individual operator contributions are
displayed.  Lower panels give the $K$ factors and $\mu_{R,F}$ uncertainties.} 
\end{figure}

\begin{figure}[tb]
 \begin{minipage}[t]{0.5\linewidth}
\centering
\includegraphics[width=.99\linewidth]{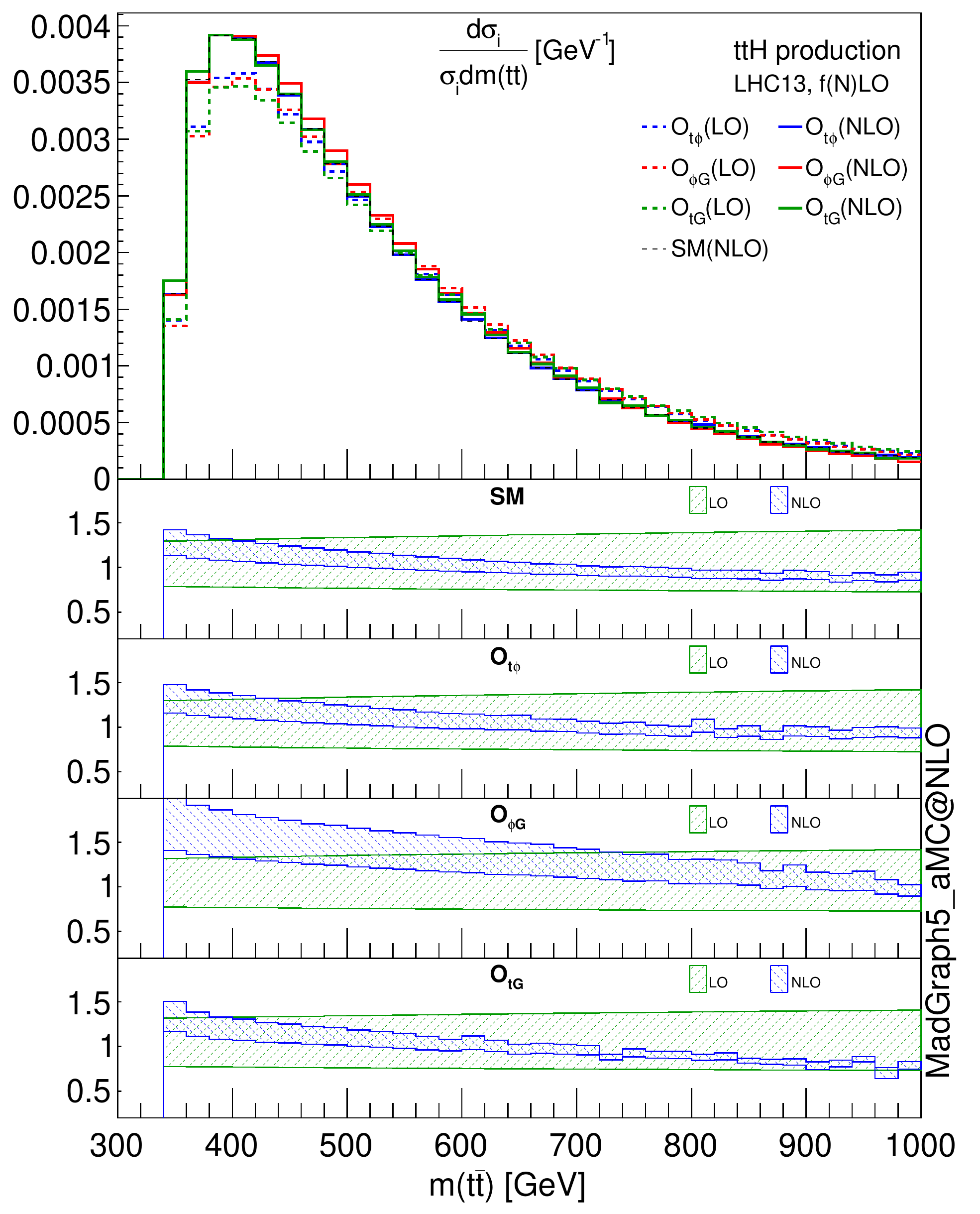}
\end{minipage}
\hspace{0.5cm}
 \begin{minipage}[t]{0.5\linewidth}
 \centering
 \includegraphics[width=.99\linewidth]{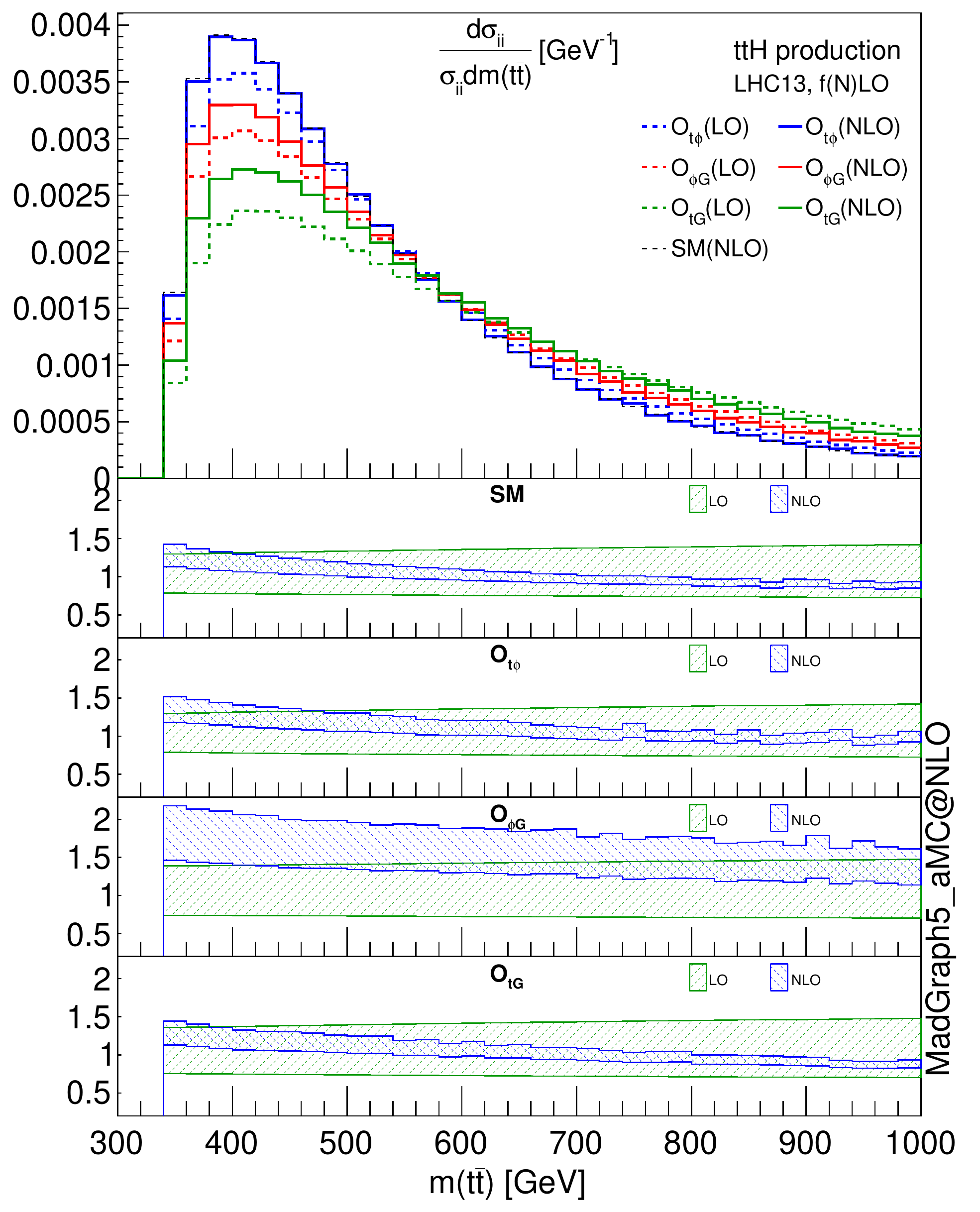}
 \end{minipage}
 \begin{minipage}[t]{0.5\linewidth}
\centering
\includegraphics[width=.99\linewidth]{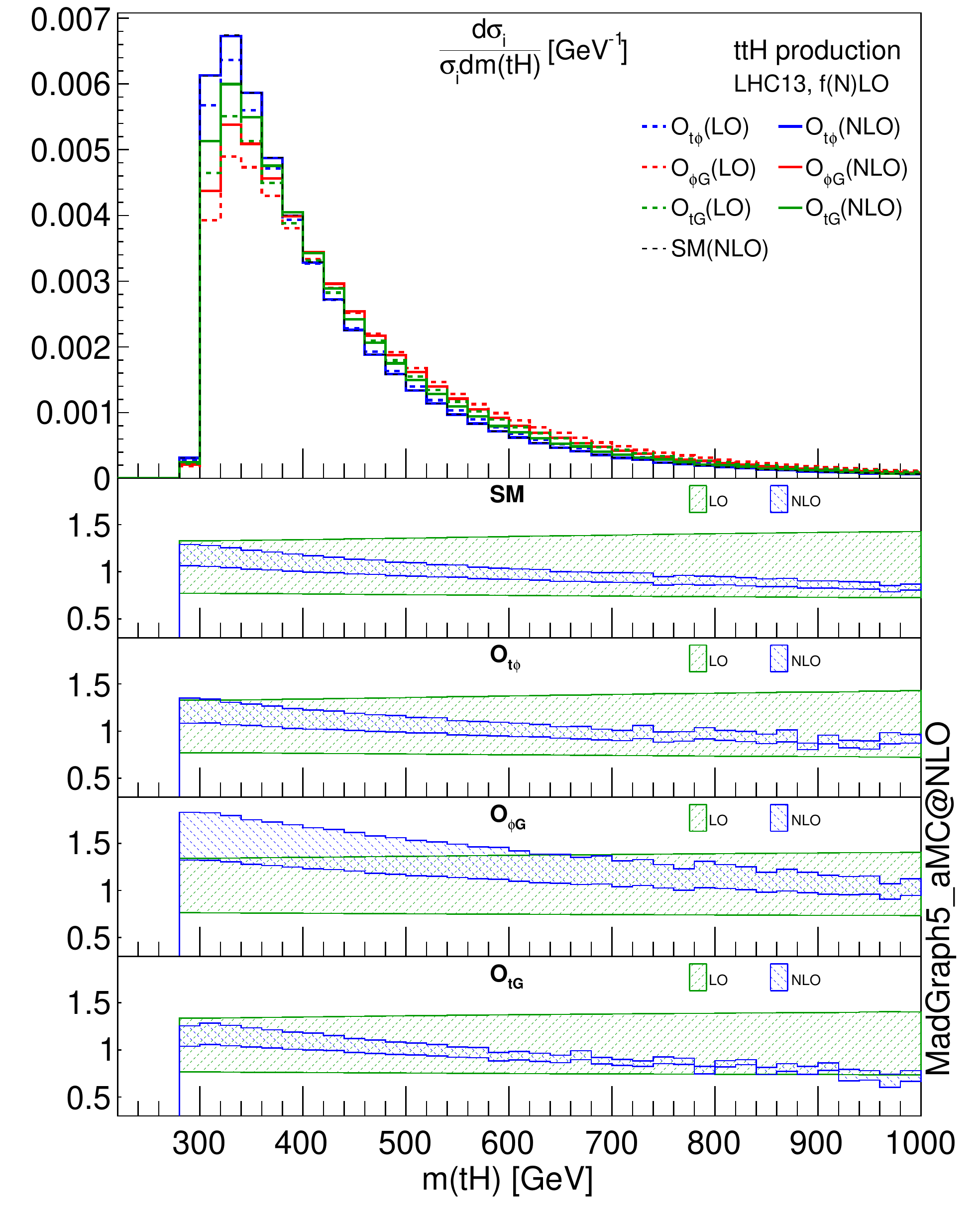}
\end{minipage}
\hspace{0.5cm}
 \begin{minipage}[t]{0.5\linewidth}
 \centering
 \includegraphics[width=.99\linewidth]{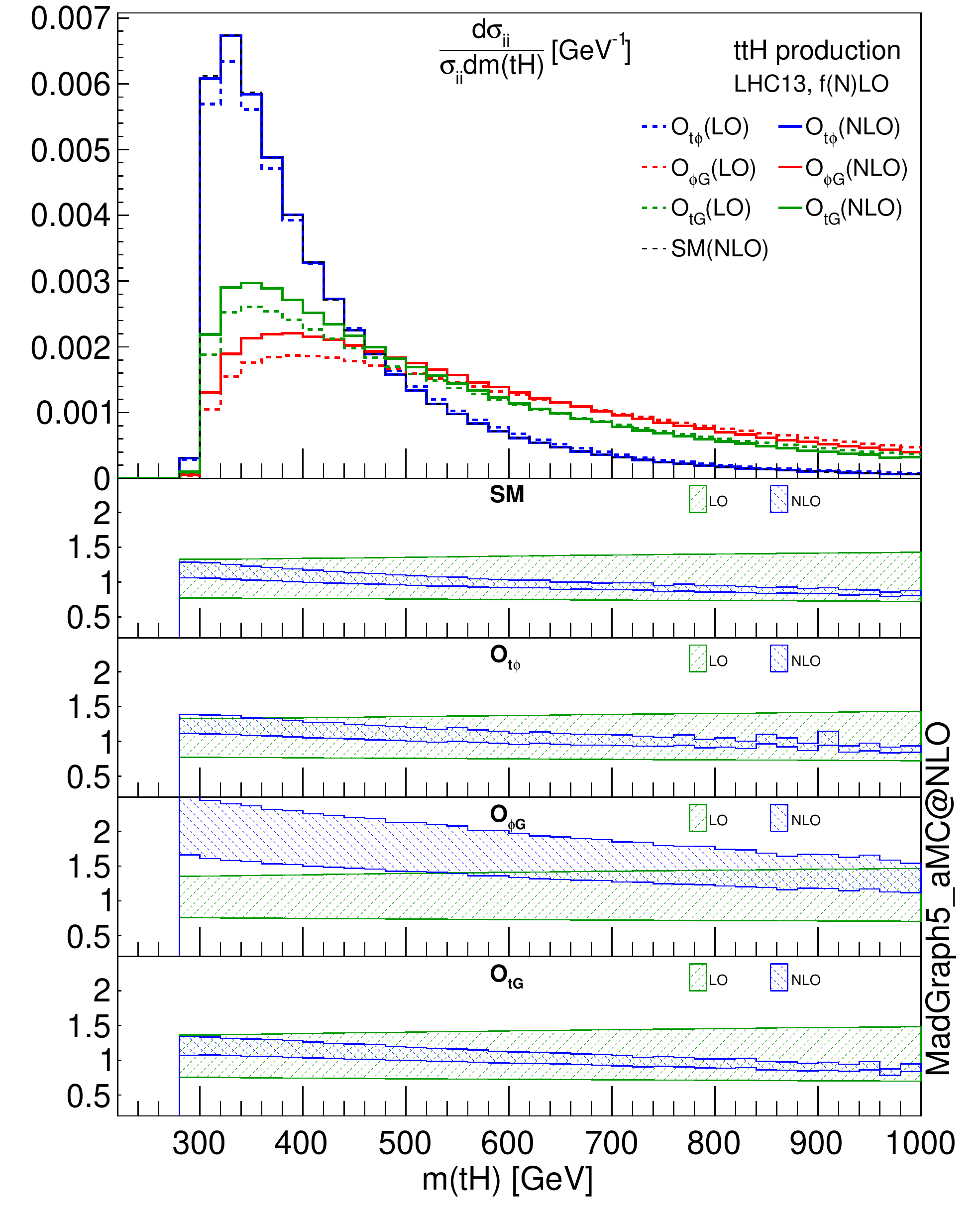}
 \end{minipage}
\caption{\label{fig:minv} 
Invariant mass distributions of the $t\bar t$ system (up) and the $t+H$ system
(down), normalised.  Left: interference contributions from $\sigma_i$. Right:
squared contributions $\sigma_{ii}$.  SM contributions and individual operator
contributions are displayed.  Lower panels give the $K$ factors and $\mu_{R,F}$
uncertainties.} 
\end{figure}

\begin{figure}[tb]
 \begin{minipage}[t]{0.5\linewidth}
\centering
\includegraphics[width=.99\linewidth]{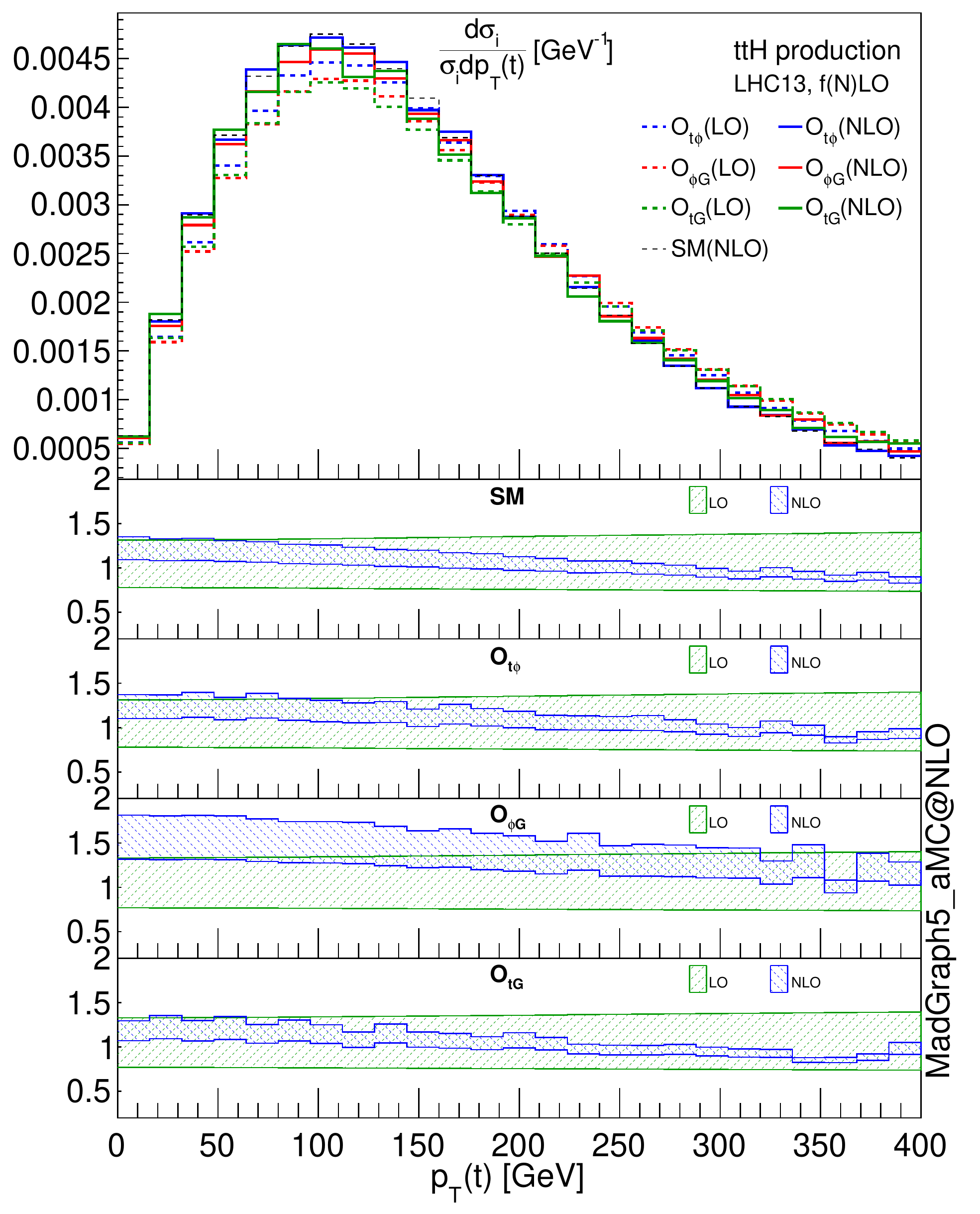}
\end{minipage}
\hspace{0.5cm}
 \begin{minipage}[t]{0.5\linewidth}
 \centering
 \includegraphics[width=.99\linewidth]{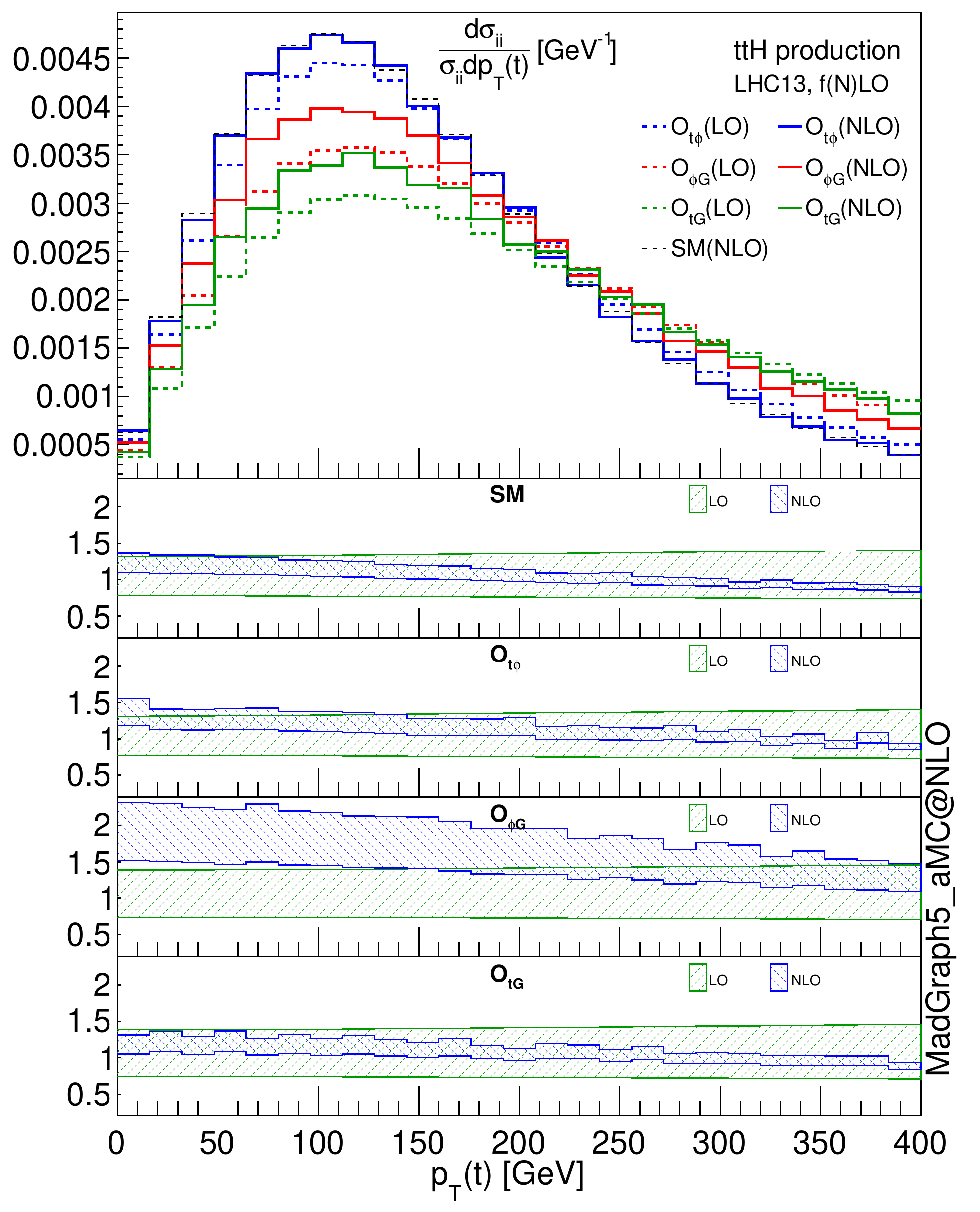}
 \end{minipage}
 \begin{minipage}[t]{0.5\linewidth}
\centering
\includegraphics[width=.99\linewidth]{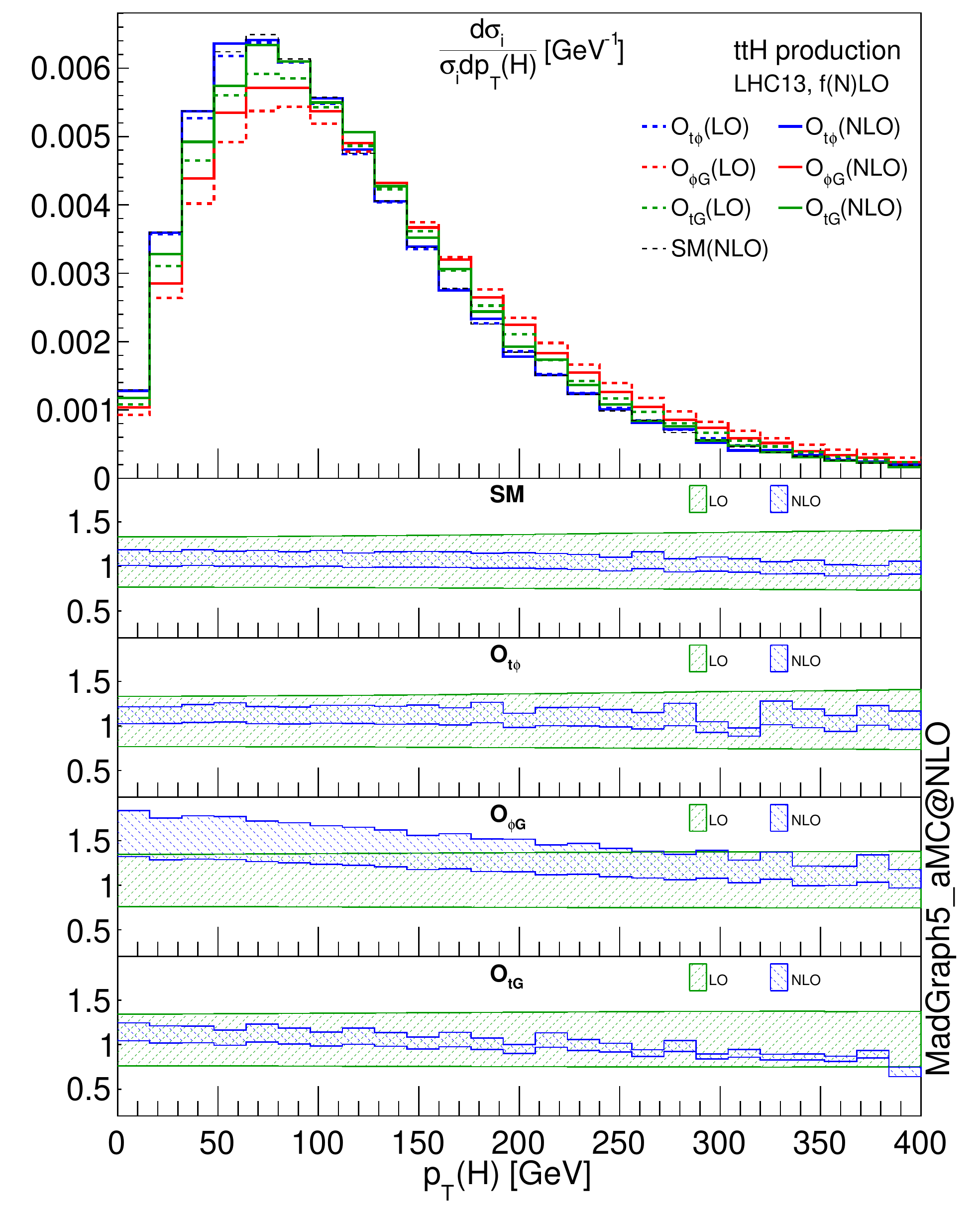}
\end{minipage}
\hspace{0.5cm}
 \begin{minipage}[t]{0.5\linewidth}
 \centering
 \includegraphics[width=.99\linewidth]{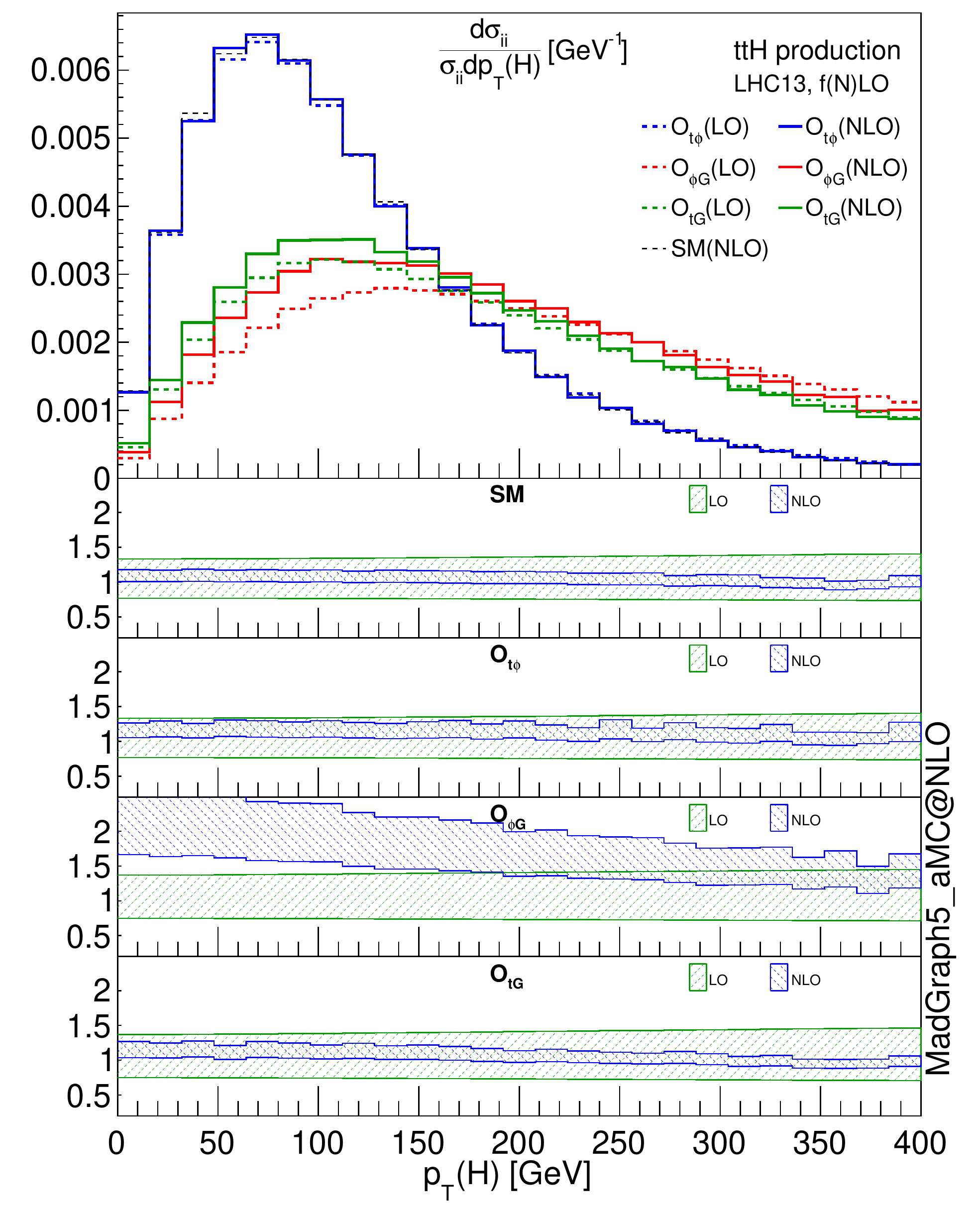}
 \end{minipage}
\caption{\label{fig:pt1} 
Transverse momentum distributions of the top quark (up) and the Higgs boson
(down), normalised.  Left: interference contributions from $\sigma_i$. Right:
squared contributions $\sigma_{ii}$.  SM contributions and individual operator
contributions are displayed.  Lower panels give the $K$ factors and $\mu_{R,F}$ uncertainties.} 
\end{figure}

\begin{figure}[tb]
 \begin{minipage}[t]{0.5\linewidth}
\centering
\includegraphics[width=.99\linewidth]{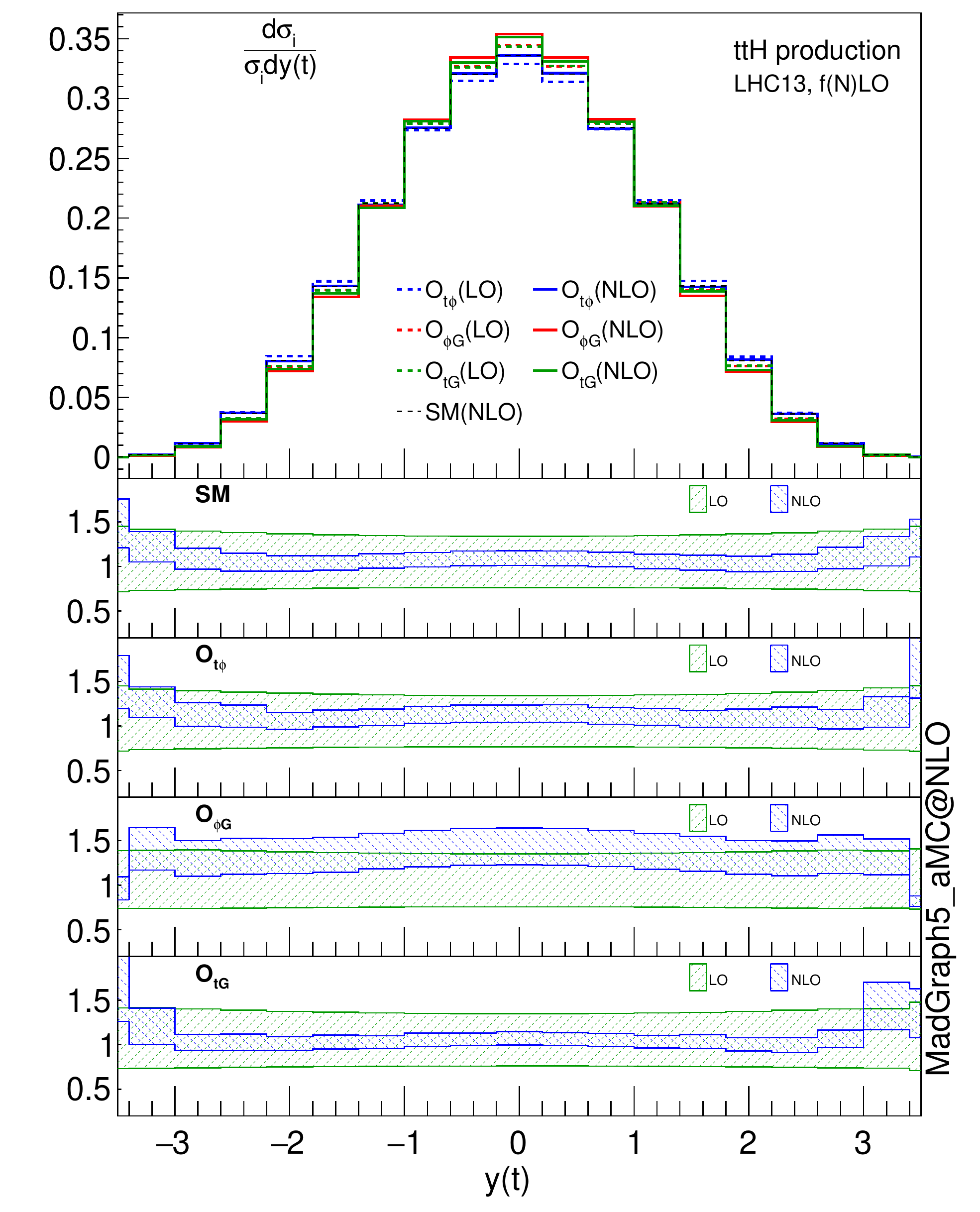}
\end{minipage}
\hspace{0.5cm}
 \begin{minipage}[t]{0.5\linewidth}
 \centering
 \includegraphics[width=.99\linewidth]{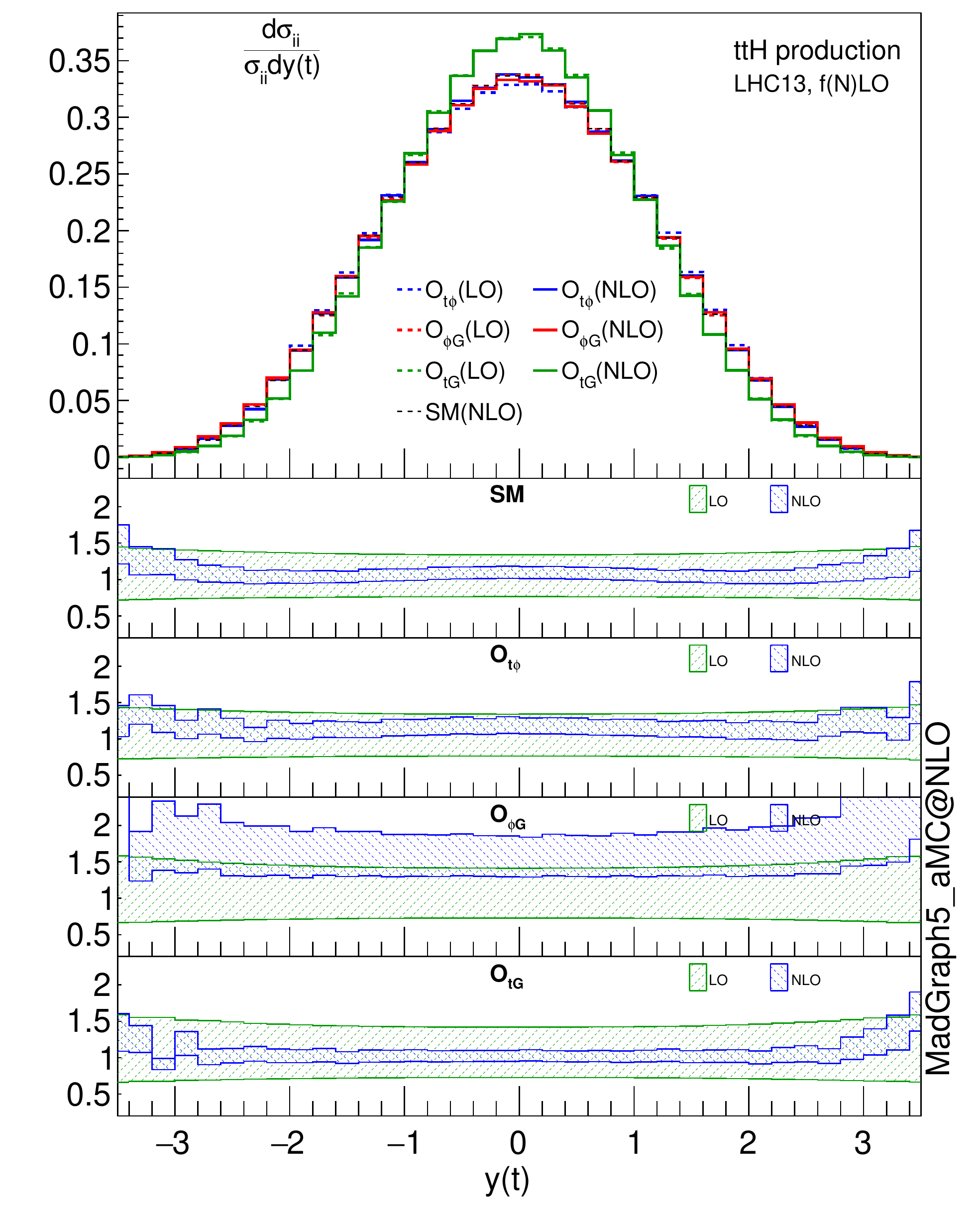}
 \end{minipage}
 \begin{minipage}[t]{0.5\linewidth}
\centering
\includegraphics[width=.99\linewidth]{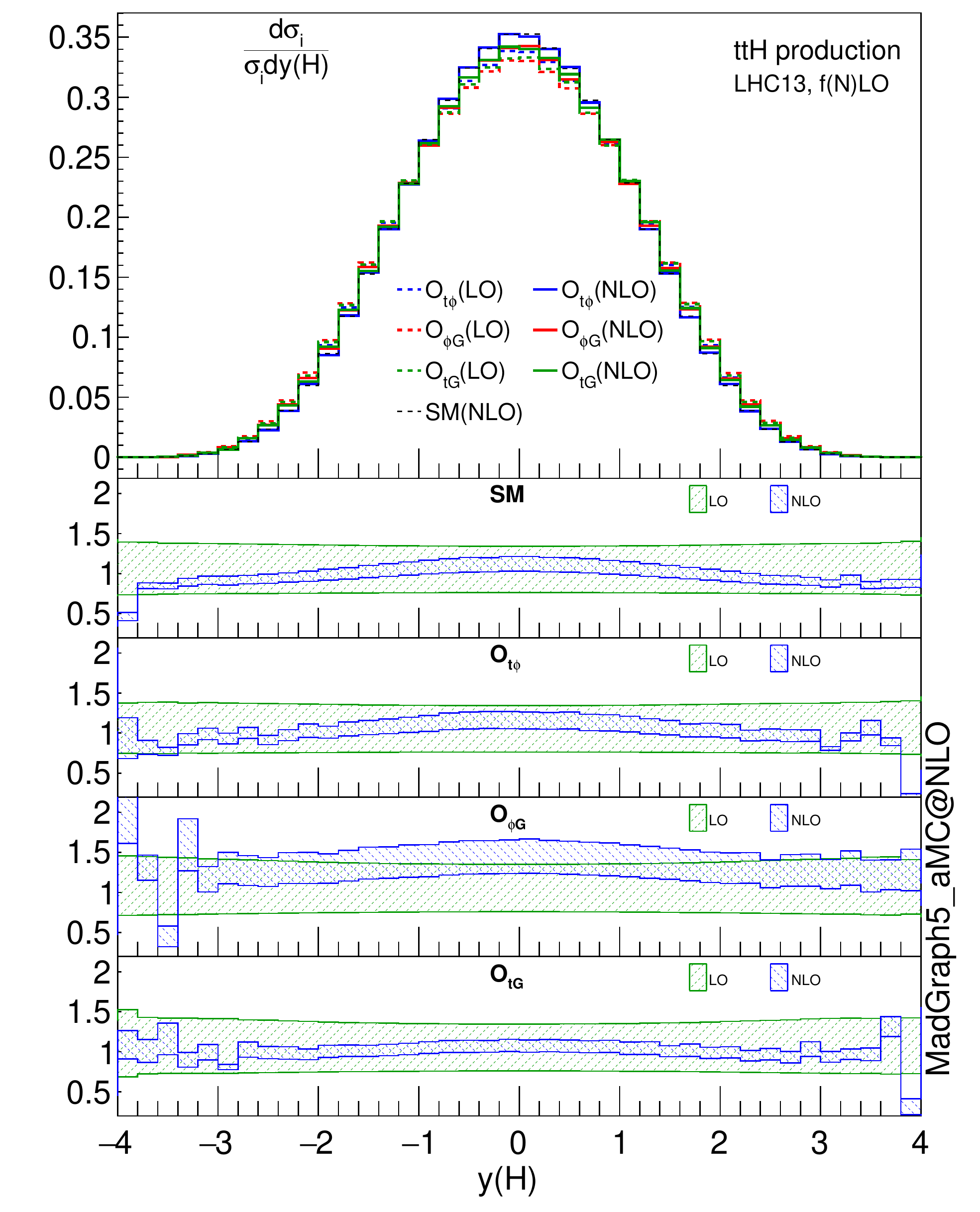}
\end{minipage}
\hspace{0.5cm}
 \begin{minipage}[t]{0.5\linewidth}
 \centering
 \includegraphics[width=.99\linewidth]{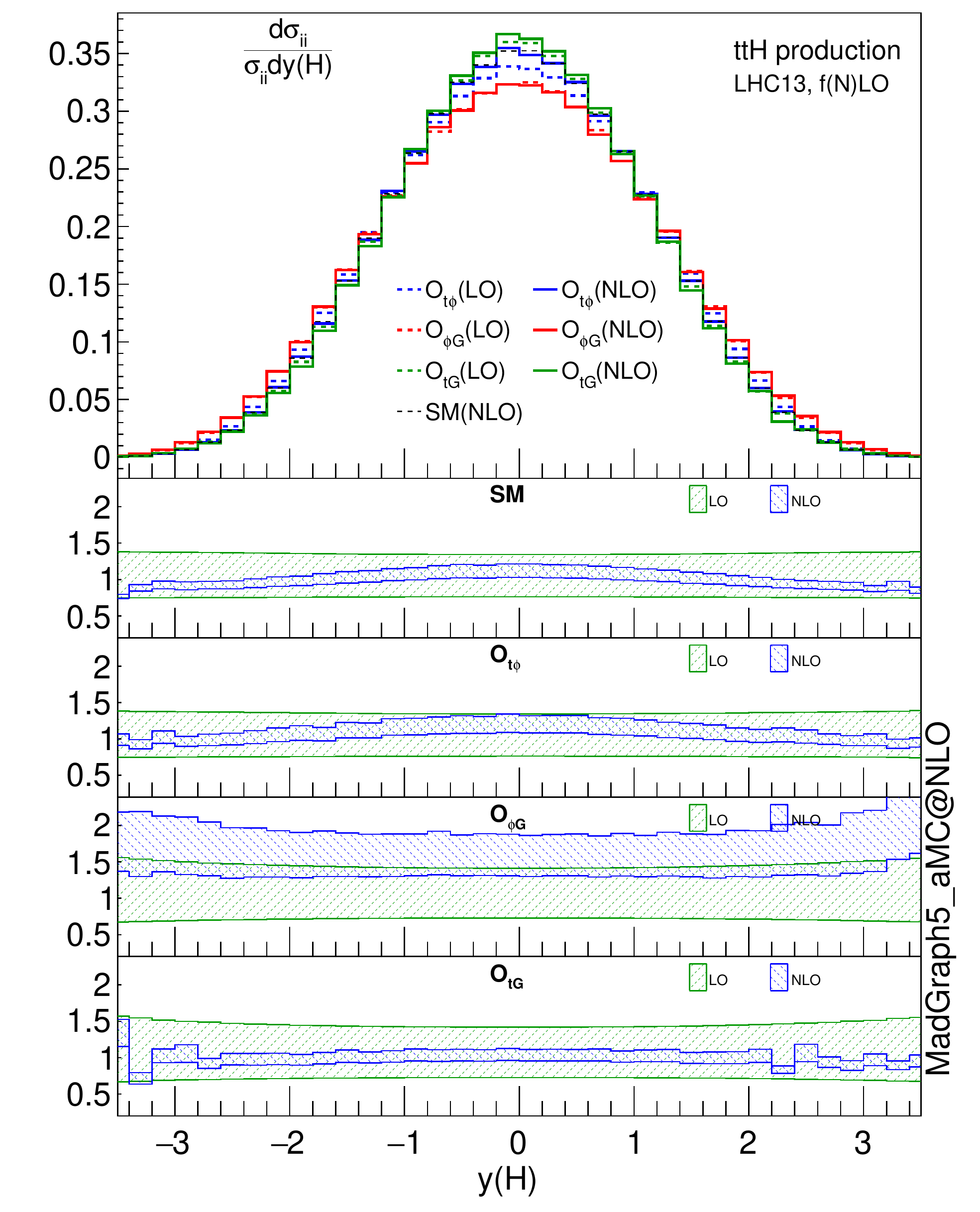}
 \end{minipage}
\caption{\label{fig:y} 
Rapidity distributions of the top quark (up) and the Higgs boson
(down), normalised.  Left:
interference contributions from $\sigma_i$. Right: squared contributions
$\sigma_{ii}$.  SM contributions and individual operator contributions are
displayed.
Lower panels give the $K$ factors and $\mu_{R,F}$ uncertainties.} 
\end{figure}

Results for $t\bar tH$ are given in Figures \ref{fig:pt2}-\ref{fig:y}.
The SM contribution as well as the individual
operator contributions, normalised, are displayed, in order to compare the
kinematic features from different operators.  The magnitudes can be read off
from the total cross section tables.  In the lower panel we give the
differential $K$ factors for each operator, together with the $\mu_{R,F}$ 
uncertainties.  Both interference and squared
contributions are shown.

\begin{figure}[H]
 \begin{minipage}[t]{0.5\linewidth}
\centering
\includegraphics[width=.99\linewidth]{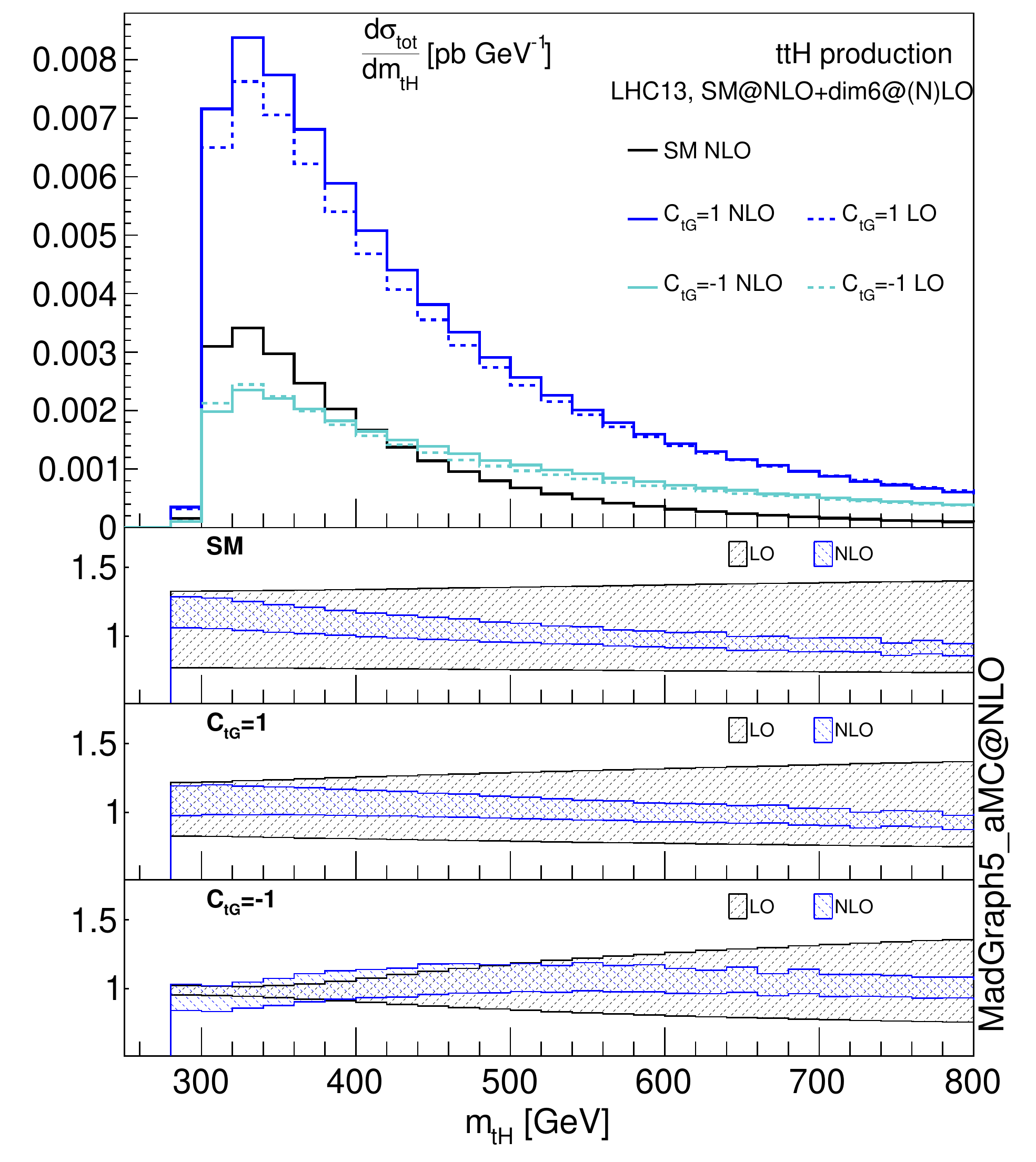}
\end{minipage}
\hspace{0.5cm}
 \begin{minipage}[t]{0.5\linewidth}
 \centering
 \includegraphics[width=.99\linewidth]{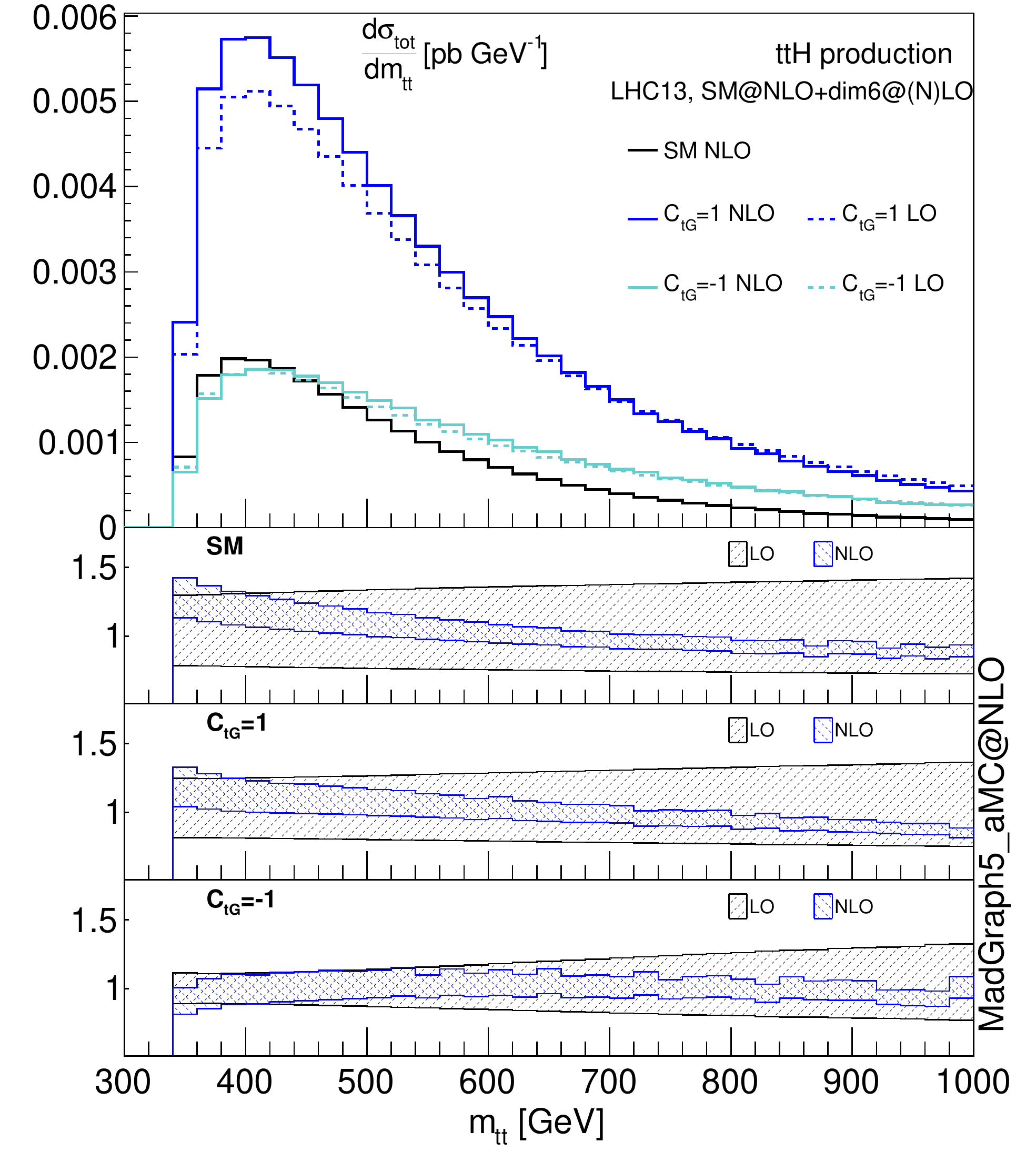}
 \end{minipage}
\caption{\label{fig:tot} 
Left: invariant mass distributions of $t+H$ system.
Right: invariant mass distributions of the top-pair system.
Results are displayed for different values of $C_{tG}$, assuming $\Lambda=1$ TeV.
Lower panels give the $K$ factors and $\mu_{R,F}$ uncertainties.} 
\end{figure}

The interference contributions are in general not sensitive to the operator,
except for a few cases such as $m(tH)$ and $p_T(H)$, which should be considered
as discriminating observables in a differential measurement.  On the contrary,
the squared contributions can be quite sensitive, and many of them can be used
to distinguish between the contributions from $O_{\varphi G}$, $O_{tG}$, and
those from SM background and $O_{t\varphi}$.  Given the current limits on the
coefficients, it is likely that the $O_{tG}$ operator still leads to observable
effects on the shape, due to large squared contributions.  As an example we
plot some total distributions for $C_{tG}=\pm1$ in Figure \ref{fig:tot}, and
one can see that in particular $C_{tG}=1$ leads to a very large deviation.  In
most cases, QCD corrections lead to nontrivial $K$ factors that are not flat
and can depend on operators, and using the SM $K$ factor is not a good
approximation.  These corrections change the shapes of differential
distribution, and missing such corrections could lead to bias in a fit where
differential information is used.

\begin{figure}[H]
 \begin{minipage}[t]{0.5\linewidth}
\centering
\includegraphics[width=.99\linewidth, trim= 2cm 6cm 0 0]{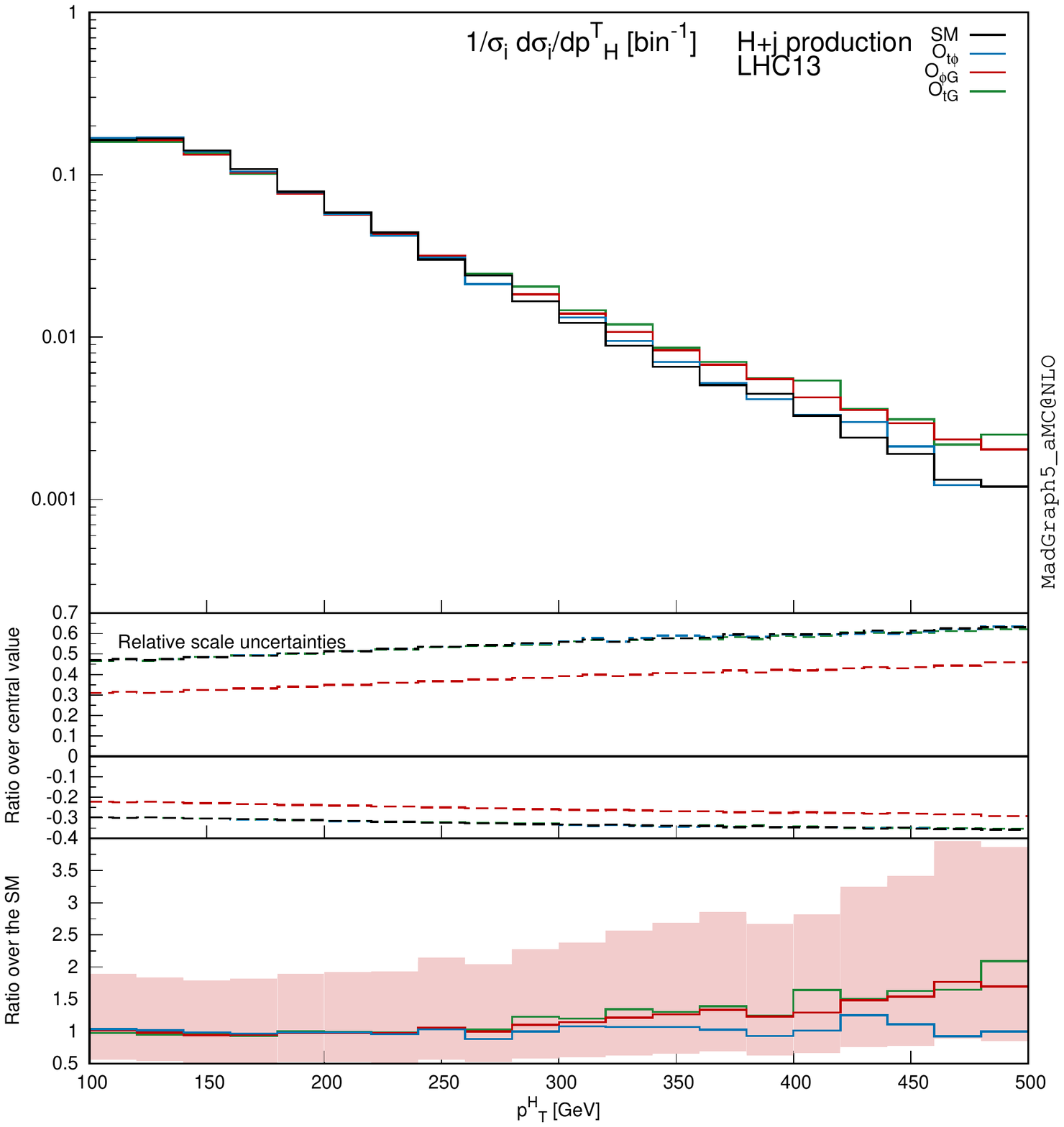}
\end{minipage}
\hspace{0.5cm}
 \begin{minipage}[t]{0.5\linewidth}
 \centering
 \includegraphics[width=.99\linewidth,trim= 2cm 6cm 0 0]{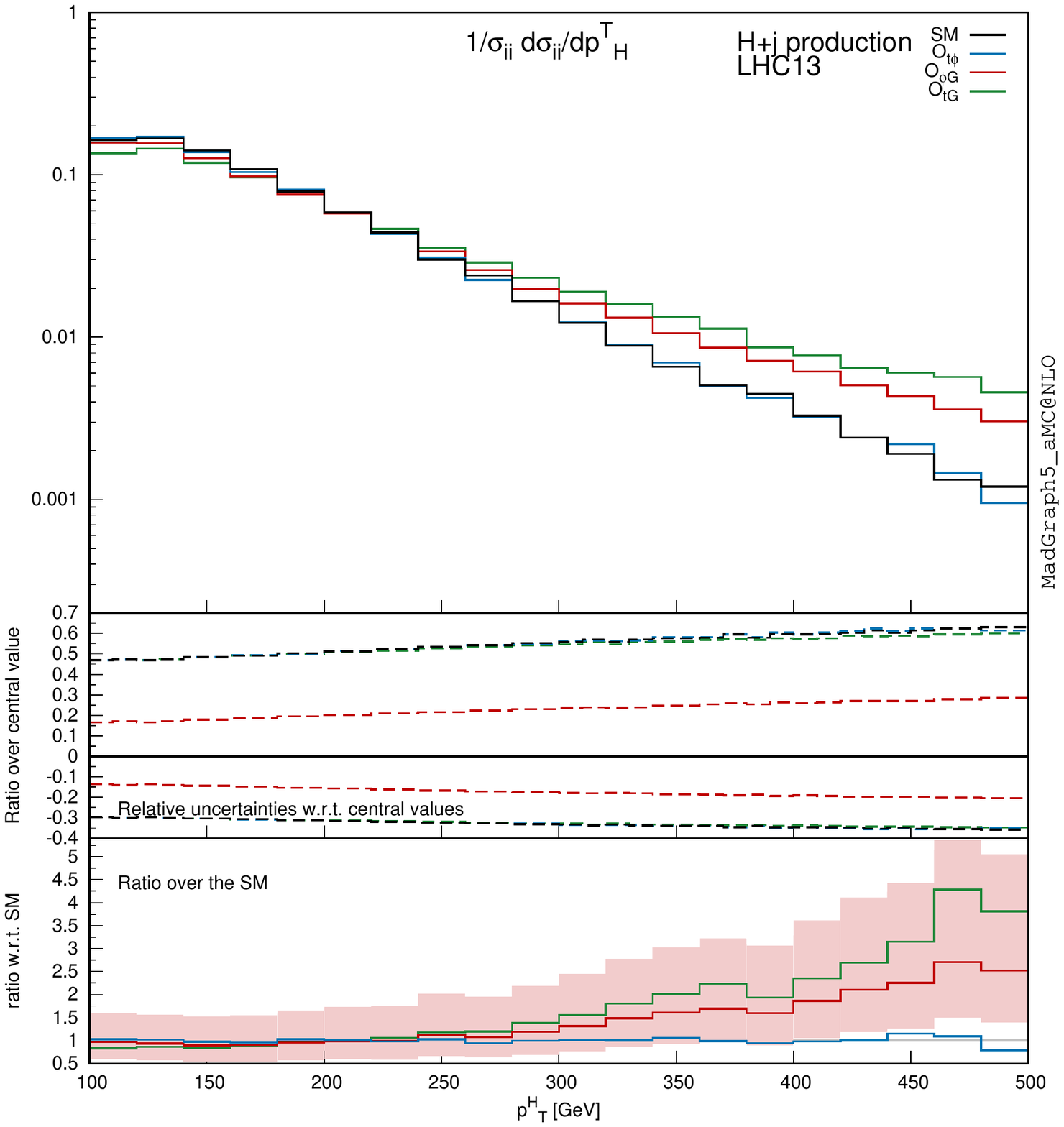}
 \end{minipage}
\caption{\label{fig:hjHpt} Higgs transverse momentum distribution in $Hj$, normalised.
Left: Interference contribution from $\sigma_i$.
Right: Squared contribution $\sigma_{ii}$. The SM and individual operator contributions are shown. 
Lower panels
give the $\mu_{R,F}$ uncertainties and the ratio over the SM.} 
\end{figure}

\begin{figure}[H]
 \begin{minipage}[t]{0.5\linewidth}
\centering
\includegraphics[width=.99\linewidth, trim= 2cm 6cm 0 0]{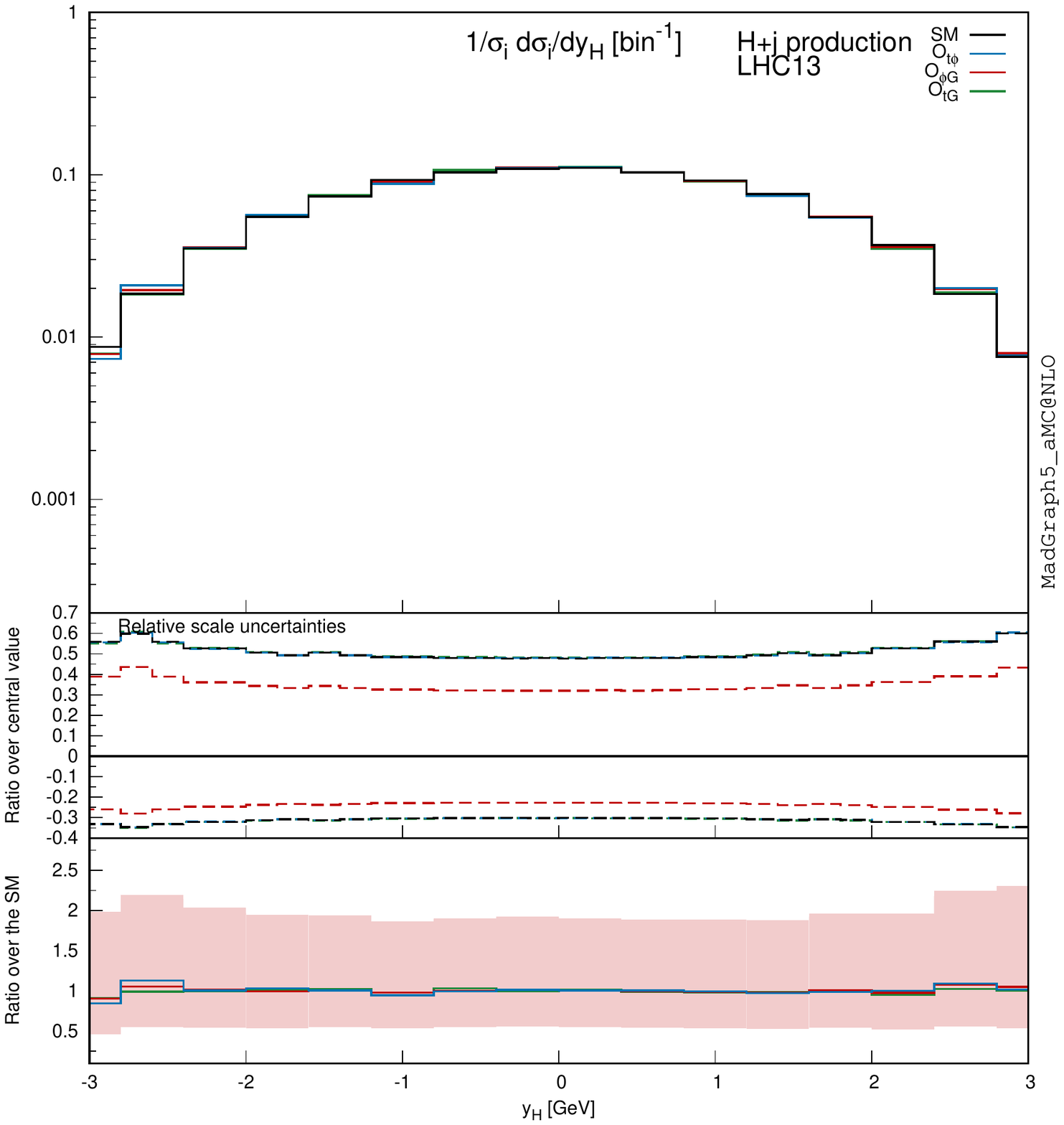}
\end{minipage}
\hspace{0.5cm}
 \begin{minipage}[t]{0.5\linewidth}
 \centering
 \includegraphics[width=.99\linewidth,trim= 2cm 6cm 0 0]{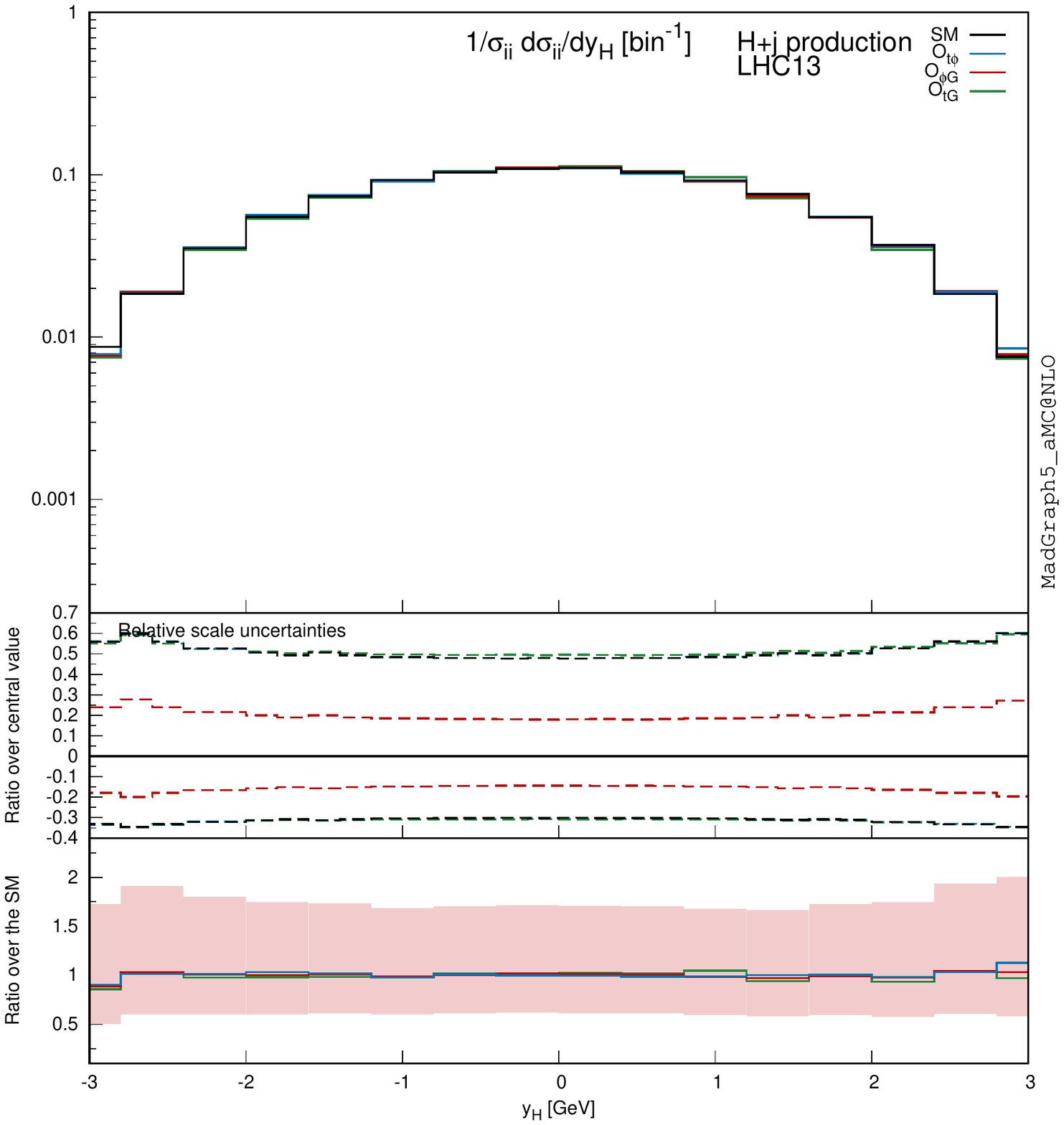}
 \end{minipage}
\caption{\label{fig:hjyH} Higgs rapidity distribution in $Hj$, normalised.
Left: Interference contribution from $\sigma_i$.
Right: Squared contribution $\sigma_{ii}$. The SM and individual operator contributions are shown. 
Lower panels
give the $\mu_{R,F}$ uncertainties and the ratio over the SM.} 
\end{figure}

\begin{figure}[H]
 \begin{minipage}[t]{0.5\linewidth}
\centering
\includegraphics[width=.99\linewidth, trim= 2cm 6cm 0 0]{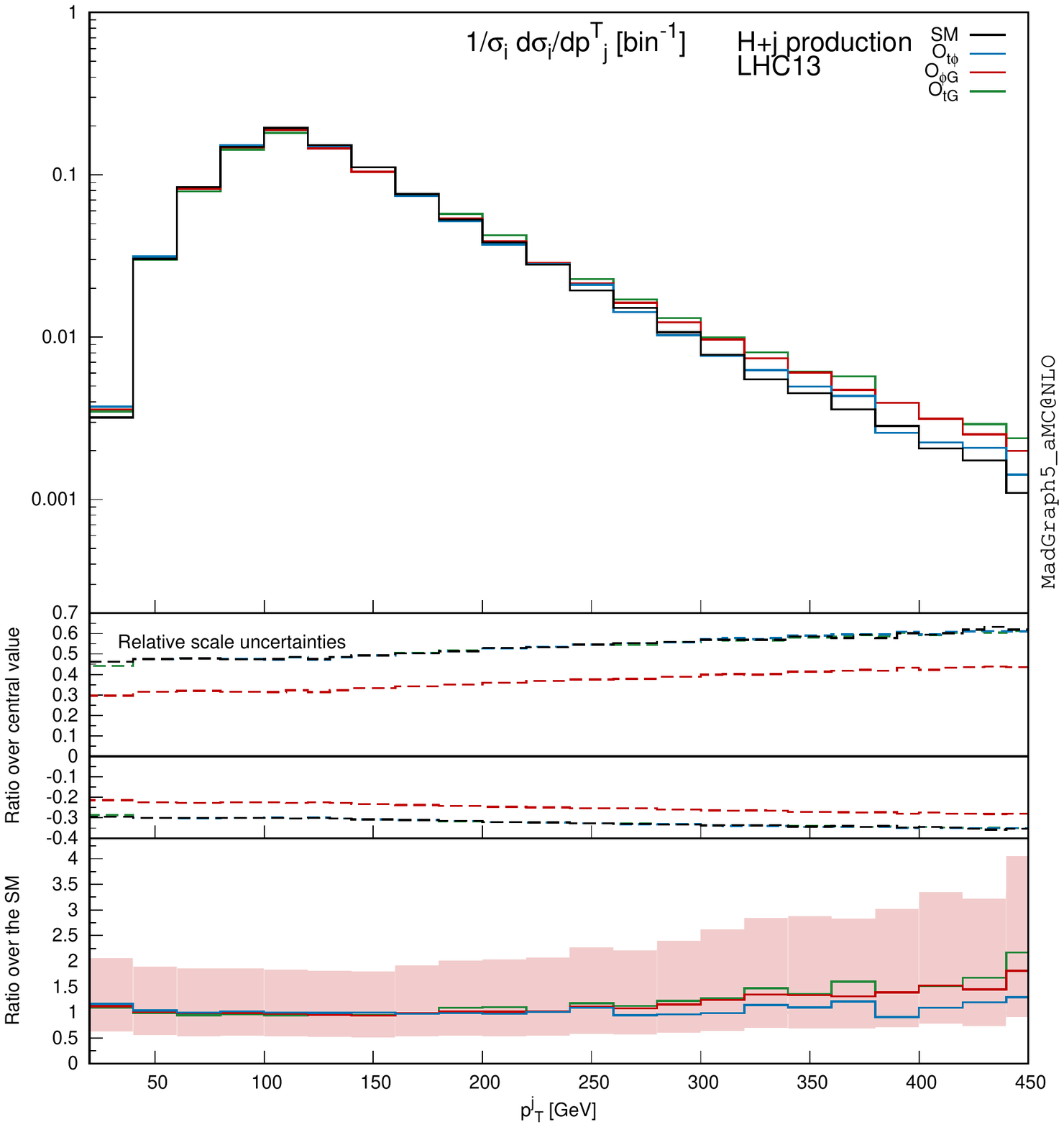}
\end{minipage}
\hspace{0.5cm}
 \begin{minipage}[t]{0.5\linewidth}
 \centering
 \includegraphics[width=.99\linewidth,trim= 2cm 6cm 0 0]{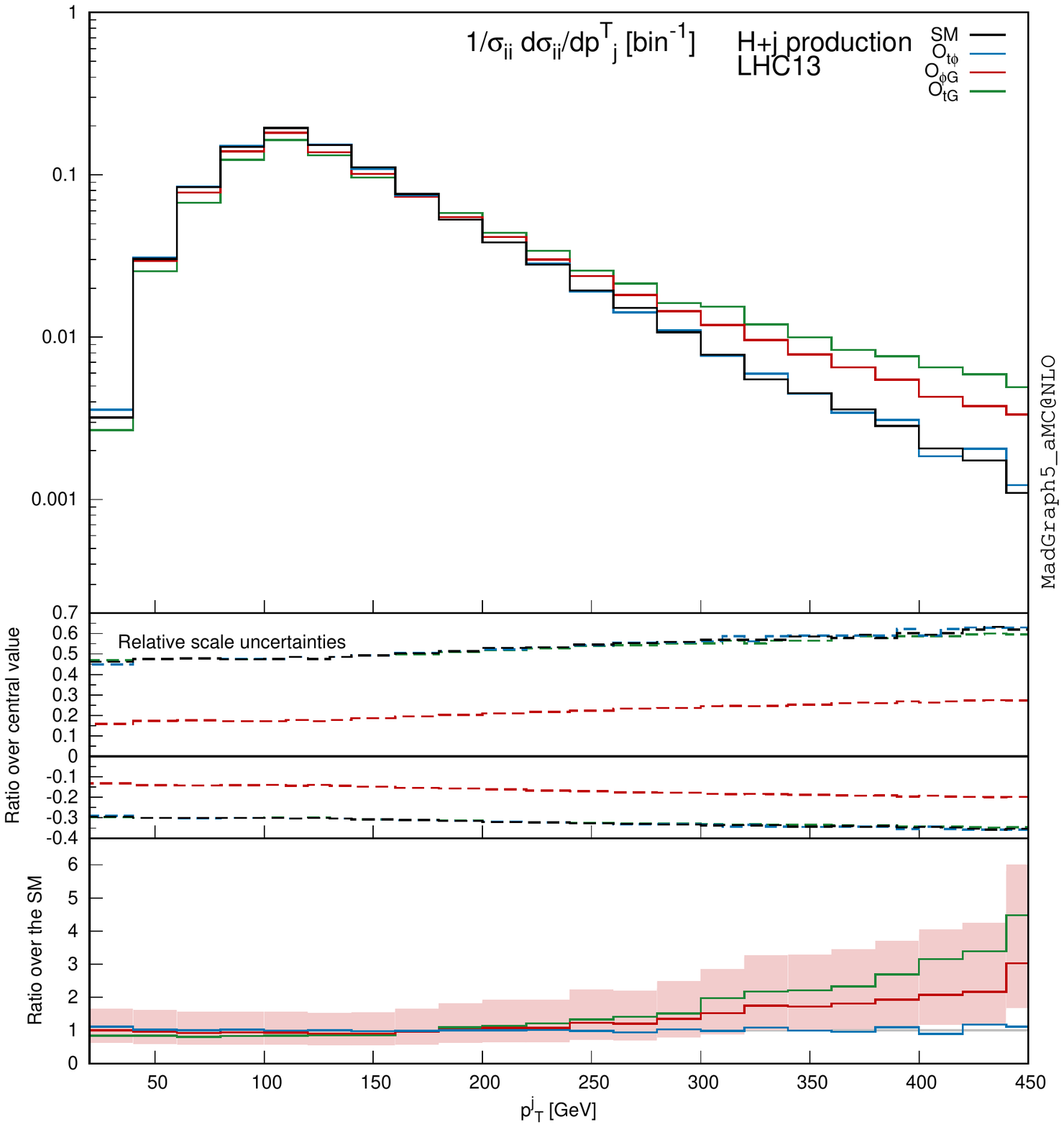}
 \end{minipage}
\caption{\label{fig:hjjpt} 
Transverse momentum distribution of the hardest jet in $Hj$, normalised.
Left: Interference contribution from $\sigma_i$.
Right: Squared contribution $\sigma_{ii}$. The SM and individual operator contributions are shown. 
Lower panels
give the $\mu_{R,F}$ uncertainties and the ratio over the SM.} 
\end{figure}

The corresponding distributions for the Higgs and jet transverse momentum and
Higgs rapidity in $Hj$ are shown in Figures \ref{fig:hjHpt}-\ref{fig:hjjpt},
where a 100 GeV cut has been imposed at parton level on the Higgs transverse
momentum. Both the linear and the quadratic terms are shown. The corresponding
scale uncertainty bands and the ratio over the SM with its scale uncertainty
are also shown in the lower panels.  

The  $O_{\phi G}$ and $O_{tG}$ give contributions which rise at high $p_T$, in
particular the squared contributions, while  $O_{t\phi}$ just gives rise to
shapes identical to those of the SM.  We note that the scale uncertainties in
the ratio over the SM are extremely small  for $O_{t\phi}$ and $O_{tG}$, and
therefore not visible in the plots, while for $O_{\phi G}$ they are larger as
we have also seen at the total cross-section level in Table \ref{tab:hh}. 

\begin{figure}[tb]
 \begin{minipage}[t]{0.5\linewidth}
\centering
\includegraphics[width=.99\linewidth, trim= 2cm 6cm 0 0]{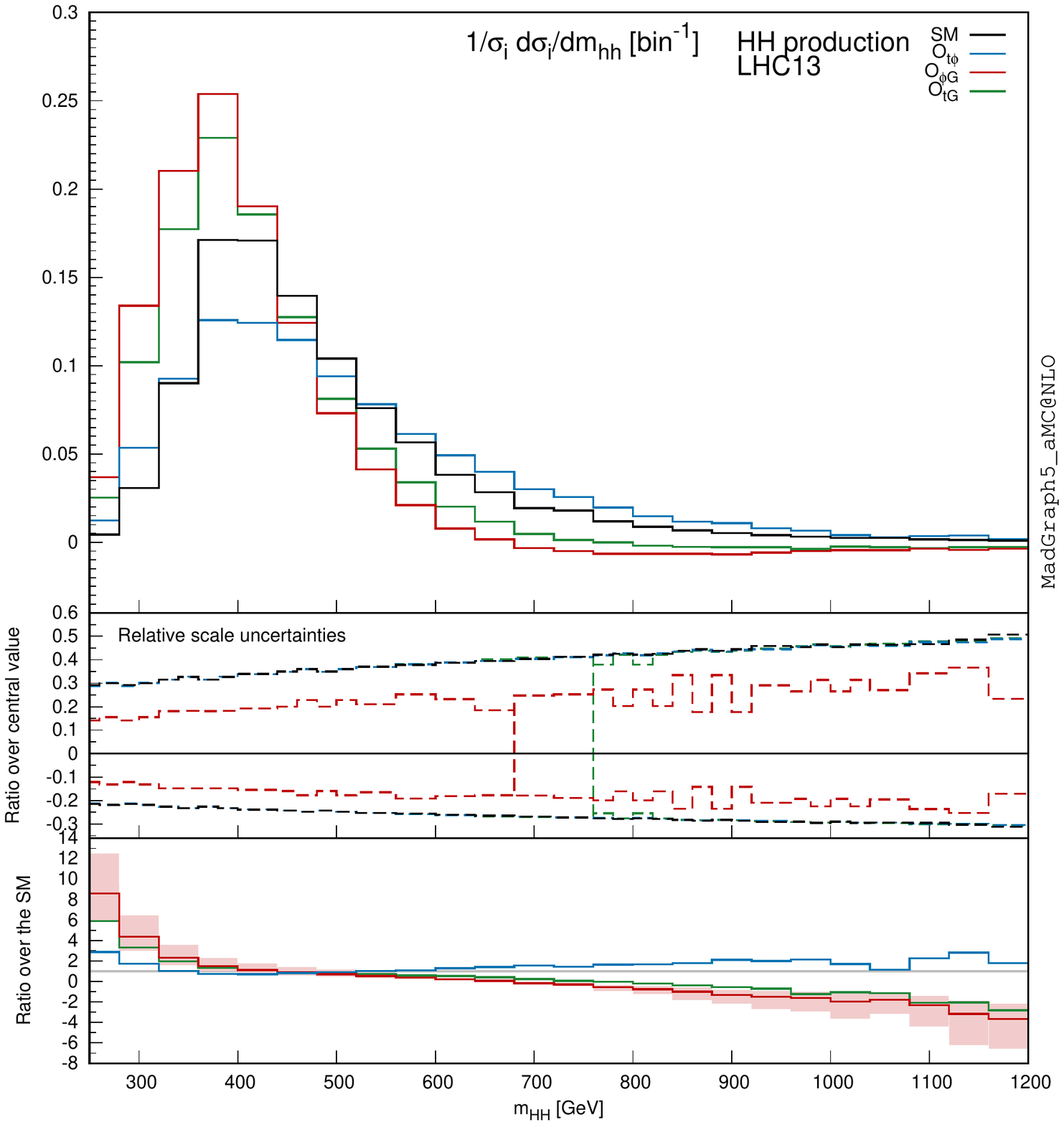}
\end{minipage}
\hspace{0.5cm}
 \begin{minipage}[t]{0.5\linewidth}
 \centering
 \includegraphics[width=.99\linewidth,trim= 2cm 6cm 0 0]{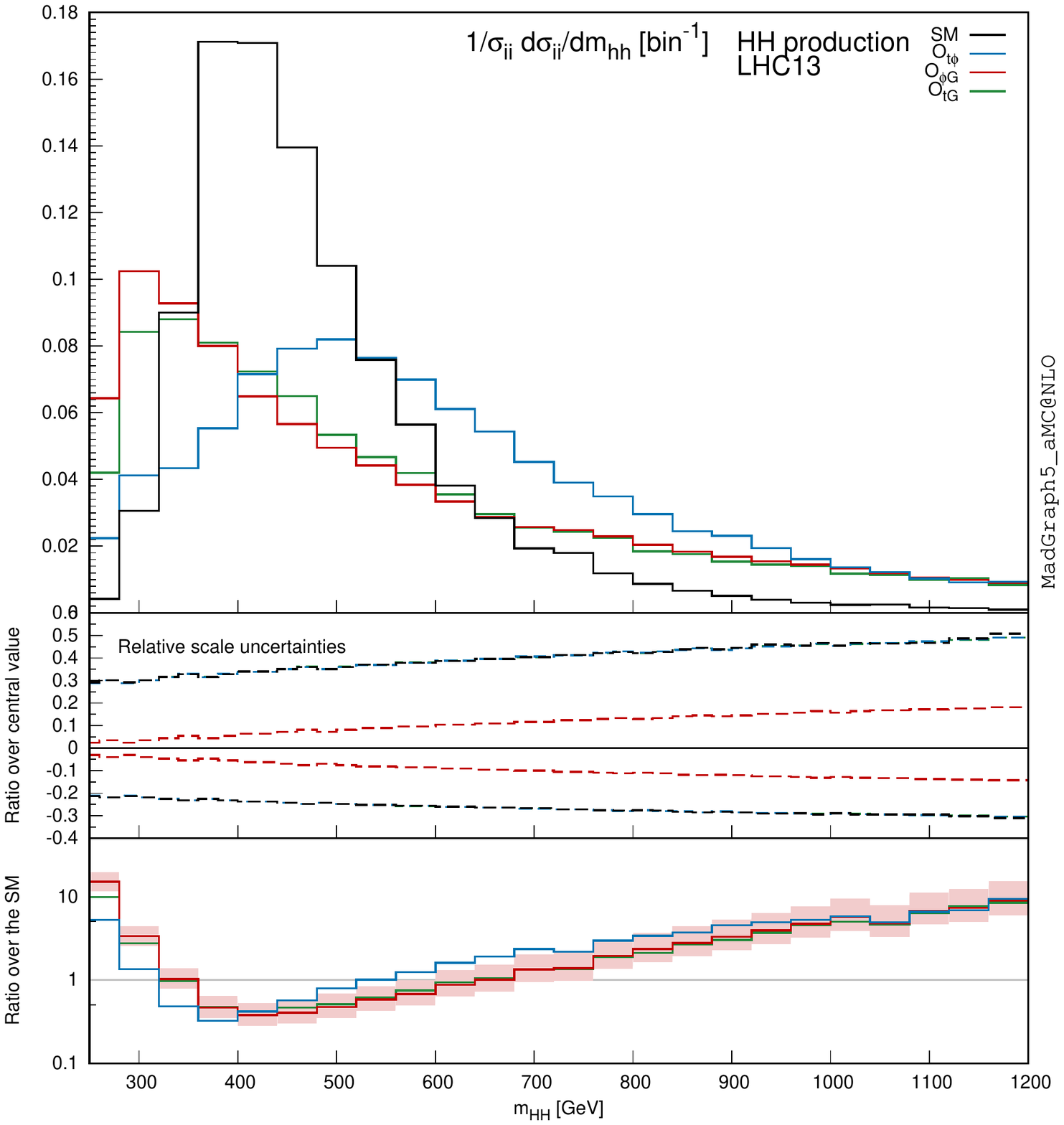}
 \end{minipage}
 \begin{minipage}[t]{0.5\linewidth}
\centering
\includegraphics[width=.99\linewidth, trim= 2cm 6cm 0 0]{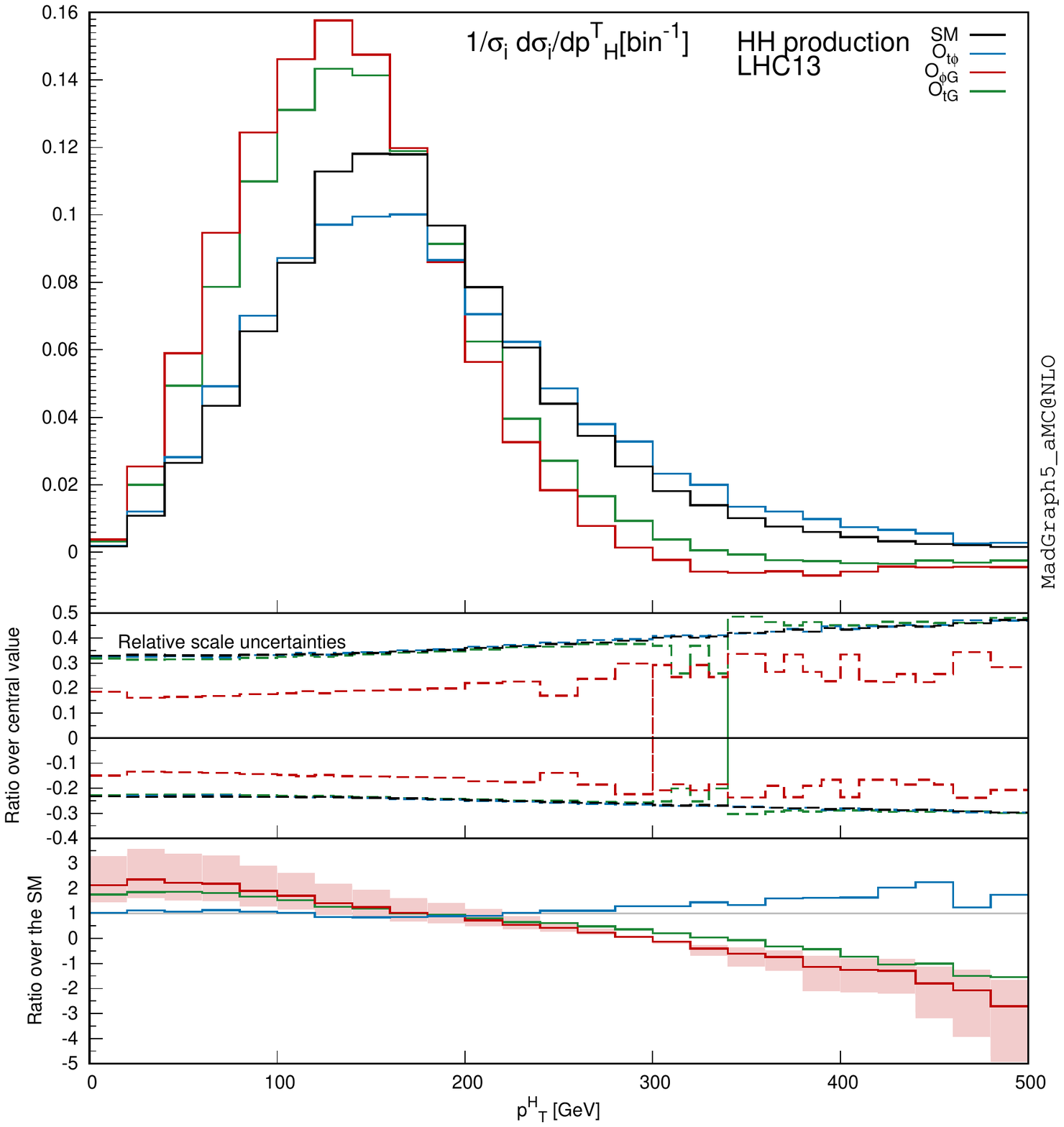}
\end{minipage}
\hspace{0.5cm}
 \begin{minipage}[t]{0.5\linewidth}
 \centering
 \includegraphics[width=.99\linewidth,trim= 2cm 6cm 0 0]{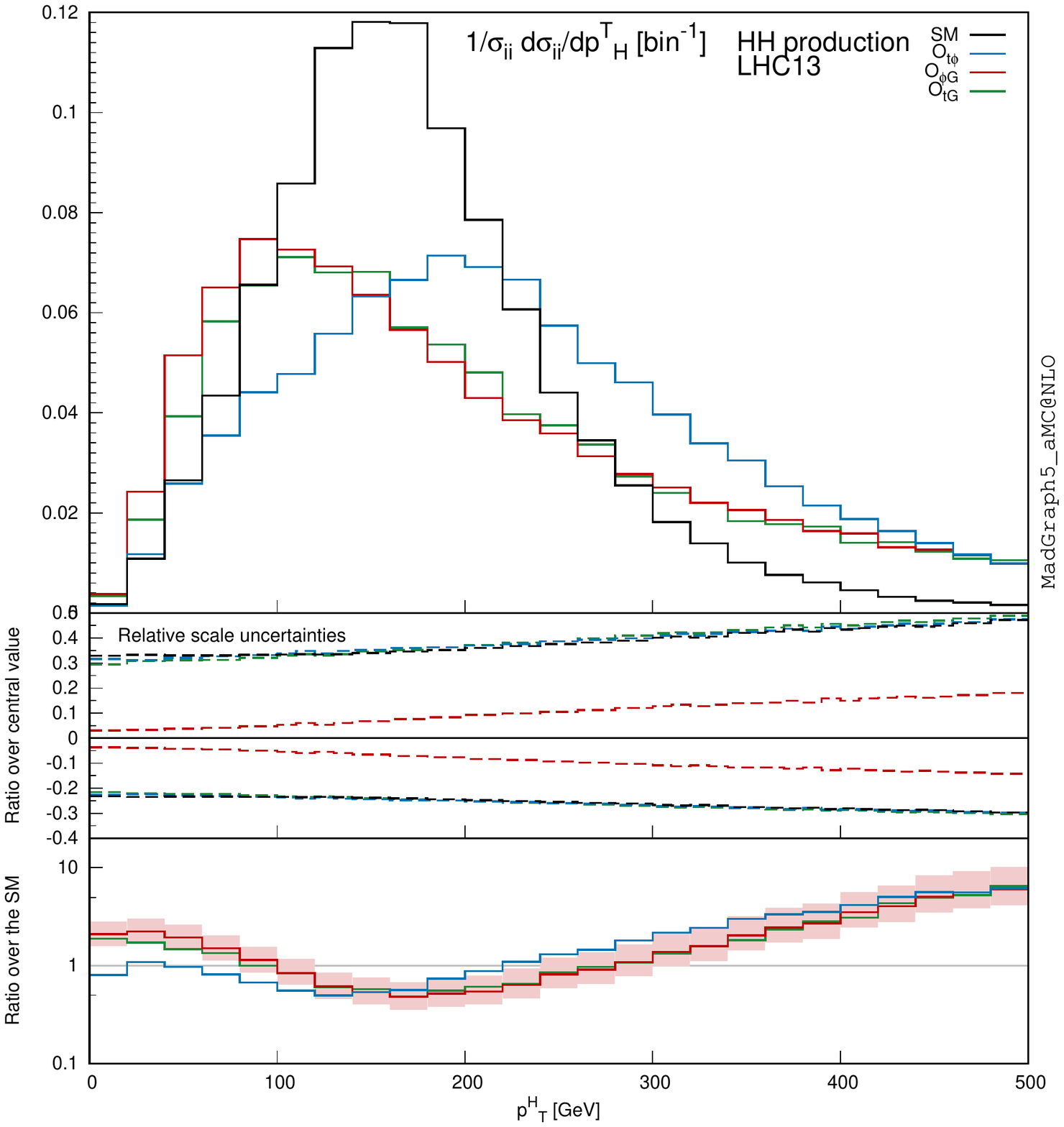}
 \end{minipage}
\caption{\label{fig:hh} 
Higgs pair Invariant mass distribution (top) and transverse momentum distribution of the hardest Higgs (bottom) in $p p \to HH$, normalised.
Left: Interference contribution from $\sigma_i$.
Right: Squared contribution $\sigma_{ii}$. The SM and individual operator contributions are shown. 
Lower panels
give the $\mu_{R,F}$ uncertainties and the ratio over the SM.} 
\end{figure}

The results for the Higgs pair invariant mass and hardest Higgs $p_T$ in $pp
\to HH$ are shown in Figure \ref{fig:hh}. All operators lead to shapes which
differ from the SM, with the squared contributions leading to distributions
rising fast with the energy. We find that the interference for $O_{\phi G}$ and
$O_{tG}$ can be destructive or constructive depending on the region of the
phase space.  

\section{RG effects}
\label{sec:rg}

In an NLO calculation in the SMEFT, the  logarithmic dependence on $\mu_{EFT}$ arises as a consequence of the running and mixing effects among
dimension-six operators.  These terms are very important for correctly interpreting the NLO results, for extracting reliable constraints from an
experimental analysis, and for estimating the theoretical uncertainties due to missing higher orders.  
On the other hand, in general, they do not provide an approximation of the complete NLO corrections.  Many discussions  on these issues are available
in the literature and we refer the reader to, e.g., Refs.~\cite{Passarino:2012cb,Passarino:2138031,Contino:2016jqw,Contino:2137947,Hartmann:2015oia,Hartmann:2015aia}. 

In this section, we present a study of  the RG effects using our NLO calculation as a concrete example.  In particular, we focus on the role of RG effects in a full NLO calculation, and on their use in the estimate of missing higher-order corrections. The impact of the $\mu_{EFT}$ dependence on the extraction of experimental constraints is further discussed in Section \ref{sec:fit}.

\subsection*{Comparing RG corrections with full NLO}

By naive power counting one might think that in an NLO calculation RG
corrections to operators dominate over the finite pieces, as they are enhanced
by factors of $\log(\Lambda/Q)$, where $Q$ is a relatively low energy scale,
at which the measurement is performed. This statement is not accurate.
First, a measurement at scale $Q$ is
designed to measure parameters defined at that same scale, i.e.~$C(Q)$, so
in practice $\log(\Lambda/Q)$ will never appear in a perturbative
expansion.  Rather, these log terms are resummed by using RG equations,
independent of processes, so they are not part of an NLO calculation, but are
contained in the definition of coefficients. The large logarithmic terms will
play a role only if one wants to relate coefficients with underlying models,
which is not the task of an NLO calculation in the EFT.

\begin{figure}[h!tb]
	\begin{center}
		\includegraphics[width=.6\linewidth]{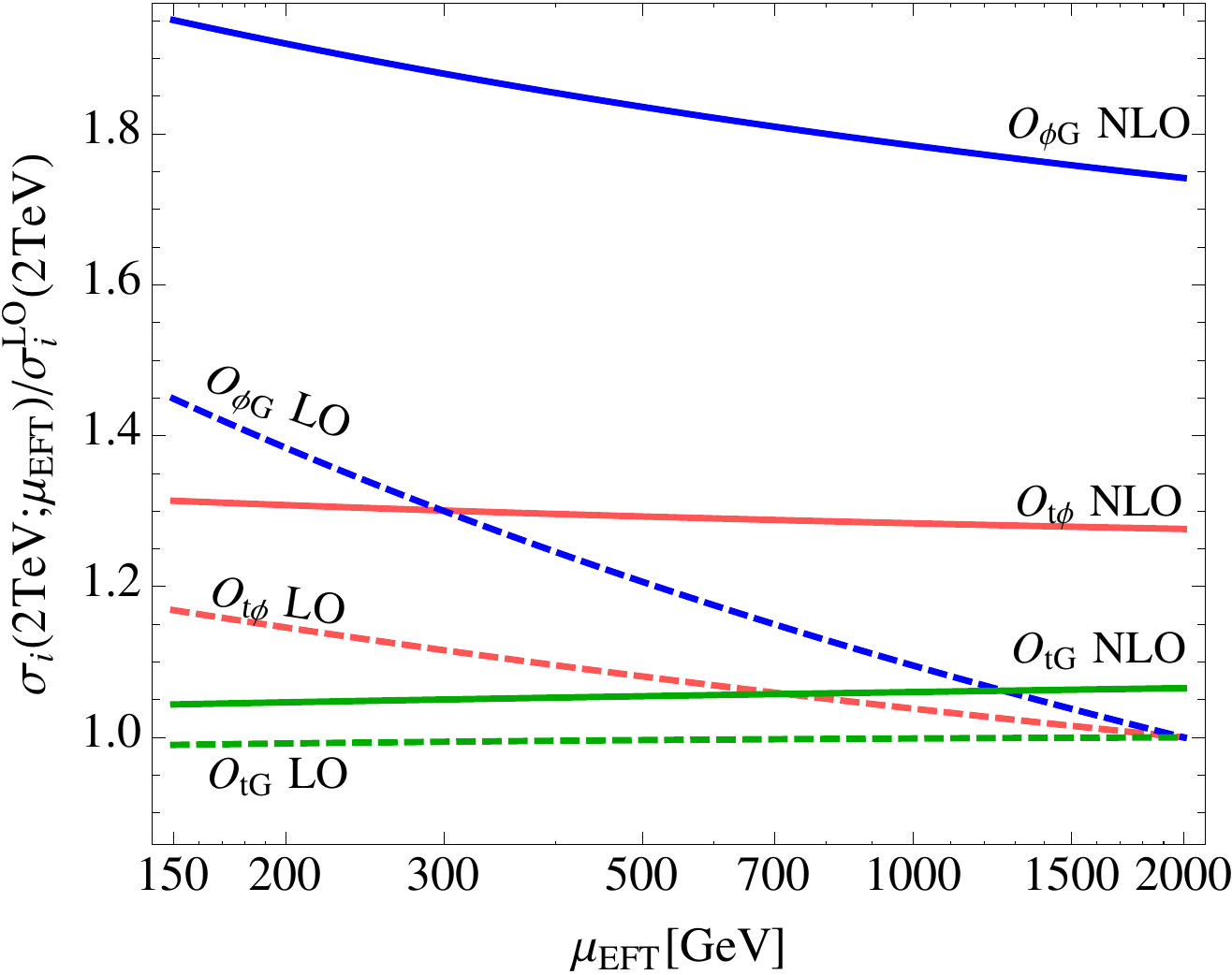}
	\end{center}
	\caption{Comparison of the RG corrections with the exact NLO results for $t\bar tH$ production.}
	\label{fig:rgnlo}
\end{figure}

Even if we do not resum these log terms, by setting $\mu_{EFT}=\Lambda$
explicitly in a perturbative calculation (which is looking for trouble and
should not be done),
the $\log(\Lambda/Q)$ terms appearing in the NLO corrections are often not the
major contribution, because at the LHC QCD corrections are typically much larger
than what one would naively estimate using $\mathcal{O}(\alpha_s/4\pi)$. It is
interesting to compare the two kinds of corrections, i.e.~RG and NLO, in the $t\bar tH$
process. In Figure~\ref{fig:rgnlo} we show the interference contribution from
three operators, $\sigma_i(\Lambda;\mu_{EFT})$ (i.e.~contributions from
$C_i(\Lambda)$, calculated with $\mu_{EFT}$) for $\Lambda=2$ TeV.  Suppose we
have an underlying theory which we match to an EFT at scale $\Lambda$ with
three coefficients $C_{t\phi}$, $C_{\phi G}$ and $C_{tG}$.  We can do a LO
calculation without running, and we normalize the results to one.  Now we may
use the RG equations to improve the results, by running the coefficients to a
lower scale near $m_t$.  The dashed lines indicate corrections from one-loop RG
only.  These corrections ranges from roughly $0$ to $40\%$.  However if we go to
NLO, the increase is much larger, depending on where the scale is, as indicated
by the solid lines.  This clearly demonstrates that RG corrections are far from
a good approximation to NLO corrections.

\subsection*{EFT scale as an uncertainty estimator}

The RG running can still be used as an estimator of missing higher order
corrections to the operators.  From Figure~\ref{fig:rgnlo} we can see that the
EFT scale dependence of the LO results roughly captures the NLO corrections at
the same order of magnitude.  On the other hand, at NLO the EFT scale
dependence becomes much smaller, indicating that the EFT scale uncertainty can
be taken under control by using the full NLO prediction.

The curves in Figure~\ref{fig:rgnlo} take into account both running and mixing
effects.  For
example, as $O_{tG}$ runs downs, it will also mix into $O_{t\phi}$ and $O_{\phi
G}$, and the green curves are the sum of all three contributions.  It is this sum
that becomes less dependent on scales at NLO.  The individual contribution
from each operator is however not physical, in particular in loop induced
processes, where the separation between tree-level and loop-level contributions
from two different operators is scale dependent.  For this reason, loop-induced
contributions are sensitive to $\mu_{EFT}$ and should be used with care when
setting bounds on individual operators. This will be further discussed in the following section 
where we set constraints on the operators.

\begin{figure}[h!tb]
	\begin{center}
		\includegraphics[width=.49\linewidth]{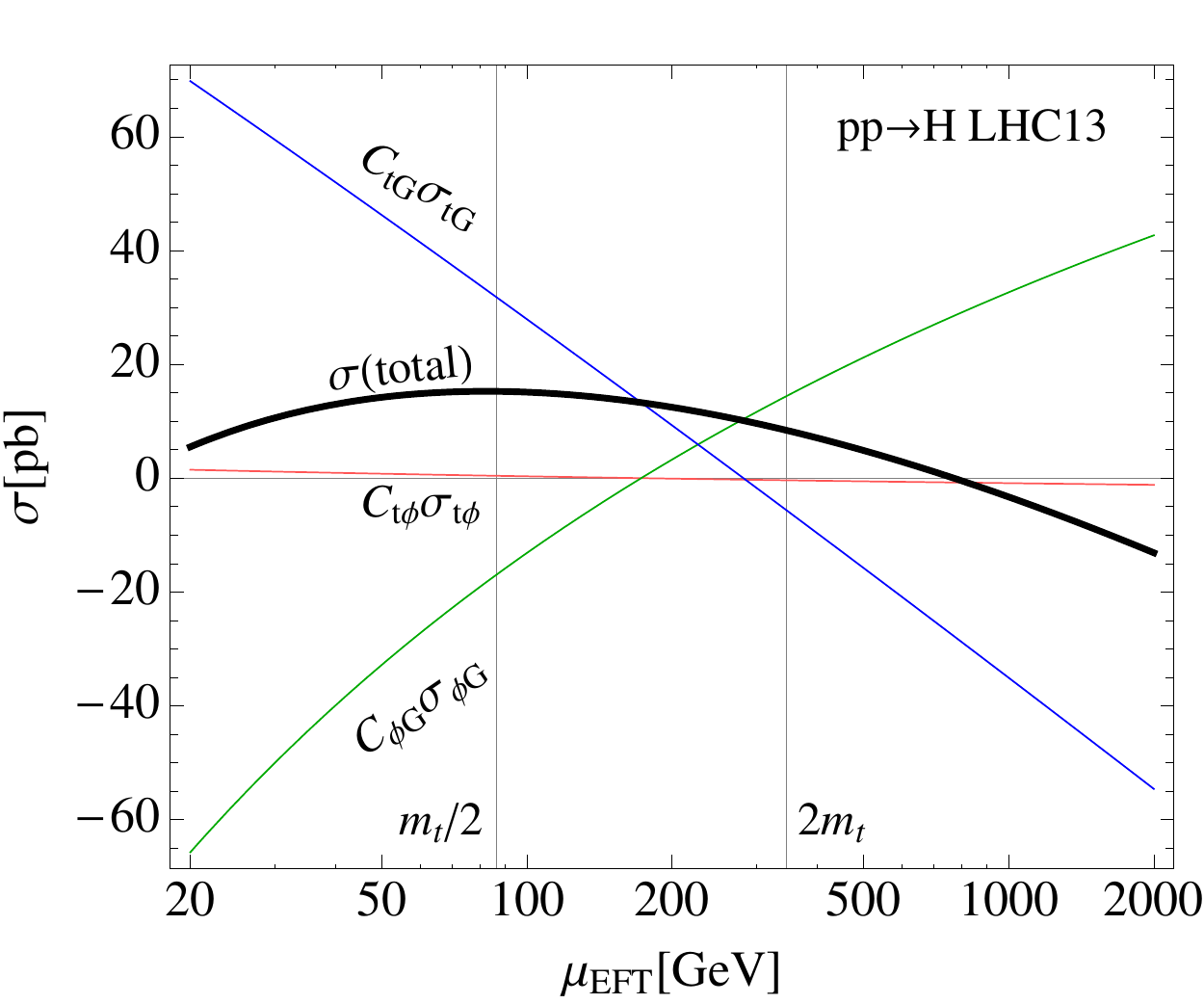}
		\includegraphics[width=.49\linewidth]{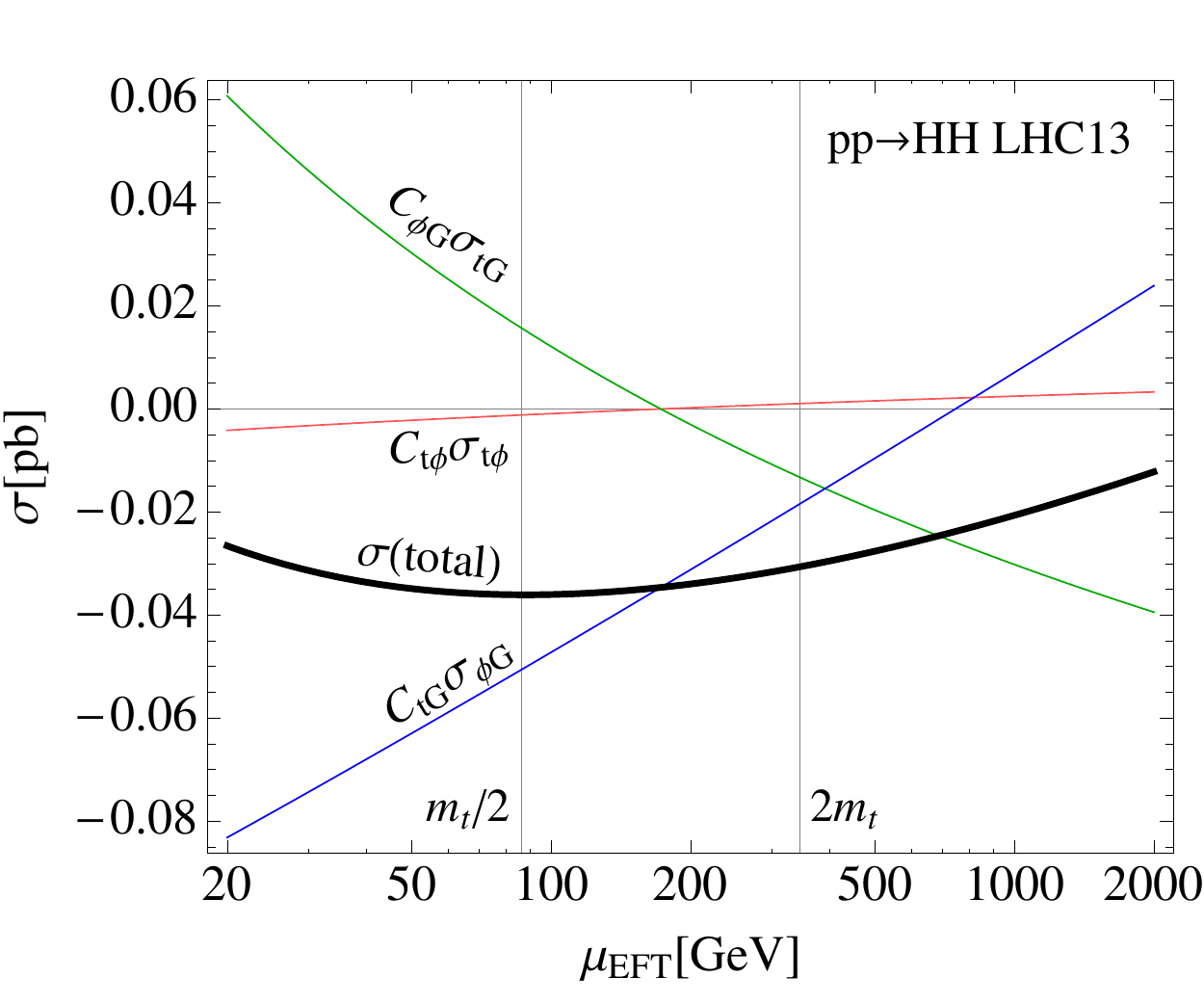}
	\end{center}
	\caption{$O_{tG}$ contribution in loop-induced processes, assuming
	$C_{tG}(m_t)=1$, $C_{t\phi}(m_t)=C_{\phi G}(m_t)=0$, and $\Lambda=1$
	TeV.  Left: $pp\to
H$. Right: $pp\to HH$. Individual linear contributions from each operator, as well
as their sum, i.e.~$\sigma_{tG}(m_t;\mu_{EFT})$, are displayed.}
	\label{fig:tgloop}
\end{figure}

For illustration, in Figure~\ref{fig:tgloop} we plot the individual contributions from the three
operators in $pp\to H$ and $pp\to HH$, as a function of $\mu_{EFT}$, assuming
$C_{tG}=1,\ C_{t\phi}=C_{\phi G}=0$ at scale $\mu_{EFT}=m_t$, and
$\Lambda=1$ TeV.  At this scale
the only contribution is from $O_{tG}$.  When $\mu_{EFT}$ deviates from $m_t$,
while the running of $C_{tG}$ is only at the percent level, its cross section
has a strong $\mu_{EFT}$ dependence as can be seen from the ``$C_{tG}\sigma_{tG}$''
curves; in the meantime, non-zero values for $C_{t\phi}$ and $C_{\phi G}$ are
induced by $O_{tG}$, in particular the latter leading to a tree level
contribution that also depends on $\mu_{EFT}$, as can be seen from
the ``$C_{\phi G}\sigma_{\phi G}$'' curves.  These dependences are canceled
out at the leading log level when all three contributions are summed,
as presented by the black curves labeled as ``$\sigma$(total)''.  This
quantity is nothing but the $\sigma_{tG}(m_t;\mu_{EFT})$ defined in
Eq.~(\ref{eq:sigma2}).  It is the physical contribution coming from
$C_{tG}(m_t)=1$, and has a weaker dependence on $\mu_{EFT}$ between $m_t/2$
and $2m_t$.  This quantity should be used as an estimation of the missing
higher-order corrections to the effective operators, and should be presented in
a perturbative EFT prediction, in the same way as the normal $\mu_{R,F}$
uncertainties are usually given.

\section{Constraints on dimension-six operators}
\label{sec:fit}
In this section we study the experimental constraints on the operator coefficients.
Both $pp\to H$ and $pp\to t \bar{t}H$ have been already measured at the LHC Run
I.  Even though the latter is still not very accurate with the current
integrated luminosity, we will see that it already provides useful information.
In addition, Run II measurements for both processes at 13 TeV have started to
appear, and we will also include them.  The simple fit we will perform is
however only for illustrative purposes, and therefore a number of
simplifications and approximations are applied. The correlations of errors
between different measurements due to common sources of uncertainties 
are ignored; uncertainties are always
symmetrised by shifting the central values, and Gaussian distribution is
assumed; theoretical uncertainties are included only for the production
process. While we have shown results only for 8 TeV, we have checked that
signal strengths at 7 TeV are identical, therefore we can use the measurements
on signal strengths reported by experimental collaborations, where 7/8 TeV data
are often combined.  We also neglect the uncertainties due to missing
dimension-eight and higher operators.  These errors are estimated to be
of order $[(2m_t+m_H)/\Lambda]^2$.  The reliability of our results will thus
depend on $\Lambda$, i.e.~the scale of new physics, and can only be assessed
as a function of $\Lambda$.

The $pp\to H$ measurements we are going use are the following: in the diphoton
channel we use \cite{Aad:2014eha,Khachatryan:2014ira} at 8 TeV and
\cite{CMS-PAS-HIG-15-005} at 13 TeV; in $WW/ZZ/\tau\tau$ channels we use
\cite{ATLAS:2014aga,Chatrchyan:2013iaa,Aad:2014eva,Chatrchyan:2013mxa,Aad:2015vsa,Chatrchyan:2014nva};
Measurements of $pp\to t\bar tH$ at 8 TeV in the diphoton channel are given in
\cite{Khachatryan:2014qaa,Aad:2014lma}, while in the multilepton and $b\bar b$
channels we use \cite{Khachatryan:2014qaa,Aad:2015iha,Aad:2015gra}.  Finally,
13 TeV measurements of $t\bar tH$ in the multilepton and $b\bar b$ channels are
also included \cite{CMS-PAS-HIG-15-008,CMS-PAS-HIG-16-004}.  Both processes depend on all
three operators.

\begin{table}[htb]
\begin{centering}
\begin{tabular}{ccccccccc}
	\hline\hline
$r_{t\phi}$
&$r_{\phi G}$
&$r_{tG}$
&$r_{t\phi,t\phi}$
&$r_{\phi G,\phi G}$
&$r_{tG,tG}$
&$r_{t\phi,\phi G}$
&$r_{t\phi,tG}$
&$r_{\phi G,tG}$
\\\hline
-0.119 & 73.4 & 0.676 & 0.00354 & 1348 & 0.114 & -4.37 & -0.0402 & 24.8
	\\\hline\hline
\end{tabular}
\caption{\label{width}	Ratio of partial Higgs width into $gg$ over the SM value,
defined as:
$\Gamma(gg)/\Gamma_{SM}(gg)\equiv
1+\sum_i\frac{1{\rm TeV}^2}{\Lambda^2}C_i
r_i +\sum_{i\leq j} \frac{1{\rm TeV}^4}{\Lambda^4}
	C_iC_jr_{ij}$.}
\end{centering}
\end{table}

The Higgs branching ratios are affected by the operators in different ways.
As an example we show in Table~\ref{width} how the Higgs partial decay width to
gluons changes for the relevant operators. In addition to $H\to gg$, 
$O_{t\varphi}$ changes the partial width of the $H\to\gamma\gamma$ decay in the following way:
\begin{equation}
\frac{\Gamma(\gamma\gamma)}{\Gamma_{SM}(\gamma\gamma)}= \left|1-0.0595C_{t \phi}\frac{1{\rm TeV}^2} {\Lambda^2} \frac{A_t}{A_t+A_W}\right|^2 =\left|1+0.0335\frac{1{\rm TeV}^2} {\Lambda^2} C_{\phi t}+0.000281 \frac{1{\rm TeV}^4}{\Lambda^4} C^2_{\phi t}\right|, 
\end{equation} where $A_t$ and $A_W$ are the top- and $W-$boson loop amplitudes entering the Higgs decay into photons. We note that the impact of $O_{t\varphi}$ is diluted in $H \to \gamma \gamma$ as this decay is dominated 
by the $W-$boson loop. All branching ratios, including those of $WW/ZZ/\tau\tau$, 
are also affected due to changes in the total width from $H\to gg$.
We include these effects at LO only (tree-level for $O_{\phi G}$, one-loop for
$O_{t \phi}$ and $O_{tG}$). For this reason different decay channels need to be
considered separately. Because the measurements are based on signal strengths,
defined as the ratio of deviation in cross sections to the SM prediction, we
prefer to have same order predictions for both the SM and operator
contributions.  For $t\bar tH$ we use our NLO predictions, while for $pp\to H$
and Higgs decay we only use LO predictions, as not all loop-induced
contributions are known at NLO.  Both production and decay rates are included
up to order $C^2/\Lambda^4$.  A $\chi^2$-fit is performed to derive the
limits. All coefficients in this section are defined with $\mu_{EFT}=m_t$
unless specifically mentioned, and all results given in this section correspond
to $95\%$ confidence level.

Top-pair production is not included in the fit.  We assume that this
degree of freedom will be used to constrain the four-fermion operators.  While
a global fit including both sectors is the only consistent way to extract
information on the dimension-six operators, this is beyond the scope of this
paper.

\begin{table}
	\centering
\begin{tabular}{cccc}
	\hline\hline
	& Individual & Marginalised  & $C_{tG}$ fixed
	\\\hline
	$C_{t\phi}/\Lambda^2\ [\mbox{TeV}^{-2}]$ & [-3.9,4.0] & [-14,31] & [-12,20]
	\\
	$C_{\phi G}/\Lambda^2\ [\mbox{TeV}^{-2}]$ & [-0.0072,-0.0063] & [-0.021,0.054] & [-0.022,0.031]
	\\
	$C_{tG}/\Lambda^2\ [\mbox{TeV}^{-2}]$ & [-0.68,0.62] & [-1.8,1.6]
	\\\hline\hline
\end{tabular}
\caption{Constraints on $C/\Lambda^2$ from the simplified fit.
	In the first column, only one operator is allowed at a time.
	In the second column, all operator coefficients are allowed to float.
	In the third column, $C_{tG}$ is set to zero while the other two
	coefficients are floated.
	\label{tab:limits}}
\end{table}

Current limits at 95\% confidence level are given in Table~\ref{tab:limits}.
The most constrained operator is $C_{\phi G}$, as it gives a tree-level
contribution to Higgs production. Individual limits (i.e.~setting other
coefficients to zero) and marginalised ones (i.e.~floating other coefficients)
are given in the first two columns.  Interestingly, the $C_{tG}$ limit is
already comparable to its current limit from $t\bar t$ production only
(assuming no four-fermion operator contributes). This is because the $t\bar tH$
cross section is more sensitive to the $O_{tG}$ operator due to the higher
partonic energies probed, and in addition the squared contribution from $C_{tG}$
given the current limits is not negligible.  Even though the current limit
from $t\bar tH$ is still weaker, given that the $t\bar tH$ measurement still has a
lot of room to improve, it will become more competitive in the near future.  In
fact, assuming $10\%$ uncertainty on $t\bar tH$ and $4\%$ uncertainty on $pp\to
H$ for 14 TeV 3000 fb$^{-1}$ \cite{atlasprojection}, we find
$-0.12<C_{tG}<0.12$ and $-1.0<C_{tG}<1.1$ ($\Lambda=$1 TeV) respectively for individual and
marginalised limits.  On the other hand, assuming a $5\%$ precision for $t\bar
t$ production at 14 TeV, the individual limit on $C_{tG}$ is
$-0.33<C_{tG}<0.33$ ($\Lambda=$1 TeV) \cite{Franzosi:2015osa}, and a factor of a few is expected
once marginalised over the four-fermion operators.

In the third column of Table~\ref{tab:limits} we show limits obtained by assuming only $C_{tG}=0$ but
floating the other two coefficients.  $C_{tG}=0$ is typically assumed in Higgs
operator analyses.  By comparing the last two columns in the table, we can
see how much more room is allowed once this operator is included.

We should also point out that, given the cross sections in Table~\ref{tab:tth1}
and the limits in Table~\ref{tab:limits}, in $t\bar tH$ production
the squared contribution from $O_{\phi G}$ is negligible, but that from the
other two operators cannot be neglected. 

	\begin{figure}[h!tb]
		\begin{center}
			\includegraphics[width=.9\linewidth]{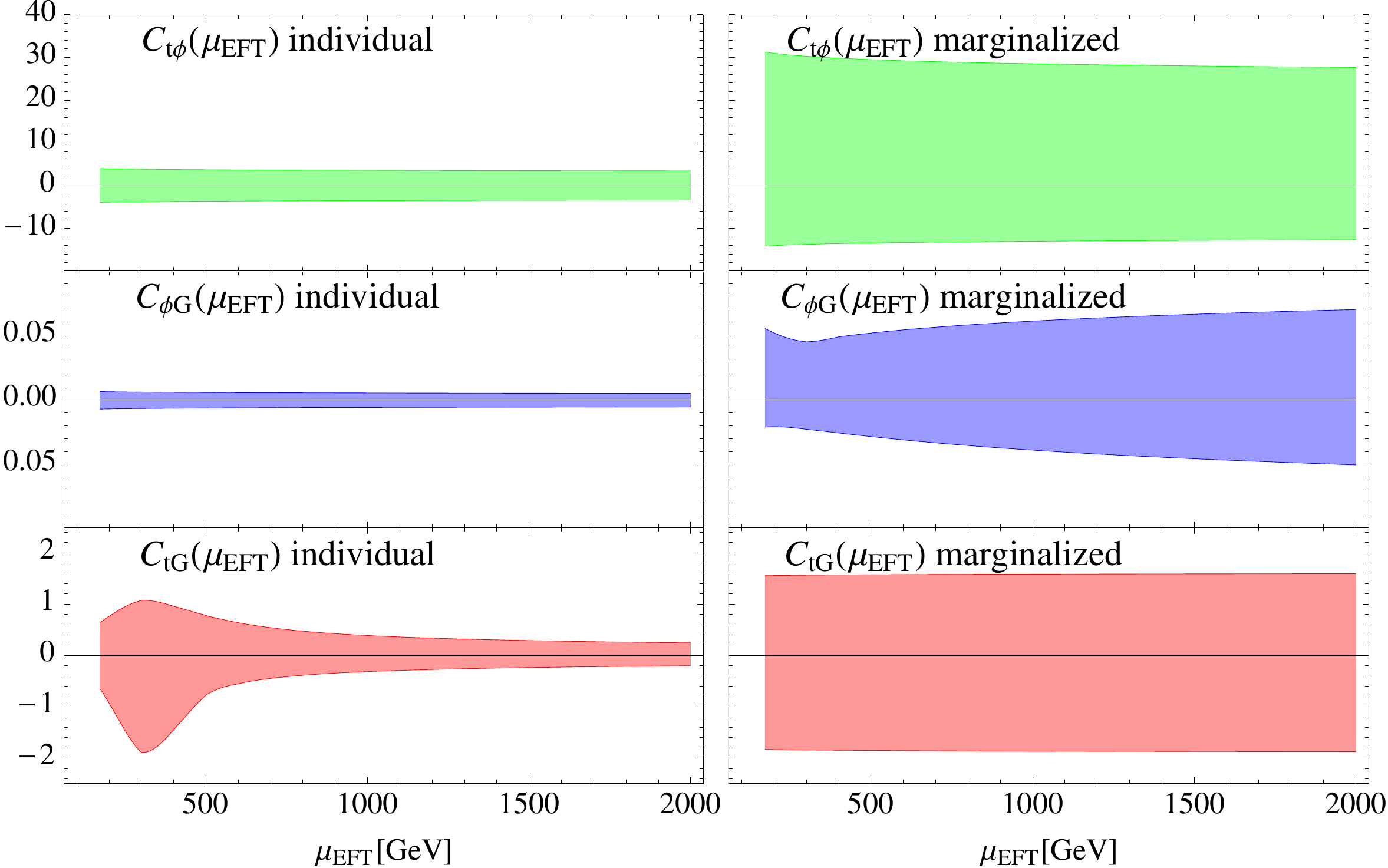}
		\end{center}
		\caption{Individual (left) and marginalised (right) limits on
			the three operator coefficients, as functions of
			$\mu_{EFT}$. $\Lambda=1$ TeV is assumed.}
		\label{fig:limits}
	\end{figure}

As we have mentioned in the previous section, limits on operator coefficients
can be sensitive to the EFT scale.  In Figure~\ref{fig:limits} we plot the
individual and marginalised bounds on the three operator coefficients as in Table \ref{tab:limits}, 
but this time as a function of $\mu_{EFT}$.  We can see that the individual
bound on $C_{tG}$
has a large dependence on scales, and does not provide valuable information.
This is because the bounds are derived by assuming $C(\mu_{EFT})=0$ for the other
two operators, which is a scale-dependent assumption.  The marginalised
bounds are more stable because they are independent of such assumptions.
	\begin{figure}[tb]
		\begin{center}
			\includegraphics[width=.48\linewidth]{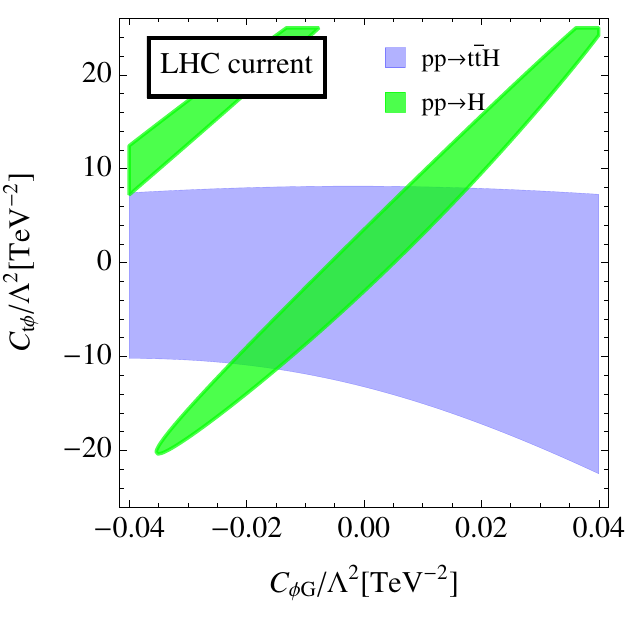}
			\includegraphics[width=.48\linewidth]{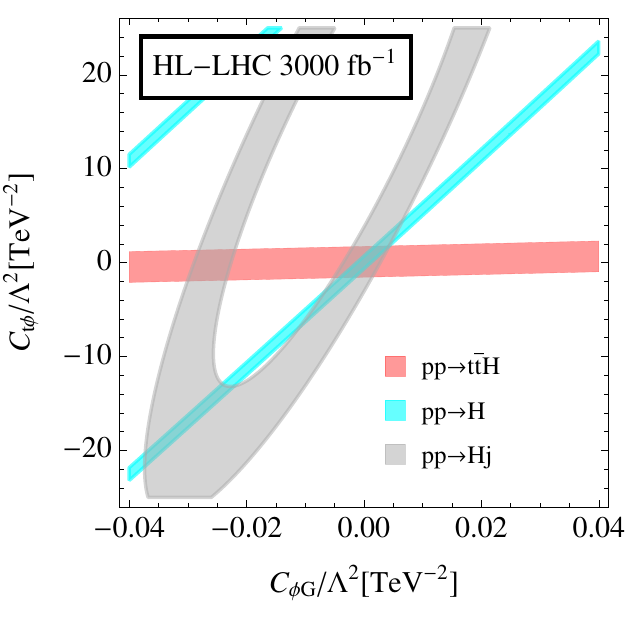}
		\end{center}
		\caption{Allowed region in $O_{t \phi}$-$O_{\phi G}$ plane at 95\% confidence level.
		Left: current constraints.  Right: future projection at HL-LHC.}
		\label{fig:tffg}
	\end{figure}
	
To better investigate the allowed region in the parameter space,
in the following we consider scenarios where two operators are allowed
at a time.  Particularly interesting is the degeneracy of the three operators
in $gg\to H$.  Through a top-quark loop, both $O_{t\phi}$ and $O_{tG}$ can
induce a $ggH$ vertex, leading to a degeneracy with the tree-level contribution
of $O_{\phi G}$.  $t\bar tH$ production is expected to break this degeneracy.
To illustrate this effect, we make plots to compare the constraints from the two
processes, each time allowing two operators to be nonzero.

We first consider the degeneracy between $O_{t\phi}$ and $O_{\phi G}$.  In
Figure~\ref{fig:tffg} left we show limits obtained from $pp\to H$ and $pp\to
t\bar tH$ separately.  We can see the flat direction in the $pp\to H$
measurement is already slightly lifted.  This is due to including branching
ratios of the diphoton and other channels, which have different dependence on
$O_{t\phi}$ and $O_{\varphi G}$.  On the other hand, the current measurement of
$t\bar tH$ cross section gives constraints in the orthogonal direction.  Even
though the precision is still not comparable to the Higgs cross section
measurement, improvements can be expected in the future.  For illustration, in
Figure~\ref{fig:tffg} right we show the 14 TeV projections for 3000 fb$^{-1}$.
For simplicity we only consider the production processes.  Estimated
experimental uncertainties on both $pp\to t\bar tH$ and $pp\to H$ are taken
from \cite{atlasprojection}.  Theoretical uncertainties are not included.

Another useful process to break this degeneracy is $p p \to Hj$ with a boosted
Higgs, as suggested in Ref.~\cite{Grojean:2013nya}.  For this reason we also
include this process in Figure~\ref{fig:tffg} right.  As an estimation for
future precision, Ref.~\cite{Grojean:2013nya} considered a $p_T(j)$ cut of 650
GeV and 10\% uncertainty on the measurement.  A large $p_T(j)$ significantly
reduces the cross section, and to be more conservative, here we consider
$p_T(j)>500$ GeV and assign a 20\% uncertainty on the measurement.  We can see
that the limit range from this measurement does cross the $pp\to H$ region as
expected, so there is some discriminating power.  The direction is however not
very ``orthogonal'', and so the discriminating power with our assumption is not
as good as $t\bar tH$, even though a more detailed analysis would be needed to
draw a final conclusion.

	\begin{figure}[h!tb]
		\begin{center}
			\includegraphics[width=.48\linewidth]{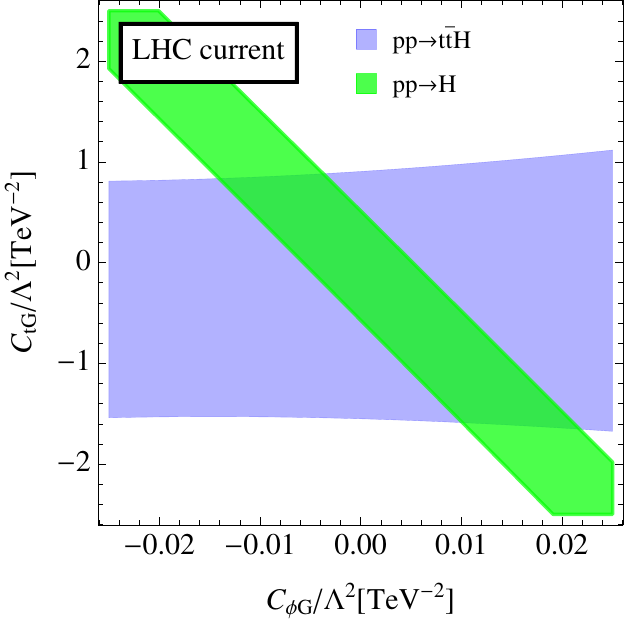}
			\includegraphics[width=.48\linewidth]{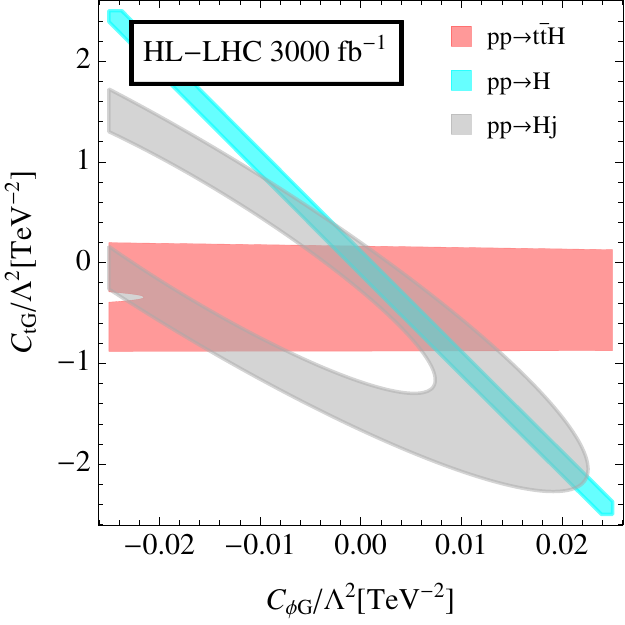}
		\end{center}
		\caption{Allowed region in $O_{tG}$-$O_{\phi G}$ plane at 95\%
	confidence level. Left: current constraints.  Right: future projection
at HL-LHC.} \label{fig:fgtg1}
	\end{figure}
	\begin{figure}[h!tb]
		\begin{center}
			\includegraphics[width=.48\linewidth]{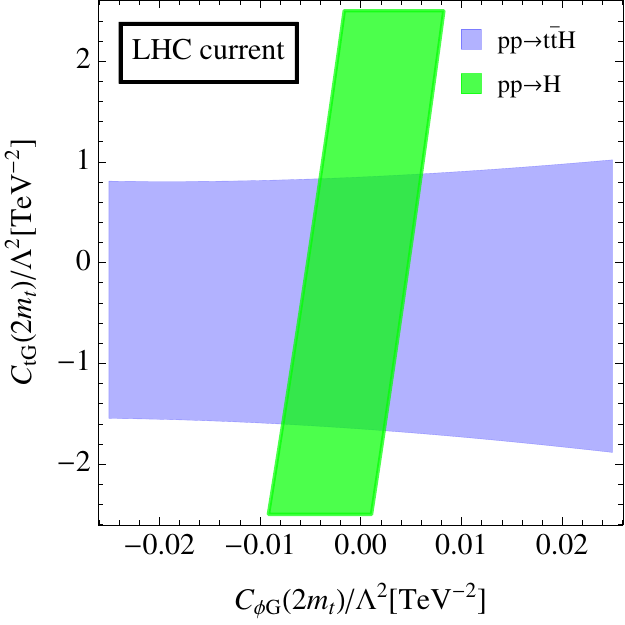}
			\includegraphics[width=.48\linewidth]{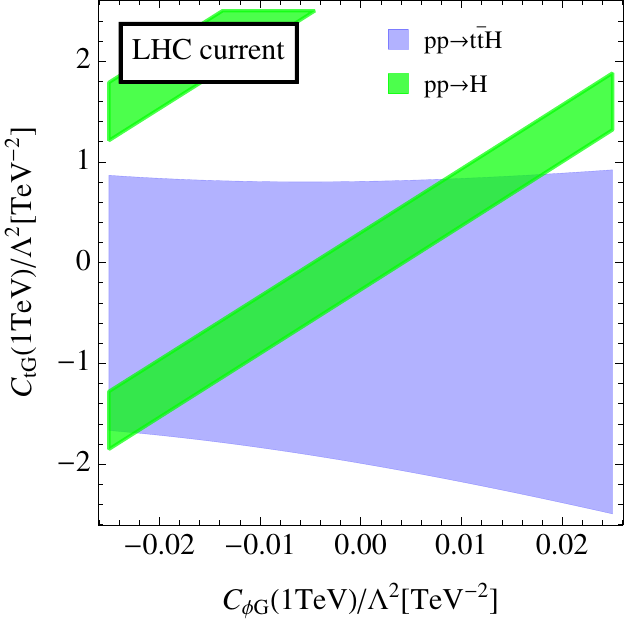}
		\end{center}
		\caption{Allowed region in $O_{tG}$-$O_{\phi G}$ plane at 95\% confidence level,
		setting $\mu_{EFT}=2m_t$ (left) and $\mu_{EFT}=1$ TeV (right).}
		\label{fig:fgtg2}
	\end{figure}

The degeneracy between $O_{tG}$ and $O_{\phi G}$, i.e.~$ttg$ and $ggH$
vertices, has been less considered in the literature \cite{Degrande:2012gr}, mainly
because $O_{tG}$ is expected to be constrained from $t\bar t$ production.
However as we have mentioned, considering future measurements, $t\bar tH$ will become
sensitive to $O_{tG}$ in its limited range, and a more reasonable strategy
could be to use $t\bar t$ to constrain four-fermion operators while leaving
$O_{tG}$ to $t\bar tH$ production.  For this reason, we plot in
Figure~\ref{fig:fgtg1} left the current limits from $pp\to H$ and $pp\to t\bar tH$.
Including Higgs decays will not lift the degeneracy in this case, but we see that
$t\bar tH$ production already gives a useful bound on $O_{tG}$.  Projections for 14
TeV 3000 fb$^{-1}$ are shown in Figure~\ref{fig:fgtg1} right.  The expected limit
on $C_{tG}$ is improved.  We can also see that $ p p \to Hj$ does not provide additional 
information in resolving $O_{tG}$ and $O_{\phi G}$.

Unlike $O_{t\phi}$, the contributions from $O_{tG}$ and $O_{\phi G}$ are sensitive
to the scale $\mu_{EFT}$, because the top loop with an $O_{tG}$ insertion
is divergent and requires a counterterm from $O_{\phi G}$.  As a result,
the dependence of $C_{\phi G}$ on $\mu_{EFT}$ due to mixing from $C_{tG}$
is expected to be canceled by the $\mu_{EFT}$ dependence in the loop.
When  the scale $\mu_{EFT}$ is changed, the change in total cross section is only
a higher order effect, but the contours in the $C_{tG}-C_{\phi G}$ plot will be
very different, as we can see by comparing with the 95\% contours in
Figure~\ref{fig:fgtg2} at two different $\mu_{EFT}$ scales, $2m_t$ and 1 TeV.
It is therefore very important to give the value of $\mu_{EFT}$ when presenting
bounds on operators.  As we can see by comparing the left plots of
Figures~\ref{fig:fgtg1} and \ref{fig:fgtg2}, even a change by a factor of two
can lead to significant difference.  In particular if $C_{\phi G}$ is set to 0
to derive the ``individual bound'' on $C_{tG}$, the result can be very
sensitive to the scale.  This is also reflected in Figure~\ref{fig:limits}.  We
conclude that the ``individual bound'' in this case does not provide useful
information.

	\begin{figure}[h!tb]
		\begin{center}
			\includegraphics[width=.48\linewidth]{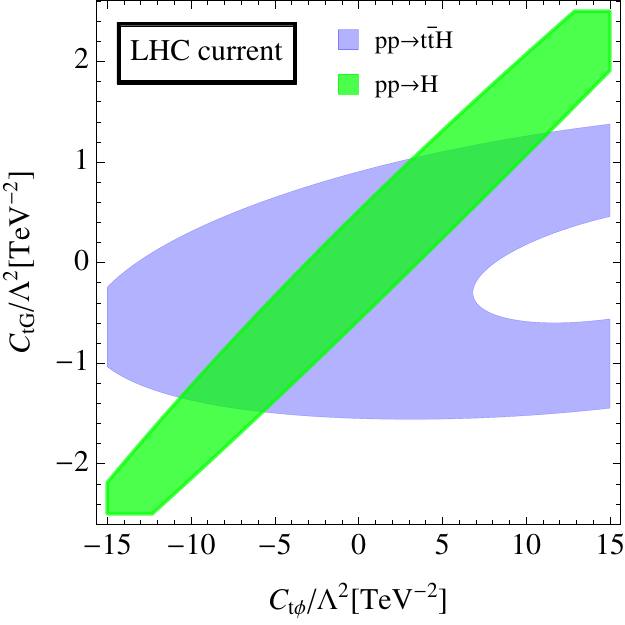}
			\includegraphics[width=.48\linewidth]{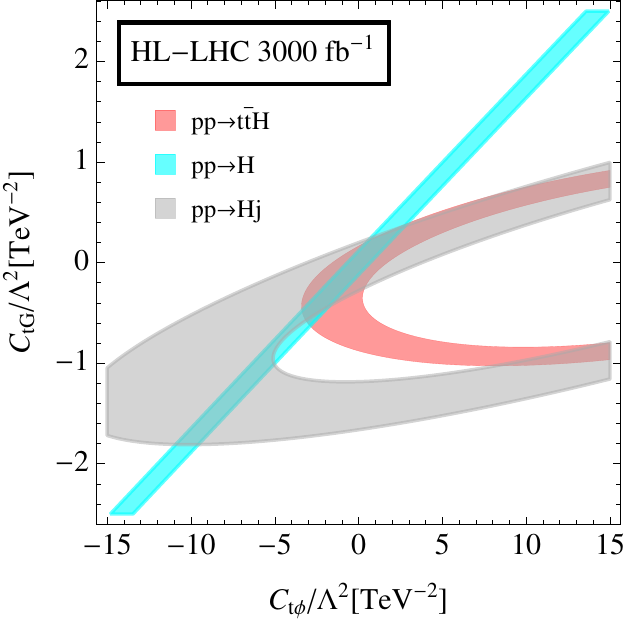}
		\end{center}
		\caption{Allowed region in $O_{tG}$-$O_{t \phi }$ plane
		at 95\% confidence level.  Left: current constraints.  
		Right: future projection at HL-LHC.}
		\label{fig:tftg}
	\end{figure}

Finally in figure~\ref{fig:tftg} we show a two operator fit for $O_{t\phi}$
and $O_{tG}$, both giving loop-induced contribution in $pp\to H$.
Figure~\ref{fig:tftg} left shows how the current $t\bar tH$ measurements help to
constrain both operators, while Figure~\ref{fig:tftg} right is the situation
for future LHC, where one can see that $t\bar tH$ and $Hj$ are equally
good in resolving the degeneracy between the two loop-induced contributions.

\section{Summary and conclusions}
\label{sec:conc}

We have presented the QCD NLO predictions for $t\bar tH$ production,
as well as results for several other loop-induced processes, $H$, $Hj$, $HH$
production, all in the SMEFT.  We have focused on the Yukawa operator, $O_{t \phi}$,
 the operator describing the interaction of gluons with the Higgs, $O_{\phi G}$, and the 
 chromomagnetic dipole moment $O_{tG}$.
Our predictions improve our understanding
of the patterns of deviations from the SM in both the top-quark and
the Higgs-boson sectors, and will play an important role in future measurements
of these two sectors.

We have shown that the QCD corrections to the $t\bar tH$ improve both the
accuracy and the precision, and in many cases lead to nontrivial modification
of the distributions of relevant observables.  $K$ factors for the inclusive cross
sections range roughly from 1 to 1.6, depending on the effective operator.
Moreover, the differential $K$ factors are not a constant and their shapes
also depend on the operator.  Using the SM $K$ factor to rescale the operator
contribution may not be a good approximation.  We have also shown that the full
NLO corrections in $t\bar tH$ can be much larger than only the log enhanced
terms which are captured by RG equations, so the RG-improved prediction
is not a good approximation to the full NLO prediction.

To assess the sensitivity of the various Higgs production mechanisms to the 
dimension-six operators, we have performed a toy fit for the three effective 
operators $O_{t\phi}$, $O_{\phi G}$ and $O_{tG}$, using $t\bar tH$ 
production and $pp\to H$.  The projection at
HL-LHC is also discussed, together with $p p \to Hj$.  We find that
the current limit on $C_{tG}$ is already comparable, even though still weaker,
to the limit obtained directly from $t\bar t$ production, and that in the
future $t\bar tH$ and $pp\to H$ could be a better approach to set limits on this
coupling. This implies that the Higgs measurements have started to become
sensitive to the chromo-dipole coupling of the top, and so this should be
included in future Higgs analysis.  Furthermore, the ratio between $t\bar tH$
and $t\bar t$ measurements is useful in decoupling the impact of potential
four-fermion operator contributions.  We have also shown how the $t\bar tH$
and $Hj$ production processes can be used to resolve the degeneracy between the
three operators in the $pp\to H$ measurement.  

We have further discussed the EFT scale dependence, that is, the scale at which
the EFT is defined, in an operator analysis.  For example we show that
individual bounds could have a strong dependence on this scale, and so should
always be interpreted with care.  This is because loop-induced processes can be
sensitive to the EFT scale even at LO, and the dependence is supposed to be
canceled by RG mixing effects.  In addition we have defined a way to estimate
the scale uncertainty induced by this additional scale, which properly takes
into account the RG mixing and running, and showed that with our NLO results
this uncertainty can be put under control.  The EFT predictions should always
be presented with this uncertainty. 

In summary, at NLO in QCD accuracy deviations from the SM in the top and the
Higgs sectors can be extracted with improved accuracy and precision, allowing
for more reliable global analyses based on the EFT approach.  Similar to our
previous works, these results are performed with the {\sc MG5\_aMC}
framework, and predictions matched to the PS are provided
in an automatic way.  For this reason NLO+PS event generation can be directly
used in a realistic simulation, and by investigating the features of potential
deviations from SM, sensitivities to EFT signals will possibly be improved with
advanced experimental techniques.

\acknowledgments
We are grateful to the LHCHXSWG for always providing us with such a stimulating
environment.  We acknowledge many illuminating discussions on EFT at NLO with
Celine Degrande, Christophe Grojean and Giampiero Passarino.  This work has
been performed in the framework of the ERC Grant No.  291377 ``LHCTheory'' and
has been supported in part by the European Union as part of the FP7 Marie Curie
Initial Training Network MCnetITN (PITN-GA-2012-315877).  C.Z.~is supported by
the United States Department of Energy under Grant Contracts DE-SC0012704.


\bibliographystyle{JHEP}
\bibliography{bib}

\end{document}